# A universal waveguide mass–energy relation for lossy one-dimensional waves in nature

Huayang Cai[1,2,3] and Bishuang Chen[4]*

[1] Institute of Estuarine and Coastal Research, School of Ocean Engineering and Technology, Sun Yat-Sen University / Southern Marine Science and Engineering Guangdong Laboratory (Zhuhai), Zhuhai, Guangdong 519082, China.

[2] State and Local Joint Engineering Laboratory of Estuarine Hydraulic Technology / Guangdong Provincial Engineering Research Center of Coasts, Islands and Reefs / Guangdong Provincial Key Laboratory of Marine Resources and Coastal Engineering / Guangdong Provincial Key Laboratory of Information Technology for Deep Water Acoustics / Key Laboratory of Comprehensive Observation of Polar Environment (Sun Yat-Sen University), Ministry of Education, Zhuhai, Guangdong 519082, China.

[3] Zhuhai Research Center, Hanjiang National Laboratory, Zhuhai, Guangdong 519082, China.

[4] School of Marine Sciences, Sun Yat-Sen University, Zhuhai 519080, China.

*Corresponding Author

Bishuang Chen: chenbsh23@mail.sysu.edu.cn

**Abstract**

Finite, lossy waveguides are ubiquitous: distributed attenuation combined with partial reflections produces feedback, resonance, delays and decay across electromagnetic, acoustic, photonic, quantum transport and electrochemical interfaces. However, a long-standing gap is that the standard tools, i.e., impedance transformations, scattering descriptions and weak-loss resonator approximations, do not yield low-dimensional invariants that remain predictive under intrinsic asymmetry and realistic boundaries, nor do they cleanly separate total absorption from the useful power that is ultimately delivered to a load. Here we establish a unified mass–energy framework for linear, single-mode, one-dimensional systems, whereby the resulting energy-like and power-flow variables $(\mathcal{U}, \mathcal{S})$ satisfy the universal invariant $\mathcal{U}^2 - \mathcal{S}^2 = |\Gamma_g|^2$, with the effective standing-wave "mass" $|\Gamma_g|$ becoming state-dependent under asymmetry. Central to this framework is the Cai–Smith chart, a bounded state-space visualization that directly maps stability, feedback proximity, and operating states, while exposing the divergence between maximum absorption and useful power delivery under loss. We further derive four fundamental laws governing power absorption and emission in such waveguides. The same invariant description is validated in multiport optics via coherent perfect absorption at exceptional points, where a basis-independent singular-value-decomposition criterion is established, reconciling quadratic and quartic near-zero scaling. Finally, we map electrochemical polarization onto this geometry by treating current density as a power-flow observable, extracting $|\Gamma_g|$ from two-mode orthogonal fits, and revealing a universal transition from storage- to transfer-dominated regimes across pH. This framework transforms waveguide theory into a unified design language for dissipative, boundary-controlled systems, applicable from photonics to electrochemistry and beyond.

## 1. Introduction

The propagation of waves along guided structures is a central mechanism for transporting energy and information across physics, engineering, biology and geoscience: electromagnetic signals travel along cables, packages and on-chip interconnects; acoustic waves are shaped by ducts and resonators; optical pulses are confined by fibres and integrated waveguides; and multiple reflections and evanescent decay govern the transport process in layered potentials and engineered barriers. Despite their disparate implementations, these systems share a one-dimensional distributed description in which phase accumulation and dissipation compete under partial reflection at finite boundaries, resulting in resonance, delays, filtering and power conversion. Transmission-line theory provides a canonical language for such dynamics and extends naturally to multiconductor and coupled structures[1]; for finite devices, the boundary conditions are summarized through impedance transformations and reflections on the complex plane, which are crystallized by Smith's transmission-line calculator and the Smith chart[2,3]. However, these tools are typically deployed as frequency-domain methods that are tied to specific circuit contexts rather than as a domain-spanning organizing principle that makes energy storage, dissipation and finite-length feedback constraints explicit.

The need for compact, transferable design rules has become sharper as guided-wave platforms proliferate: programmable and integrated photonics for AI and neuromorphic computing[4-8], wireless systems above 100 GHz where material losses and scattering are central[9-11], and nonreciprocal thermal/electromagnetic platforms whose functions are inseparable from direction-dependent transport and asymmetric attenuation[12]. Across these settings, the same structural questions recur: how much resonance is compatible with the distributed loss, which boundary combinations maximize the absorbed power, and how does directionality reshape feedback and decay? Closely related constraints appear in wave analogies to quantum transport, including transmission-line treatments of resonant tunnelling[13] and the connection between evanescent wave propagation and tunnelling time delays[14]. Superconducting circuits provide a sharp laboratory where quantization and escape dynamics in Josephson junctions expose the interplay among resonance, dissipation and macroscopic tunnelling in a single guided wave-like degree of freedom[15,16]. Electrochemical systems offer complementary arenas where boundary-controlled interfacial kinetics are directly measurable: porous-electrode theory[17], impedance models for porous structures[18], modern impedance spectroscopy[19], and polarization curves exhibiting strong electrolyte-dependent asymmetries in hydrogen electrocatalyses[20,21]. Beyond electrochemistry, analogous distributed descriptions are standard in neuronal cable theory[22] and its transmission-line parallels[23-26], acoustics in ducts and resonators[27-29], arterial pulse-wave models[30-32], Liénard-type reductions for dissipative transmission-line dynamics[33-35], wave-like heat conduction processes under Maxwell–Cattaneo relaxation[36,37], and quasicrystal dynamics combining phonon wave equations with phason telegraph-type equations[38,39], highlighting the broad scope of the underlying problem.

A crucial missing component is a low-dimensional, invariant description of finite, lossy waveguides that remains applicable under intrinsic asymmetry and realistic boundary conditions. The existing approaches often rely on weak-loss approximations, use scattering matrices without yielding simple constraints, or require domain-specific modelling schemes that obscure universal principles. More specifically, they do not simultaneously (i) provide closed-form feasibility and performance bounds under distributed losses and passivity; (ii) separate the objectives of maximum absorption and maximum useful delivery in a way that remains predictive for finite devices; and (iii) unify steady-state resonance, decay dynamics and dispersive delays within a single state space.

Here, we develop a geometry-first framework in which a boundary-composed generalized reflection coordinate and a single intrinsic asymmetry parameter serve as state variables, enabling resonance, decay and optimal power-transfer limits to be expressed in a unified language across electromagnetic, acoustic, optical, quantum and electrochemical waveguides. Building on Miller's modal radiation framework[12], we further derive four fundamental laws that govern power absorption and emissions in finite, lossy waveguides. In addition to one-port matching and finite-length feedback limits, we show that the same invariant geometry extends to genuinely multiport scattering: coherent perfect absorption at exceptional points can be formulated and classified using a basis-independent singular value decomposition (SVD) metric and a waveguide-equivalent single-loop feedback representation, turning protocol dependence into a predictable and experimentally controllable effect. This waveguide-invariant framework bridges a long-standing conceptual gap, as it enables the application of resonance-aware, waveguide-inspired matching principles to systematic electrochemical performance analysis and optimization processes. These formulations parallel mass–energy relations[40] while being written entirely in measurable guided-wave quantities and are designed to support both physical interpretation and optimization processes in experimentally relevant finite systems.

## 2 Results and Discussion

### 2.1 Universal waveguide mass–energy relation and Cai–Smith geometry

The core of our framework is illustrated in **Fig. 1**. For a given angular frequency $\omega$ (and the corresponding dimensionless frequency $\Omega$), any finite, linear, single-mode one-dimensional guide of length $L$ can be described by a complex propagation constant $\gamma(\omega)$ and the associated complex electrical length $K(\Omega) = \gamma(\omega)L = R(\Omega) + i\Phi(\Omega)$, where $R$ and $\Phi$ are the accumulated attenuation and phase, respectively. We parameterize the position by a normalized axial coordinate $s \in [0,1]$ and define the dimensionless axial state $\zeta = sK(\Omega)$. The boundary reflections at the source and load are denoted by $\Gamma_S(\Omega)$ and $\Gamma_L(\Omega)$, respectively, and they are combined into a single boundary-composed generalized reflection coefficient through the following bilinear (Möbius) map:

$$\bar{\Gamma}_g(\Omega) = \frac{\Gamma_L(\Omega)e^{-2K(\Omega)} + \Gamma_S(\Omega)}{1 + \Gamma_S(\Omega)\Gamma_L(\Omega)e^{-2K(\Omega)}}. \tag{1}$$

Under strictly lossy propagation and passive terminals, this composition is contractive, and the admissibility condition $|\bar{\Gamma}_g(\Omega)| < 1$ holds (see Supplementary Method S2). From the normalized forwards/backwards fields, we construct a dimensionless energy-like density $\mathcal{U}(\zeta)$ and a power-flow-like density $\mathcal{S}(\zeta)$ and obtain the universal invariant

$$\mathcal{U}^2(\zeta) - \mathcal{S}^2(\zeta) = |\Gamma_g(\zeta)|^2. \tag{2}$$

This invariant, which can be equivalently stated as $\mathcal{U}^2 = \mathcal{S}^2 + |\Gamma_g|^2$, organizes all reachable states as "mass shells" in the $(\mathcal{S}, \mathcal{U})$ plane (**Fig. 1a**), with the effective standing-wave "mass" given by $|\Gamma_g(\zeta)|$. **Equation (2)** is formally identical to Einstein's relativistic mass–energy relation[40] $E_{tot}^2 = (Pc)^2 + (m_0c^2)^2$, where $E_{tot}$ denotes the total energy, $P$ is the momentum, $c$ represents the speed of light, and $m_0$ signifies the remaining mass; this correspondence holds exactly within the single-mode, lossy, boundary-controlled waveguide reduction scheme. The intrinsic asymmetry is captured by a single coefficient $\xi \in [0,1]$ that promotes the mass term to an axial state variable via radial rescaling (see Supplementary Methods S2 and S3),

$$\Gamma_g(\zeta) = \hat{\Gamma}_g(\Omega)\exp\left[-2(2\xi - 1)\Re(\zeta)\right]. \tag{3}$$

Here, $\xi = 1/2$ recovers the symmetric reference (constant mass), while $\xi < 1/2$ and $\xi > 1/2$ yield systematic mass growth and mass decay processes along $\Re(\zeta)$, respectively. This deformation paradigm is visualized in **Fig. 1a** for $\xi = \{0.1, 0.5, 0.9\}$, where the exact shells shift as the effective mass decreases from $|\Gamma_g| \approx 0.73$ to $0.53$ to $0.38$ under the same accumulated attenuation, while the dashed curves show the high-speed (low-mass) approximation $\mathcal{U} \approx |\mathcal{S}| + |\Gamma_g|^2/(2|\mathcal{S}|)$, which is accurate away from $\mathcal{S} = 0$ and quantify how a small standing-wave component corrects the travelling-wave limit (see Supplementary Method S4).

Because **Eqs. (1) – (3)** promotes $\Gamma_g$ to a minimal complex state coordinate whose magnitude directly sets the effective standing-wave "mass", it is natural to represent the full-state evolution process by tracking the admissible locus of $\Gamma_g$ in the complex plane under the passivity bound $|\Gamma_g| < 1$. In this representation, **Eq. (3)** becomes a geometry on the complex $\Gamma_g$ unit disk (**Fig. 1b**), which we term the Cai–Smith chart. This chart extends the traditional Smith chart[2,3] by incorporating the attenuation of the load, source, and propagation line while also accounting for asymmetry parameters, thus providing a more comprehensive representation of the system dynamics. The background colour represents the effective mass magnitude $|\Gamma_g|$, and the white dashed circles indicate constant-mass contours. For a representative boundary pair $(\Gamma_S, \Gamma_L) = (0.2, 0.6)$ and a phase sweep $K = R + i\Phi$ (here, $R = 0.10$), **Eq. (1)** traces a trajectory $\Gamma_g(\Omega)$ whose radius corresponds to the round-trip feedback strength (mass) and whose angle denotes the round-trip phase. Intrinsic asymmetry acts primarily as a radial deformation of this same trajectory by the factor $\exp[-2(2\xi - 1)R]$, expanding it towards the unit circle for $\xi < 1/2$ and contracting it towards the origin for $\xi > 1/2$. Therefore, proximity to the unit circle serves as an immediate visual indicator of strong energy recirculation, whereas asymmetry-driven mass decay pulls the trajectories inwards and suppresses the feedback-amplified energy storage. In combination, the two panels link energy flow states to a stability-constrained disk geometry, allowing the resonance proximity and asymmetry-induced deformation to be directly readable.

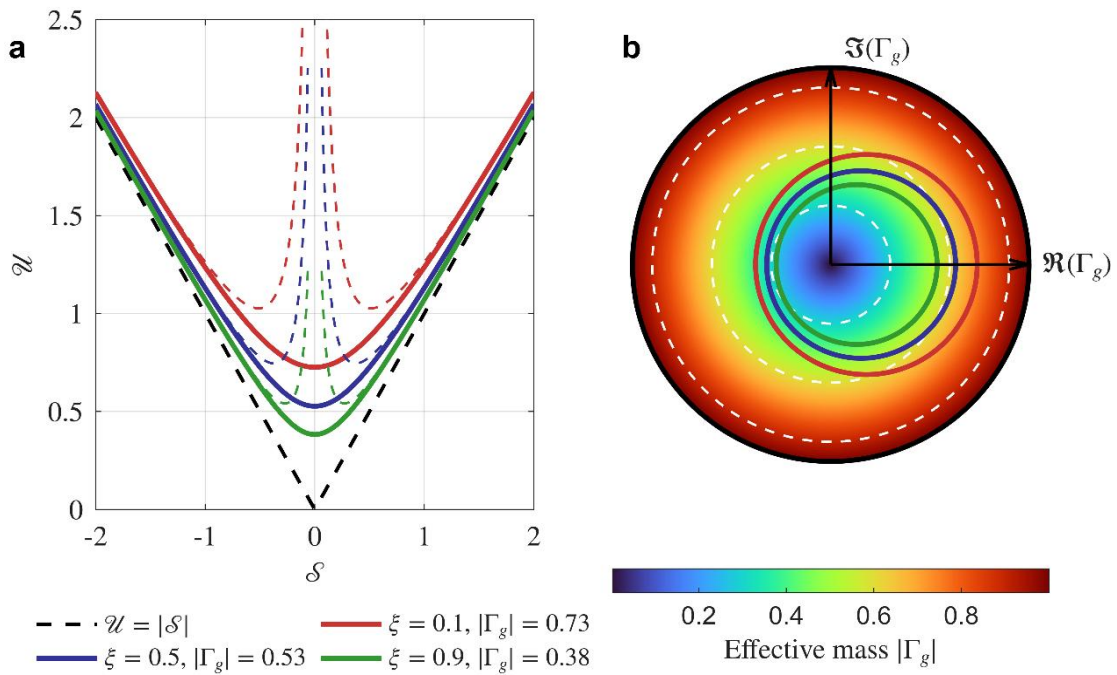

**Fig. 1 | Generalized waveguide mass–energy relation and Cai–Smith representation with intrinsic asymmetry. a.** State-dependent mass–energy shells given by $\mathcal{U}^2 = \mathcal{S}^2 + |\Gamma_g|^2$. The effective mass is modulated by the asymmetry coefficient $\xi$ through $|\Gamma_g(\zeta)| = \bar{\Gamma}_g(\Omega)\exp\{-2(2\xi - 1)R\}$ (here, $R = 0.2$), as shown for $\xi = 0.1, 0.5, 0.9$. The solid curves denote the exact relation, and the dashed curves show the high-speed approximation $\mathcal{U} \approx |\mathcal{S}| + |\Gamma_g|^2/(2|\mathcal{S}|)$; the black dashed line indicates the massless limit $\mathcal{U} = |\mathcal{S}|$ (i.e., $|\Gamma_g| \to 0$) as a reference bound. **b.** Cai–

Smith chart in the complex $\Gamma_g$ plane. The background colormap represents the effective mass magnitude $|\Gamma_g|$ over the unit disk, with dashed circles indicating constant $|\Gamma_g|$ levels. Coloured trajectories are generated from the boundary-composed reflection coefficient $\bar{\Gamma}_g(\Omega) = \frac{\Gamma_L e^{-2K} + \Gamma_S}{1 + \Gamma_S \Gamma_L e^{-2K}}$ using $\Gamma_S = 0.2$, $\Gamma_L = 0.6$, and $K = R + i\Phi$ with $R = 0.10$ and $\Phi \in [-\pi/2, \pi/2]$, followed by the use of asymmetry-induced radial rescaling to obtain $\Gamma_g(\zeta)$ for $\xi = 0.1, 0.5, 0.9$. The axes are suppressed for clarity; the arrows indicate $\Re(\Gamma_g)$ and $\Im(\Gamma_g)$.

## 2.2 Intrinsic asymmetry and state-dependent generalized reflection

Intrinsic asymmetry enters the theory through the complex electrical length $K(\Omega) = R(\Omega) + i\Phi(\Omega)$ and a single partition coefficient $\xi \in [0,1]$ that assigns direction-dependent attenuation to the forwards and backwards travelling components. For the resistance, inductance, conductance, capacitance (RLGC) class, a convenient dimensionless representation is obtained by writing the electrical length as $K(\Omega) = \sqrt{(\delta_R + i\Omega)(\delta_G + i\Omega)}$, where $\Omega$ is the dimensionless frequency and $(\delta_R, \delta_G)$ denotes dimensionless loss budgets (series-like and shunt-like budgets, respectively). The resulting dispersion classification, as summarized in **Extended Data Fig. 1**, shows that the attenuation $\alpha L = \Re[K(\Omega)]$ and phase $\beta L = \Im[K(\Omega)]$ exhibit a universal crossover effect at $\Omega_c = \sqrt{\delta_R \delta_G}$. At this crossover location, where $\Re[K(\Omega_c)] = \Im[K(\Omega_c)]$, the behaviour separates a dissipation-dominated regime from a phase-dominated regime, while the reference point $\Omega = 1$ provides a common normalization scheme across different systems. This classification system identifies when finite-length feedback is primarily controlled by exponential decay (large $\Re[K]$), by phase winding (large $\Im[K]$), or by their competition near $\Omega_c$ and therefore determines how strongly boundary reflections can recirculate energy for a given loss budget (see Supplementary Method S1).

The same asymmetry parameter $\xi$ promotes generalized reflection to a state-dependent quantity along the guide. Writing the normalized axial state as $\zeta = sK(\Omega)$ with $s \in [0,1]$, the boundary-composed generalized reflection $\Gamma_g$ acquires a systematic radial deformation along $\Re(\zeta)$ so that the effective "mass" magnitude $|\Gamma_g(\zeta)|$ increases for $\xi < 1/2$ and decreases for $\xi > 1/2$ (**Fig. 1 and Extended Data Fig. 2**). This deformation has an immediate physical signature in the spatial structures of instantaneous fields. For a fixed boundary composition $\rho = |\overline{\Gamma}_g|$ and fixed $(\delta_R, \delta_G, \Omega)$, the voltage and current profiles in an electrical-like field ($v(s,t)$ and $i(s,t)$, respectively) evolve over phase snapshots but remain confined within phase-independent reachable envelopes. The maxima and minima of these envelopes are set by the competing forwards-decay and backwards-amplification factors, which are controlled by $\xi$. In practice, increasing $\xi$ shifts the field energy towards the direction that is associated with stronger net decay, shrinking the reachable envelope and suppressing the standing-wave component, whereas decreasing $\xi$ expands the envelope and enhances the degree of feedback-driven contrast. These envelope constraints provide a direct, experimentally interpretable statement of "how the asymmetric loss reshapes what the guide can do" before any resonance condition is invoked (see Supplementary Method S3).

A key benefit of the state-dependent formulation is that it yields controlled approximations with universal error laws. In the low-mass (high-speed) regime quantified by $\chi = |\mathcal{S}|/|\Gamma_g|$, the mass–energy identity admits systematic expansions whose accuracy depends only on $\chi$ (not on a microscopic implementation). The relative errors of the key variables, including $\mathcal{U}$, the velocity-like ratio $\beta_w = \mathcal{S}/\mathcal{U}$, and the Lorentz factor-like quantity $\gamma_w = \mathcal{U}/|\Gamma_g|$, together with the rapidity (log-Doppler) error induced for $\ln D^+ = \text{atanh}(\beta_w)$, are shown in **Extended Data Fig. 3**. Both the first- and second-order truncations exhibit universal power-law error decay with $\chi$, providing quantitative thresholds (e.g., $\chi_{th} \gtrsim 10$) for when "low-mass" reductions are reliable and when the full state-dependent mass must be retained (see Supplementary Method S4).

Taken together, the asymmetry-controlled state dependence exposes a kinematic structure in $(\mathcal{U}, \mathcal{S})$ space that makes composition and multisection aggregation transparent. Because $\beta_w = \mathcal{S}/\mathcal{U}$ is bounded by $|\beta_w| \leq 1$ as a direct consequence of $\mathcal{U}^2 - \mathcal{S}^2 = |\Gamma_g|^2$, physically meaningful composition rules must respect this bound. As shown in **Extended Data Fig. 4**, Lorentz velocity addition enforces $|\beta_w| \leq 1$, whereas naive (Galilean) addition can violate it, motivating the rapidity parameterization $\beta_w = \tanh(\phi)$ in which cumulative effects are summed in a linear manner: $\phi_{\text{tot}} = \phi_1 + \phi_2 + \cdots$ (**Extended Data Figs. 4a and 4b**). For a multisection guide, the section-specific mass terms $|\Gamma_{g,k}|$ therefore map to section rapidities that sum to a cumulative $\beta_w^{\text{cum}}$ (**Extended Data Fig. 4c**), and the associated state transformations correspond to Lorentz boosts (hyperbolic rotations) that preserve the invariant hyperbolae $\mathcal{U}^2 - \mathcal{S}^2 = |\Gamma_g|^2$ (**Extended Data Fig. 4d**). Overall, **Extended Data Figs. 1-4** establish the fact that intrinsic asymmetry is not an ad hoc correction. Instead, it functions as a single control parameter that (i) classifies dispersion through $K(\Omega)$, (ii) determines the reachable spatial envelopes of finite-length fields, (iii) sets the universal approximation accuracy through $\chi = |\mathcal{S}|/|\Gamma_g|$, and (iv) induces consistent composition kinematics for concatenated lossy sections (see Supplementary Method S5).

**2.3 Resonance, critical coupling, useful power and modal power constraints**

The Cai–Smith chart becomes predictive once the boundary-composed generalized reflection coefficient $\Gamma_g$ is tied to the absorbed power through the same round–trip feedback factor that controls the finite-length recirculation process. At a fixed dimensionless frequency $\Omega$, we write the complex electrical length as $K(\Omega) = R(\Omega) + i\Phi(\Omega)$ and express the round-trip feedback in polar form:

$$F(\Omega) = \Gamma_S(\Omega)\Gamma_L(\Omega)e^{-2K(\Omega)} = \kappa(\Omega)e^{i\Theta(\Omega)}, \tag{4}$$

where $\kappa(\Omega) = |\Gamma_S\Gamma_L|e^{-2R}$ is the surviving feedback magnitude after undergoing distributed attenuation and $\Theta(\Omega) = \arg(\Gamma_S\Gamma_L) - 2\Phi$ is the round-trip phase detuning (the associated definitions and derivations are presented in Supplementary Method S6.1). This feedback factor compresses boundary reflections and propagation losses into two scalar parameters that directly predict resonance proximity, absorption limits, and power partitioning on the $(\kappa, \Theta)$ plane (**Fig. 2**; **Extended Data Figs. 5 and 6**).

The response is partitioned into universal interference classes by $\Theta(\Omega)$, with constructive build-up occurring near the odd-$\pi$ lines $\Theta \simeq (2m+1)\pi$ under the $D = 1 + F$ convention and recirculation suppression occurring near $\Theta \simeq 2m\pi$. This dependence is compactly captured by the feedback denominator. Writing the build-up (feedback proximity) as $\Lambda(\Omega) = 1/|1 + F(\Omega)|$, the loci of the constant $\Lambda$ in the $(\kappa, \Theta)$ plane obey $|1 + \kappa e^{i\Theta}|^2 = \Lambda^{-2}$, i.e., $1 + \kappa^2 + 2\kappa\cos\Theta = \Lambda^{-2}$, so that at a fixed $\kappa$, the maximum build-up occurs at $\Theta = \pi$ with $\Lambda_{\max} = 1/|1-\kappa|$, whereas the minimum build-up occurs at $\Theta = 0$ with $\Lambda_{\min} = 1/(1+\kappa)$ (see Supplementary Method S6).

In finite-length, lossy waveguides, the performance is determined not only by the total absorptivity but also by how the absorbed power is partitioned between a useful load and unavoidable internal dissipation. We compress boundary and propagation effects into the dimensionless complex feedback factor $F = \kappa e^{i\Theta}$, which maps the system state onto the $(\kappa, \Theta)$ plane (**Figs. 2a–d**). This representation determines the boundary-composed reflection $\bar{\Gamma}_g$ and hence the one-port absorptivity $\mathcal{A} = 1 - |\bar{\Gamma}_g|^2$, and it yields a constructive decomposition of the absorbed power into a useful load share $\mathcal{J}_L$ and an internal loss share $\mathcal{J}_{\text{int}} = \mathcal{A} - \mathcal{J}_L$, together with the useful power efficiency $\eta_{\text{use}} = \mathcal{J}_L/\mathcal{A}$. The grey region indicates the passive feasibility constraint $|\Gamma_L| \leq 1$; equivalently, $\kappa \leq \kappa_{\max} = |\Gamma_S|e^{-2R}$, indicating that increasingly strong coupling can become inadmissible even when the absorptivity would otherwise continue to increase. Along the resonant cut $\Theta = \pi$, $\mathcal{J}_L$ is maximized at $\kappa_{\text{opt}} = |\Gamma_S|^2 e^{-4R}$ (**Fig. 2a**), whereas the critical coupling point $\kappa_{\text{cc}} = |\Gamma_S|^2$ indicates a distinct reorganization of the partition landscape; in the present example, $(R = 0.1, |\Gamma_S| = 0.7)$, $\kappa_{\max} = 0.57$, $\kappa_{\text{opt}} = 0.33$ and $\kappa_{\text{cc}} = 0.49$ all remain feasible (**Fig. 2e**). Fixing $\kappa = \kappa_{\text{opt}}$ and the scanning phase (**Fig. 2f**) reveals strong $\Theta$-selectivity: near $\Theta \approx \pi$, the system simultaneously achieves high absorptivity and useful-load capture rates, whereas detuning rapidly suppresses $\mathcal{J}_L$ and redirects the absorbed energy to internal dissipation, thereby cleanly separating the "maximum absorption" from the "maximum useful power" within the same invariant geometry.

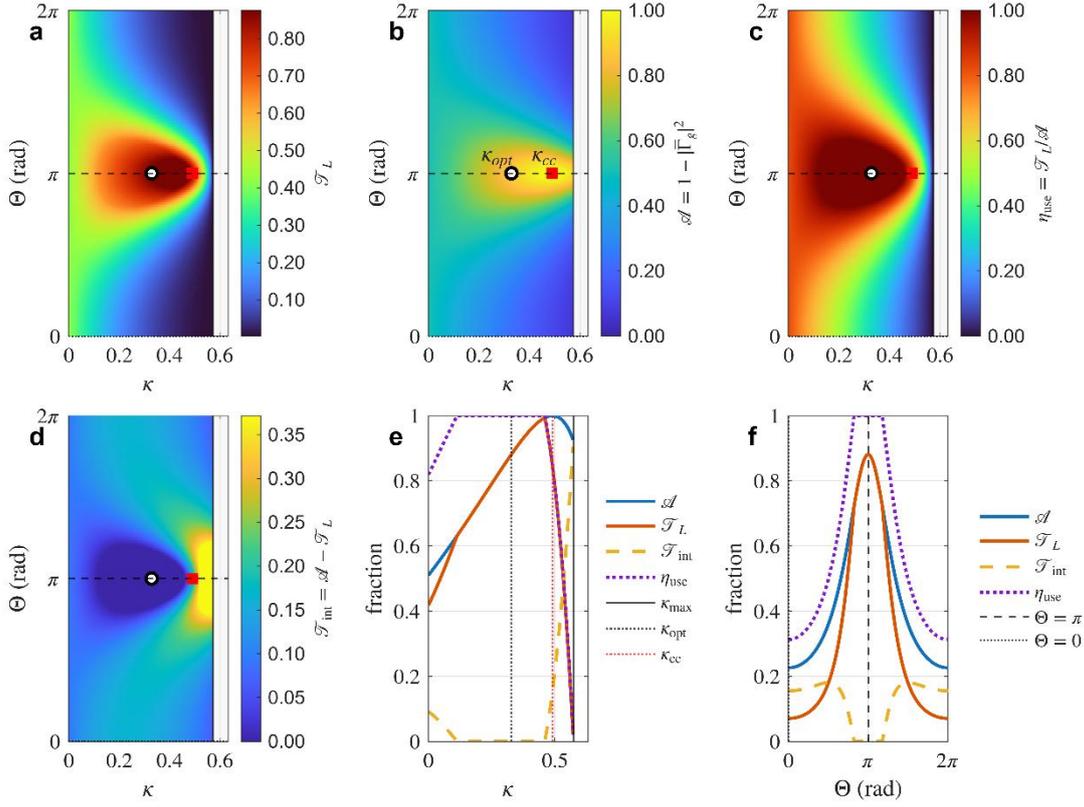

**Fig. 2 | Feedback control of the absorbed power partition on the ($\kappa, \Theta$) plane.** A finite, lossy single-mode waveguide is parameterized by the complex feedback factor $F = \kappa e^{i\Theta}$, where $\kappa$ and $\Theta$ control the round-trip magnitude and phase, respectively. The boundary-composed reflection is $\bar{\Gamma}_g = (\Gamma_S + F/\Gamma_S)/(1 + F)$, yielding one-port absorptivity $\mathcal{A} = 1 - |\bar{\Gamma}_g|^2$. The panels show the partitioning of the absorbed power into useful load power and internal dissipation: **a**. useful load fraction $\mathcal{T}_L = \mathcal{A}\,\eta_{\text{use}}$; **b**. absorptivity $\mathcal{A}$; **c**. useful power efficiency $\eta_{\text{use}} = \mathcal{T}_L/\mathcal{A}$ (defined for $\mathcal{A} > 0$); **d**. internal dissipation fraction $\mathcal{T}_{\text{int}} = \mathcal{A} - \mathcal{T}_L$. The grey strip indicates the passive feasibility bound $|\Gamma_L| \leq 1$; equivalently, $\kappa \leq \kappa_{\text{max}} = |\Gamma_S|e^{-2R}$. **e**. Resonant cut at $\Theta = \pi$, highlighting how $\mathcal{A}$, $\mathcal{T}_L$, $\mathcal{T}_{\text{int}}$ and $\eta_{\text{use}}$ vary with $\kappa$; vertical lines mark $\kappa_{\text{max}}$, the useful power optimum $\kappa_{\text{opt}} = |\Gamma_S|^2 e^{-4R}$, and the critical coupling point $\kappa_{\text{cc}} = |\Gamma_S|^2$. **f**. Phase cut at $\kappa = \kappa_{\text{opt}}$ showing the strong $\Theta$-selectivity of the absorption partition, with reference lines at $\Theta = 0$ and $\Theta = \pi$. Parameters: $\Omega = 1$,

$\delta_R = \delta_G = 0.1$ gives $R = 0.1$ ($e^{-2R} = 0.82$); $|\Gamma_S| = 0.7$ so that $\kappa_{\max} = 0.57$, $\kappa_{\mathrm{opt}} = 0.33$ and $\kappa_{\mathrm{cc}} = 0.49$ (feasible).

A resonance-aware absorption principle is followed by expressing the normalized absorbed power metric directly in terms of $(\kappa, \Theta)$ through the same denominator that governs finite-length feedback:

$$P_{\mathrm{abs}}(\Omega) \propto \frac{1}{1 + \kappa(\Omega)^2 + 2\kappa(\Omega)\cos\Theta(\Omega)}. \tag{5}$$

**Equation (5)** yields two immediate design rules. First, the resonance corresponds to phase alignment with the constructive interference class, i.e., $\Theta(\Omega_{\mathrm{res}}) \simeq (2m+1)\pi$, under $D = 1 + F$. Second, on this resonant line, the absorbed power is maximized at the critical coupling $\kappa(\Omega_{\mathrm{res}}) = |\Gamma_S|^2 \simeq 1$, which defines a narrow "critical coupling corridor" on the Cai–Smith chart; operating points with $\kappa < 1$ are undercoupled (insufficient feedback), whereas those with $\kappa > 1$ are overcoupled (excess feedback relative to the distributed loss) and are increasingly sensitive to detuning through the $\cos\Theta(\Omega)$ term (see Supplementary Method 6).

At fixed dimensionless frequency $\Omega$, the Cai–Smith feedback variables $(\kappa, \Theta)$ make the absorbed power constraint predictive by linking boundary composition to the generalized reflection coefficient $\Gamma_g = (\Gamma_L e^{-2K} + \Gamma_S)/(1 + \Gamma_S \Gamma_L e^{-2K})$. In the constructive (odd-$\pi$) interference class $\Theta = \pi$, maximum absorption occurs at the critical coupling where the numerator cancels, i.e., $\Gamma_L e^{-2K} + \Gamma_S = 0$, yielding $\Gamma_g = 0$ and the resonance-aware bound $P_{\mathrm{abs,max}}^{(\mathrm{cc})}(R) = e^{-4\xi R}$ (**Extended Data Fig. 5a**; here, $\xi = 0.5$). However, passivity enforces $|\Gamma_L| \leq 1$, so the critical coupling requirement $|\Gamma_L|_{\mathrm{cc}} = |\Gamma_S|e^{2R}$ is feasible only for $R \leq R_{\mathrm{cc}} = \frac{1}{2}\ln(1/|\Gamma_S|)$; beyond this threshold, the optimum saturates at the passive load boundary, and the achievable absorbed power decreases sharply (**Extended Data Fig. 5b**). A comparison with an amplitude-matched but phase-mismatched trajectory ($\Theta = 0$, **Extended Data Fig. 5b**) further reveals that a strong feedback magnitude alone is insufficient: the correct interference class is essential for approaching the critical coupling corridor where the waveguide laws impose the tightest power bounds.

To separate useful power delivery from total absorption, we introduce the load-delivered power fraction

$$T_L(\Omega) = \frac{(1 - |\Gamma_S|^2)(1 - |\Gamma_L|^2)e^{-2R}}{|1 + \Gamma_S\Gamma_L e^{-2K}|^2}, \tag{6}$$

which explicitly represents the load coupling $(1 - |\Gamma_L|^2)$ from the resonance-sensitive feedback denominator $D = 1 + \Gamma_S\Gamma_L e^{-2K}$. In the constructive (odd-$\pi$) resonance class, the absorbed power partition becomes a two-parameter design problem in terms of attenuation $R$ and load reflectivity $|\Gamma_L|$. Mapping the finite, lossy system through the feedback factor $F = \Gamma_S\Gamma_L e^{-2K}$ reveals a sharply structured landscape for the useful load power fraction $\mathcal{T}_L(R, |\Gamma_L|)$ (**Extended Data Fig. 6a**): the maximum useful capture ridge follows the analytic optimum $|\Gamma_L|_{\text{opt}} = |\Gamma_S|e^{-2R}$, whereas the condition for perfect absorption (critical coupling) requires the opposite trend $|\Gamma_L|_{\text{cc}} = |\Gamma_S|e^{2R}$ and is therefore cut off by the passivity bound $|\Gamma_L| \leq 1$ beyond $R_{\text{cc}} = \frac{1}{2}\ln(1/|\Gamma_S|)$. Consequently, increasing the distributed loss quickly separates the "maximum absorption" and "maximum useful power" operating regimes. The resulting upper envelope of the useful load power, i.e., $\mathcal{T}_{L,\text{max}}(R) = \frac{(1-|\Gamma_S|^2)e^{-2R}}{1-|\Gamma_S|^2 e^{-4R}}$, provides a closed-form performance bound at a fixed attenuation rate and shows that the critical coupling branch (when feasible) does not, in general, coincide with the useful power optimum (**Extended Data Fig. 6b**), underscoring the notion that a resonance-aware design must target $\mathcal{T}_L$ rather than absorptivity alone.

Mapping $T_L$ onto the Cai–Smith chart reveals that its optimum does not, in general, coincide with the one-port maximum-absorptivity corridor defined by $\mathcal{A} = 1 - |\bar{\Gamma}_g|^2$ (**Extended Data Fig. 7a**). Even when the phase is tuned to the nearest odd-$\pi$ resonance ($\Theta = \pi$), the competition between cavity build-up and coupling imposes a distinct optimum at $|\Gamma_L|_{\text{opt}} = |\Gamma_S| e^{-2R}$ (**Extended Data Fig. 7b**), whereas perfect one-port matching requires the critical coupling magnitude $|\Gamma_L|_{\text{cc}} = |\Gamma_S| e^{2R}$. For sufficiently lossy guides, $|\Gamma_L|_{\text{cc}} > 1$ becomes unattainable for passive loads, so the system can approach but never reach $\bar{\Gamma}_g = 0$, underscoring the fact that resonance-enhanced absorption and maximum useful power transfer are fundamentally different optimization objectives.

These statements, however, are macroscopic: the critical coupling specifies an optimal condition for the total absorbed power of a finite lossy waveguide. In general, in guided-wave systems (multiport devices, multimode structures, or radiative couplers), the total power is carried and exchanged through multiple independent modal channels. This raises a natural question: How should the macroscopic resonance-aware optimum be distributed among the independent modes of the target system? We answer this by switching to a modal viewpoint in which absorption and emission are quantified for each channel and by showing that the resulting allocation scheme is constrained by universal conservation and symmetry relations, i.e., the modal expression of the same macroscopic invariant constraints. The four fundamental laws that follow (**Eqs. 7–10**) are rigorously derived in Supplementary Method S7 and numerically verified in **Extended Data Fig. 8**; they extend the Cai–Smith mass–energy geometry to channel-resolved power bookkeeping.

Specifically, on the basis of the intrinsic mode converter, the $p$-th input channel is characterized by a power absorption coefficient $\alpha_{Mp}$, and the corresponding output channel is characterized by a power emission coefficient $\varepsilon_{Mp}$. We then obtain four fundamental laws governing power absorption and emissions in linear, lossy, one-dimensional waveguides, closely paralleling the universal modal radiation laws for thermal emitters[12].

**Law 1 (modal absorption–emission equivalence):**

$$\alpha_{Mp} = \varepsilon_{Mp}, p = 1,2,\dots. \tag{7}$$

On the intrinsic (eigenchannel) basis, each channel behaves as an effective one-dimensional guide with its own generalized reflection state on the Cai–Smith chart; thus, absorption and emission are two directional views of the same mass–energy shell constraint $\mathcal{U}^2 - \mathcal{S}^2 = |\Gamma_{g,p}|^2$. Geometrically, exchanging absorption with emission corresponds to reversing the net flux direction ($\mathcal{S} \to -\mathcal{S}$) at a fixed $|\Gamma_{g,p}|$, leaving the shell invariant and enforcing equal channel-resolved coefficients.

**Law 2 (equal power profile equivalence):** If a normalized input mode $|i\rangle$ and a normalized output mode $|o\rangle$ possess identical spatial power distributions, then

$$\alpha_i = \varepsilon_o. \tag{8}$$

Modes with the same power profile induce the same internal energy/flux partition along the guide and therefore correspond to the same trajectory (or the same set of $(\mathcal{U},\mathcal{S})$ states) on the Cai–Smith chart. Consequently, they sample the same mass–energy shell geometry and must exhibit identical integrated absorption/emission properties when projected onto the corresponding input/output channels.

**Law 3 (global modal power balance):** For any complete orthonormal sets of input modes $\{|i_n\rangle\}$ and output modes $\{|o_m\rangle\}$,

$$\sum_n \alpha_{i_n} = \sum_m \varepsilon_{o_m}. \tag{9}$$

This is the basis-independent conservation statement: the total absorptivity (the trace over a complete input basis) equals the total emissivity (the trace over a complete output basis), regardless of how the field is decomposed into modes. In the Cai–Smith domain, while different modal bases redistribute weight across different points/trajectories on the disk, the aggregate budget is fixed by the same macroscopic invariant (the accessible shell structure set by boundaries and the distributed loss).

**Law 4 (reciprocal directionality):** In a reciprocal waveguide, the absorption of any input mode equals the emission into its phase-conjugated (backwards) version:

$$\alpha_i = \varepsilon_{i^*}. \tag{10}$$

Reciprocity imposes direction-reversal symmetry on the Cai–Smith chart: the phase-conjugated channel corresponds to the conjugate (mirror) state, which preserves $|\Gamma_g|$ and hence the mass–energy shell. Therefore, any decay/delay or absorption constraint derived from pole proximity and the shell geometry applies equally to the backwards (phase-conjugated) channel, enforcing symmetric directional bounds at the modal level.

Together, the macroscopic critical coupling optimum constrains the achievable total absorption level through $(\kappa, \Theta)$, while the four laws (**Eqs. 7–10**) enforce how that total must be partitioned across independent channels via per-mode equivalences, sum rules, and (when applicable) reciprocity. **Extended Data Fig. 8** summarizes and numerically verifies each law, making explicit the fact that these modal constraints are not additional assumptions but rather the channel-resolved form of the same conservation structure that underlies the Cai–Smith mass–energy geometry.

The separation between total absorption and useful delivery in finite lossy waveguides anticipates the electrochemical mapping below: polarization curves provide a steady-state "input−output" response, but the device-relevant objective is not the total activity implied by current alone. In Section 2.6, we show that a polarization-derived generalized reflection magnitude $\rho(\eta_{\text{eff}})$ plays the role of a feedback/recirculation strength so that the deliverable (useful) throughput naturally acquires a penalty factor $(1 - \rho^2)$ that is analogous to the useful load share in **Eq. (6)** and the related partitions.

**2.4 Decay dynamics, time delays and resonance tunnelling analogy**

The steady-state resonance and critical coupling optima are the static footprints of a finite-length feedback process; dynamically, the same boundary-mediated recirculation scheme controls how long energy dwells in the guide (the ring-down, linewidth and quality factor) and how rapidly the response phase winds with the frequency (group delay). In our reduction, these dynamical signatures are governed by the complex round–trip feedback factor $F(\Omega) = \kappa(\Omega)e^{i\Theta(\Omega)}$ and the universal feedback denominator $D(\Omega) = 1 + F(\Omega)$. Because $|F| < 1$ under passivity, the frequency response admits a discrete-echo representation $D^{-1} = \sum_{n \geq 0} (-F)^n$, making explicit the fact that the measured spectrum is a coherent sum of repeated round trips that are weighted by $\kappa^n$. The pole geometry $F \approx -1$ (equivalently, $|D| \ll 1$) therefore provides a compact control parameter that simultaneously governs the resonant build-up and the dynamic dwell time. We quantify pole proximity in the frequency domain by $\Lambda(\Omega) = 1/|D(\Omega)|$ (**Fig. 3b**): sharp peaks occur when $\Theta(\Omega) \approx \pi$, and the trajectory approaches the unit circle such that $F(\Omega)$ grazes the pole direction $F = -1$. This same condition is visualized geometrically on the Cai–Smith chart of $F$ (**Extended Data Fig. 9a**), where a narrow detuning corridor around the odd-$\pi$ constructive interference line is highlighted by $\exp[-(\Delta\Theta/\Theta_\Sigma)^2]$ with $\Delta\Theta = \text{wrap}(\Theta - \pi)$, making the resonant build-up direction explicit in the $(\kappa, \Theta)$ plane.

Within this unified $(\kappa, \Theta)$ geometry, four canonical resonance archetypes serve as reference points (**Fig. 3a and Extended Data Fig. 9a**). The quarter-wave archetype corresponds to a configuration that can achieve a large feedback magnitude while satisfying the odd-$\pi$ phase condition at isolated frequencies; its trajectory therefore produces sparse but strong near-pole encounters, yielding pronounced peaks in $\Lambda(\Omega)$ and long ring-down effects (**Figs. 3b and 3c**). The Fabry–Pérot archetype represents strong reflections at both terminations, so its phase accumulation process revisits the odd-$\pi$ line repeatedly as $\Omega$ sweeps, generating multiple near-pole encounters (a mode comb in $\Lambda$) at essentially the same $\kappa$-controlled lifetime scale. The Helmholtz-type archetype captures the intrinsic boundary asymmetry that reduces the attainable $\kappa$ for a given detuning corridor; the corresponding reference point lies deeper inside the disk, producing broader spectral features, shorter $\tilde{\tau}_A$ values and lower $Q$ values. Finally, the tunnelling-like archetype corresponds to trajectories that remain mainly inside the disk for most frequencies (moderate $\kappa$ values, large effective attenuation) but can still graze the odd-$\pi$ direction closely enough to yield strong phase winding and delay effects; this provides a waveguide-level interpretation of "large delay without near-unity transmission" within the same feedback pole picture.

A compact decay law follows by converting the discrete round-trip attenuation into an exponential envelope. Let $T_{\text{RT}}(\Omega)$ denote the round-trip time. After $n$ round trips, the amplitude is scaled as $\kappa^n$, yielding the amplitude ring-down time $\tau_A$ and its dimensionless form $\tilde{\tau}_A = \tau_A/T_{\text{RT}}$,

$$\tilde{\tau}_A(\Omega) = \frac{1}{-\ln \kappa(\Omega)} = \frac{1}{-\ln(1-\delta(\Omega))}, \delta(\Omega) = 1 - \kappa(\Omega), \tag{11}$$

with the universal asymptotic scaling scheme $\tilde{\tau}_A \sim \delta^{-1}$ for $\delta \ll 1$ (**Fig. 3c and Extended Data Fig. 9b**). The resonance sharpness follows in a one-to-one manner from the same $\kappa$ budget: the linewidth scales as $\Delta\omega \propto \tau_A^{-1}$, and the quality factor increases proportionally to $\tilde{\tau}_A$, $Q = (\omega t_{\text{rt}}/4)\tilde{\tau}_A$ (**Extended Data Fig. 9b**), so $Q \sim (\omega t_{\text{rt}}/4)\delta^{-1}$ in the small-$\delta$ regime (see Supplementary Method S6).

The same pole geometry also explains the dispersive delay and its tunnelling-like manifestations. Writing the group-delay proxy directly in terms of the feedback denominator, i.e.,

$$\tau_g(\Omega) = -\frac{d}{d\Omega}\arg D(\Omega) = -\text{Im}\left[\frac{d}{d\Omega}\ln D(\Omega)\right], \tag{12}$$

shows that a large delay arises from the rapid phase winding of $D(\Omega)$, which occurs when a frequency sweep moves $F(\Omega)$ tangentially past the odd-$\pi$ pole direction so that $D(\Omega)$ rotates quickly in the complex plane. This mechanism is geometric rather than modal- or domain-specific: it is set by the same $(\kappa, \Theta)$ variables that control $\dot{\tau}_A$ and $Q$ on the Cai–Smith chart (**Extended Data Fig. 9**) and the pole proximity factor $\Lambda(\Omega)$ (**Fig. 3b**). Importantly, a tunnelling-like delay does not require $\kappa \to 1$: the trajectories can remain well inside the disk (moderate $\kappa$) yet graze the odd-$\pi$ direction closely enough to yield pronounced $|\tau_g|$ peaks (**Fig. 3d**), reconciling large delays with substantial attenuation within a single-feedback pole framework. Taken together, **Fig. 3** and **Extended Data Fig. 9** reveal that decay dynamics, the linewidth/quality factor and time delays are direct consequences of the same finite-length feedback geometry that organizes steady-state resonance and power-transfer limits. In **Extended Data Fig. 10**, we use four textbook resonator archetypes to illustrate how the waveguide-feedback geometry $F(\Omega)$ simultaneously controls the enhancement process and constrains dissipation/emission budgets. Across quarter-wave, Fabry–Pérot, Helmholtz and tunnelling settings, the resonant features appear as corridor-local maxima of $\Lambda(\Omega) = 1/|1 + F(\Omega)|$, whereas the accompanying $\kappa(\Omega) = |F(\Omega)|$ quantifies round-trip retention. An evaluation of a passive reciprocal two-port realization at the selected $\Omega_{\text{res}}$

confirms the corresponding waveguide laws with compact insets (Laws 1-4), thereby linking the familiar resonance phenomenology (matching, high-$Q$ interference, localized storage and barrier-enabled transmission) to universal, basis-independent absorption/emission constraints that are encoded by $\mathbf{A_b} = \mathbf{I} - \mathbf{S^\dagger S}$ and $\mathbf{E_m} = \mathbf{I} - \mathbf{SS^\dagger}$ (Supplementary Method S7).

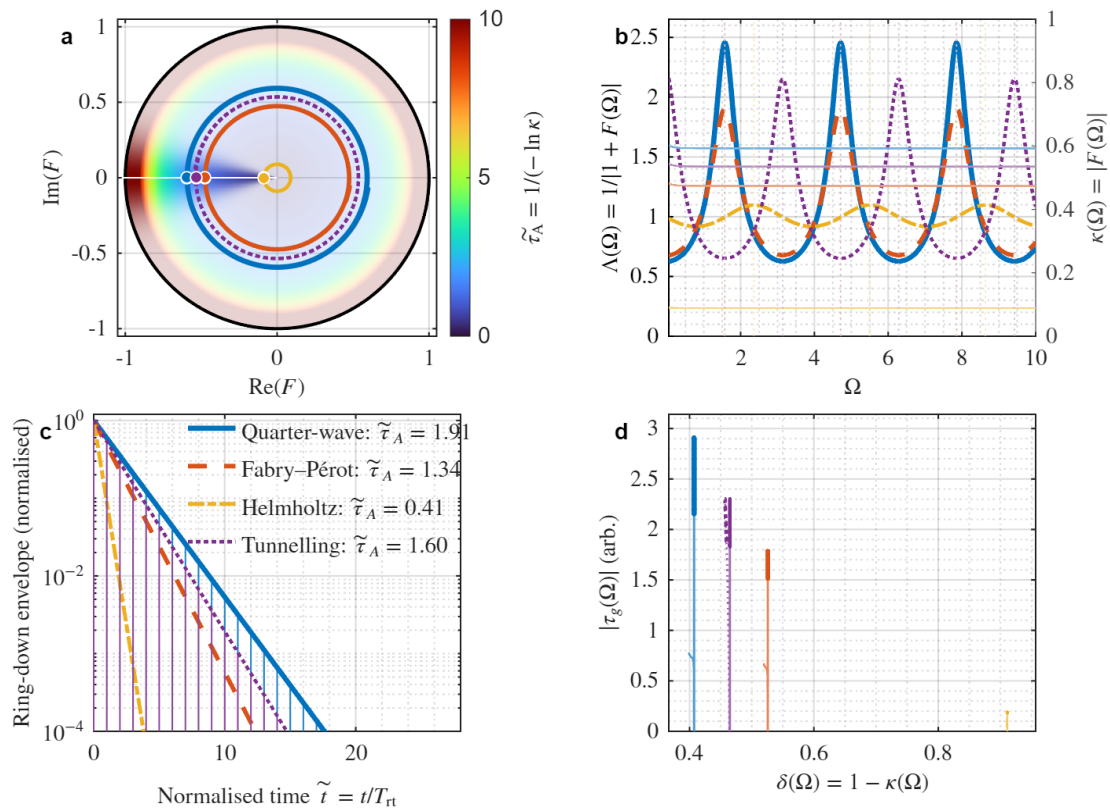

**Fig. 3 | Unified resonance geometry on the Cai–Smith disc. a.** Feedback-plane (complex $F$) representation of finite-waveguide recirculation. The background encodes the dimensionless ring-down time $\tilde{\tau}_A = 1/(-\ln \kappa)$ with $\kappa = |F|$, and the shaded corridor highlights the resonant direction $\Theta = \arg F \approx \pi$ where the pole condition is approached. The coloured trajectories show four canonical resonance archetypes as $\Omega$ is swept: quarter-wave, Fabry–Pérot, Helmholtz-type and tunnelling-like archetypes. White markers indicate representative near-resonant points along $\Theta \approx \pi$. **b.** Frequency-domain pole proximity quantified by $\Lambda(\Omega) = 1/|1 + F(\Omega)|$ (left axis)

together with the loop-magnitude budget $\kappa(\Omega) = |F(\Omega)|$ (right axis). Peaks in $\Lambda$ occur when the trajectory passes near $F \approx -1$; i.e., $D(\Omega) = 1 + F(\Omega)$ decreases. **c.** Time-domain ring-down envelopes produced by discrete round-trip echoes $\kappa^n$ and their continuous-time limit $\exp(-\tilde{t}/\tilde{\tau}_A)$ with $\tilde{t} = t/T_{\rm rt}$, showing longer dwell times for larger values of $\kappa$. **d.** Group-delay proxy $|\tau_g(\Omega)| = |-\operatorname{d}\arg D/\operatorname{d}\Omega|$ plotted against the radial deficit $\delta(\Omega) = 1 - \kappa(\Omega)$ for near-resonant points ($|\Theta - \pi| < \Theta_\Sigma$). Large delays arise from grazing the pole direction in the complex $D$ plane and do not require a near-unity loop magnitude, providing a waveguide-level interpretation of tunnelling-like time-delay anomalies.

**2.5 Coherent perfect absorption at exceptional points revealed by waveguide-invariant mapping**

Coherent perfect absorption (CPA) is often introduced as the time-reversed analogue of lasing, in which a specially phased multiport excitation drives the outgoing field to zero[41]. Yet the central obstacle in realizing and interpreting CPA at exceptional points (CPA-EPs) is not simply achieving $\mathbf{S}(\omega_0)\mathbf{a} = \mathbf{0}$ at a real frequency $\omega_0$ for a given unit-norm coherent input $\mathbf{a}$, but reconciling three intertwined challenges that have hindered a unified, experiment-ready description: (i) the observed near-zero collapse can depend strongly on how the coherent input is prepared, leading to apparent discrepancies between quadratic and quartic scaling; (ii) the classification into nongeneric (symmetric, i.e., $\delta_R = \delta_G$) versus generic (asymmetric, i.e., $\delta_R \neq \delta_G$) CPA-EPs is usually expressed in internal temporal coupled-mode theory (TCMT) parameters that are not uniquely identifiable in complex scatterers; and (iii) practical CPA-EP design requires a reversible mapping from measured scattering to a small set of control variables that remain meaningful across platforms. Our waveguide-invariant framework addresses these difficulties by re-expressing CPA-EP physics in terms of an operational, basis-independent channel metric defined by the SVD of the measured two-port scattering matrix and a single-loop feedback geometry whose state-space trajectories expose the underlying universality class (**Extended Data Figs. 11-13**).

For a general passive two-port scatterer with scattering matrix $\mathbf{S}(\omega)$, a unit-norm coherent input $\mathbf{a}$ produces output power $P_{\text{out}}(\omega) = \|\mathbf{S}(\omega)\mathbf{a}\|^2$ and absorbed fraction $\mathcal{A}(\omega) = 1 - P_{\text{out}}(\omega)$. Maximizing absorption over all coherent inputs is therefore equivalent to minimizing the output norm, which is attained by the smallest singular value:

$$P_{\text{out}}^*(\omega) = \min_{|\mathbf{a}|=1} \|\mathbf{S}(\omega)\mathbf{a}\|^2 = \sigma_{\min}^2(\omega), \mathcal{A}_{\max}(\omega) = 1 - \sigma_{\min}^2(\omega). \quad (13)$$

This immediately yields an experimentally accessible CPA criterion that does not depend on any internal modal parameterization: CPA at a real frequency occurs if $\sigma_{\min}(\omega_0) = 0$, and the required coherent excitation is the right singular vector $\mathbf{a}_{\text{CPA}} = \mathbf{v}_{\min}(\omega_0)$. In our waveguide language, the eigenchannel "mass" is therefore $\rho_{\text{svd}}(\omega) = \sigma_{\min}(\omega) = \sqrt{P_{\text{out}}^*(\omega)/P_{\text{in}}}$, and CPA corresponds to the maximally absorbing eigenchannel reaching the disk center $\rho_{\text{svd}}(\omega_0) = 0$, the multiport analogue of the one-port critical-coupling corridor $\overline{\Gamma}_g(\Omega_0) = 0$ in Sec. 2.3.

Crucially, the SVD formulation also disentangles the long-standing protocol dependence near CPA-EPs. We distinguish a fixed coherent setting, in which the input is frozen as $\mathbf{a}_0 = \mathbf{v}_{\min}(\omega_0)$ and held constant as the detuning $\Delta\omega = \omega - \omega_0$ is swept, from an eigenchannel-tracking (SVD-probe) setting, in which the input is continuously re-optimized as $\mathbf{a}^*(\omega) = \mathbf{v}_{\min}(\omega)$ to follow the minimum-output channel. The two protocols probe different physical limits: fixed inputs reveal leakage induced by coherent-channel drift with detuning, whereas SVD-probe measurements isolate the intrinsic collapse of the absorption channel itself. **Extended Data Fig. 11a-c** quantifies this distinction by plotting the channel mass $\rho(\Delta\omega) = \sqrt{P_{\text{out}}(\Delta\omega)/P_{\text{in}}}$ and the corresponding one-channel output $P_{\text{out}}$ under both protocols, together with near-zero scaling fits. In the nongeneric (symmetric) CPA-EP, fixed excitation generically yields an apparent quadratic floor $P_{\text{out}} \propto |\Delta\omega|^2$ because the optimal singular vector rotates linearly with detuning, producing first-order coherent mismatch; the SVD-probe removes this extrinsic leakage and recovers the higher-order suppression associated with the underlying absorbing degeneracy. By contrast, in the generic (asymmetric) CPA-EP the absorbing channel is intrinsically defective and enforces a quartic universality $P_{\text{out}} \propto |\Delta\omega|^4$ that is visible under both protocols (**Extended Data Fig. 11c**).

Beyond the effective one-channel view, the same framework predicts the full two-port redistribution under CPA-EP driving. Writing the outgoing amplitudes as $\mathbf{b}(\Delta\omega) = \mathbf{S}(\Delta\omega)\mathbf{a}(\Delta\omega)$, the port-resolved outputs $P_j(\Delta\omega) = |\mathbf{b_j}(\Delta\omega)|^2$ and total output $P_{\text{tot}} = \sum_j P_j$ exhibit protocol-dependent leakage and mode conversion that remain consistent with the eigenchannel laws (Eqs. 7–10) (**Extended Data Fig. 11d-f**). To visualize the driven channel on the Cai–Smith chart, we further evaluate a complex waveguide-state proxy $\Gamma(\Delta\omega) = \mathbf{a}(\Delta\omega)^\dagger \mathbf{S}(\Delta\omega)\mathbf{a}(\Delta\omega)$; the resulting unit-disk trajectories provide a compact geometric diagnostic of how the system approaches (or departs from) the disk center and how fixed-versus-optimal preparation reshapes the admissible path under passivity (**Extended Data Fig. 11g, h**).

To convert this channel-based diagnosis into a platform-agnostic design language, we next construct an explicit, reversible mapping from a two-resonator TCMT realization to an equivalent two-port waveguide scattering form with a single-loop feedback denominator. Specifically, we seek effective, frequency-dependent waveguide parameters $\Gamma_S^{\text{eff}}(\omega)$, $\Gamma_L^{\text{eff}}(\omega)$ and $K_{\text{eff}}(\omega)$ such that the waveguide-form matrix $S_{\text{WG}}(\omega)$ reproduces $S_{\text{TCMT}}(\omega)$ elementwise, with a shared loop closure $D(\omega) = 1 + \Gamma_S^{\text{eff}}(\omega)\Gamma_L^{\text{eff}}(\omega)z(\omega)$ and $z(\omega) = e^{-2K_{\text{eff}}(\omega)}$. Operationally, at each frequency we invert the measured/transcribed coefficients $[r_1(\omega), r_2(\omega), t(\omega)]$ to a unique round-trip factor $z(\omega)$ (and hence a continuous $K_{\text{eff}}$), then back-substitute to obtain $\Gamma_{S,L}^{\text{eff}}(\omega)$; this mapping, which solves a single scalar equation and performs closed-form back-substitution, fully absorbs the underlying TCMT parameterization into a set of effective waveguide boundaries and thereby achieves a numerically exact reconstruction of the complete scattering response. **Extended Data Fig. 12** verifies this equivalence across a broad band by comparing singular values of $\mathbf{S}_{\text{TCMT}}$ and the reconstructed $\hat{\mathbf{S}}_{\text{WG}}(\omega)$, and by plotting the inferred $z(\omega)$, $K_{\text{eff}}(\omega)$ and boundary trajectories $\Gamma_{S,L}^{\text{eff}}(\omega)$, thereby exposing CPA-EP optics as a concrete instance of the same single-loop feedback geometry used throughout this work.

Finally, the waveguide representation clarifies why "protocol dependence" is not an experimental nuisance but a predictable consequence of coherent-channel drift. In practice, many optical CPA measurements prepare the input at the CPA frequency and then scan detuning without retuning the coherent state, whereas the SVD-probe protocol retunes both amplitude ratio and phase continuously. **Extended Data Fig. 13** isolates this distinction in a local neighborhood around $\Omega_0$ by comparing (i) a fixed-at-$\Omega_0$ coherent excitation and (ii) a per-frequency optimal excitation: panel-wise comparisons of $\sigma_{\min}/\sigma_{\max}$, the effective output norm under fixed excitation, the detuning-dependent optimal amplitude ratio $|\mathbf{a}_1/\mathbf{a}_2|$ and phase difference $\Delta\phi$, and the resulting port-resolved and total outputs show how a frozen coherent state converts the intrinsic CPA-EP collapse into an apparent lower-order floor, whereas per-frequency optimization recovers the eigenchannel-defined universality class. In this way, CPA-EP optics becomes a canonical demonstration of our cross-domain scattering language: the same basis-independent SVD metric and waveguide-invariant disk geometry that diagnose absorbing degeneracies here will later be used to map electrochemical polarization into the interplay between energy storage and power transfer (Sec. 2.6), providing a unified route from measured responses to optimal operation and universality-class identification.

**2.6 Waveguide-invariant mapping reveals the interplay between energy storage and power transfer in electrochemical polarization settings**

A central challenge with regard to interpreting electrochemical polarization curves is to disentangle concurrent energy storage processes (for example, intermediate accumulation and double-layer charging) from net power transfers manifested as faradaic current[42-47]. Although coupled kinetic–transport models can reproduce the measured $j(\eta)$, they rarely provide a state-resolved, geometry-based principle that reveals how an operating point migrates between storage-dominant and transfer-dominant regimes under finite dissipation and bidirectional feedback conditions. Here, we treat the overpotential as the external driving coordinate and the measured current density as an experimentally accessible power flow-like observable from which waveguide-compatible exchange and transport budgets can be constructed. Accounting for the Ohmic loss induced with an effective series resistance $R_\Omega \geq 0$, we work with the effective interfacial overpotential $\eta_{\text{eff}} = \eta - R_\Omega j$ and fit each condition using a two-channel orthogonal decomposition scheme (Supplementary Method S9), which yields a unique polarization-derived state trajectory $\Gamma_g(\eta_{\text{eff}})$ on the Cai–Smith chart.

The fitted orthogonal decomposition scheme provides unsaturated directional branch fluxes for each channel $J_{\text{ox},p}$ and $J_{\text{red},p}$, whose sums $J_{\text{ox}}^{\text{tot}}$ and $J_{\text{red}}^{\text{tot}}$, respectively, act as counterpropagating power streams in a lossy waveguide. From these, we form the transfer proxy $\mathcal{S} = J_{\text{ox}}^{\text{tot}} - J_{\text{red}}^{\text{tot}}$ and the exchange proxy $\mathcal{U} = J_{\text{ox}}^{\text{tot}} + J_{\text{red}}^{\text{tot}}$ so that $\mathcal{U}$ captures the bidirectional exchange that is invisible to $j$ alone, while $\mathcal{S}$ captures the net transport. We then construct a bounded generalized reflection amplitude $\rho(\eta_{\text{eff}}) \in [0,1]$ and a continuous coherence angle $\varphi(\eta_{\text{eff}}) \in [0,\pi]$ from the branch-balancing metric $\beta_w = (J_{\text{ox}}^{\text{tot}} - J_{\text{red}}^{\text{tot}})/(J_{\text{red}}^{\text{tot}} + J_{\text{ox}}^{\text{tot}})$, thus defining the polarization-derived complex generalized reflection factor $\Gamma_g(\eta_{\text{eff}}) = \rho\, e^{i\varphi}$ without the need for impedance spectroscopy. This construction enforces $|\Gamma_g| \leq 1$ via passivity and turns the full polarization curve into a compact state-space trajectory on a universal geometric disk.

To connect the mapped trajectory to practical catalytic operation under a finite driving cost, we define a useful output proxy that applies simultaneously to both cathodic reduction and anodic oxidation systems:

$$\Pi_{\text{use}}(\eta_{\text{eff}}) = |j_{\text{tot}}(\eta_{\text{eff}})|[1 - \rho^2(\eta_{\text{eff}})], \tag{14}$$

where $(1 - \rho^2)$ discounts the bidirectional competition (reflection losses) that suppresses the net useful output as $\rho \to 1$. We then define the useful output density as

$$\Pi_{\text{dens}}(\eta_{\text{eff}}) = \frac{\Pi_{\text{use}}(\eta_{\text{eff}})}{|\eta_{\text{eff}}| + \eta_0}, \eta_0 > 0, \tag{15}$$

which avoids the formal divergence at $|\eta_{\text{eff}}| \to 0$ and is equivalent to imposing a small numerical cut-off in practice (Supplementary Method S9.8). The density-optimal operating point is $\eta_{\text{eff}}^* = \arg\max_{\eta_{\text{eff}}<0} \Pi_{\text{dens}}(\eta_{\text{eff}})$, together with $\Pi_{\text{use}}^* = \Pi_{\text{use}}(\eta_{\text{eff}}^*)$ and $\Pi_{\text{dens}}^* = \Pi_{\text{dens}}(\eta_{\text{eff}}^*)$.

Applying this mapping to the HER on Au(111) across pH[46,47] (**Fig. 4; Extended Data Figs. 14-18**), the Cai–Smith trajectory $\Gamma_g(\eta_{\text{eff}})$ provides a direct, state-resolved map of how electrochemical operations navigate the fundamental trade-off between energy storage (high exchange rate, strong bidirectional competition) and power delivery (nearly unidirectional transfer). In the waveguide language, a large $\rho$ corresponds to strong bidirectional feedback and exchange-dominant operations, whereas a decreasing $\rho$ indicates weakening of the feedback and the emergence of transfer-dominant throughput. Viewed on the disk, the full polarization sweep evolves from a high-reflection, storage-like regime through a feedback-weakened transition into a low-reflection, transfer-like regime, making the "hidden" exchange content of the curve explicit and comparable across different pH values in a single geometric space. The accompanying $\mathcal{U}$ and $|\mathcal{S}|$ traces (**Fig. 4d**) quantify this transition: exchanges and transfers vary continuously with $\eta_{\text{eff}}$ and the pH rather than being inferred indirectly from $j$ alone, thereby separating "how much activity is circulating" from "how much is delivered".

A key consequence is a new physical reading of the apparent pH dependence of macroscopic Tafel behaviour. Rather than attributing all the pH trends to a single changing slope or a single rate-determining step, the mapping reveals a systematic rebalancing of two orthogonal modes with distinct intrinsic asymmetry parameters that are coupled through a global feedback strength encoded by $\rho$ (**Extended Data Fig. 17**). In this view, changes in the measured slopes and curvature arise from a shifting modal mixture and a shifting position along the universal storage transfer manifold, which is precisely the information needed for function-specific optimization purposes.

The density metric in **Eq. (15)** converts the state trajectory into an operational design rule. The extracted $\eta_{\text{eff}}^*$ shows a pronounced pH dependence: at pH=1.0, $\eta_{\text{eff}}^* = -0.53$ V yields $\Pi_{\text{dens}}^* = 92.9$ mA,cm$^{-2}$V$^{-1}$; at pH=2.5, $\eta_{\text{eff}}^* = -0.65$ V provides the lowest value at $\Pi_{\text{dens}}^* = 12.9$ mA,cm$^{-2}$V$^{-1}$; within pH=3.0–5.0, the optimum shifts to much more negative driving ($\eta_{\text{eff}}^* \approx -1.45$ to $-1.50$ V); and at pH=13.0, $\eta_{\text{eff}}^* = -1.14$ V recovers $\Pi_{\text{dens}}^* = 37.8$ mA,cm$^{-2}$V$^{-1}$ (complete values are given in **Extended Data Table 3**). These trends separate performance into distinct facets: $\Pi^*_{use}$ quantifies the usable throughput after discounting bidirectional competition via $(1-\rho^2)$, while $\Pi^*_{dens}$ quantifies the output per unit driving cost, explicitly revealing where a large throughput is achievable only at disproportionately large $|\eta_{eff}|$ values and is therefore density-inefficient.

Overall, the results of the electrochemical case study demonstrate that a polarization curve is not merely a kinetic fitting target but also a trajectory of the system state evolution process in a universal geometric space: $\Gamma_g(\eta_{eff})$, together with $(\mathcal{U}, \mathcal{S})$ and **Eq. (15)**, provides a direct and quantitative map of how operations move between exchange-dominant "storage-like" behaviour and "transfer-dominant power-delivery" behaviour under finite dissipation and feedback mechanisms. This resolves a long-standing interpretative gap in the polarization analysis field: the inability to separate circulating exchange activity from net useful transfer using $j(\eta)$ alone, even when the curve is perfectly fitted. Importantly, the same pipeline transfers to Pt(111) across different pH values[46,47] (**Extended Data Figs. 19-24; Extended Data Tables 4**), supporting the claim that the mapping elevates the polarization analysis process from phenomenological curve fitting to a physically grounded, geometry-based design language. Beyond the HER, the construction relies only on steady-state polarization data and passivity-compatible budgets and is therefore directly applicable to other electrocatalytic reactions (ORR/OER/$CO_2$RR), battery and corrosion polarization and, more broadly, to any driven dissipative system in which competing bidirectional pathways coexist with a net transport objective.

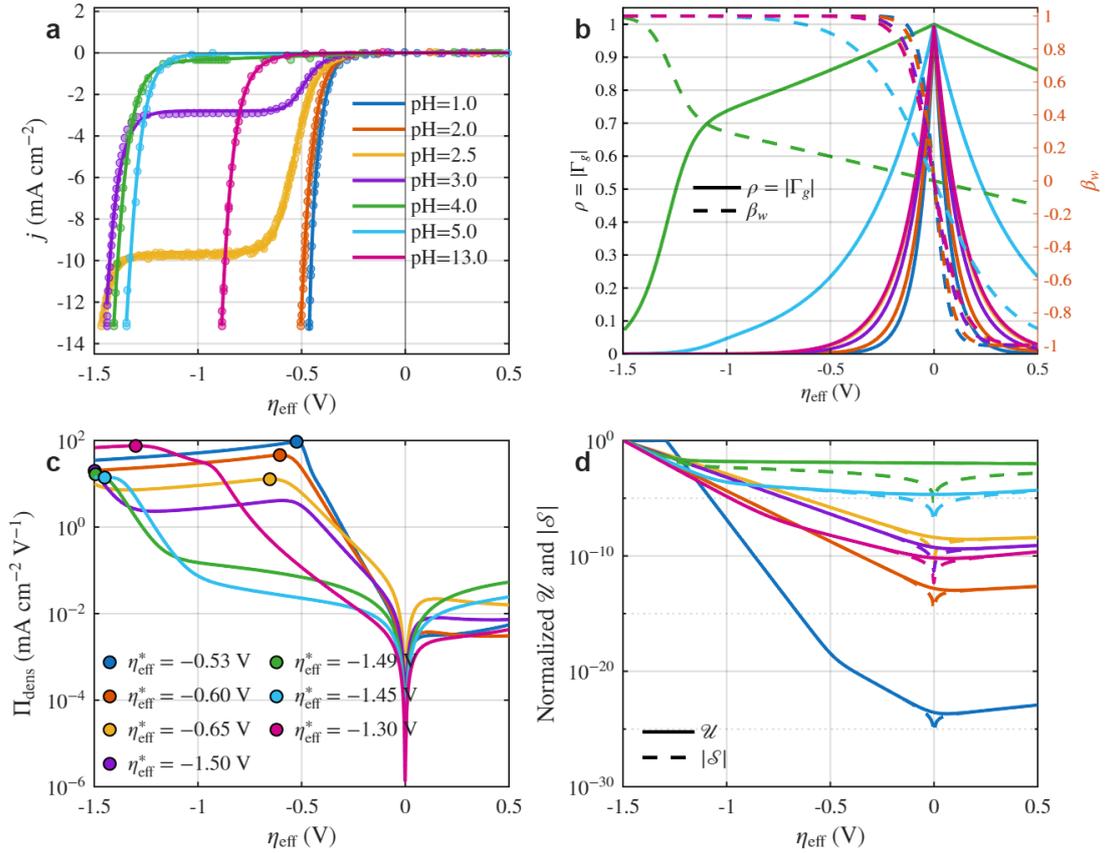

**Fig. 4 | Waveguide-invariant mapping of electrochemical polarization kinetics on the Au(111) surface across different pH values (the analogous Pt(111) mapping is shown in Extended Data Fig. 19). a.** Measured[46,47] (Copyright 2013 and 2026 Springer Nature) polarization curves $j(\eta)$ (markers) and the fitted two-channel waveguide model (solid lines) for Au(111) across pH values ranging from 1–13, with the driving coordinate expressed as the effective interfacial overpotential $\eta_{\text{eff}}$ (including ohmic correction). **b.** Polarization-derived waveguide state diagnostics versus $\eta_{\text{eff}}$: the generalized reflection amplitude $\rho = |\Gamma_g|$ (solid; left axis) quantifying the bidirectional feedback strength and the coherence diagnostic $\beta_w = (J_{\text{ox}}^{\text{tot}} - J_{\text{red}}^{\text{tot}})/(J_{\text{red}}^{\text{tot}} + J_{\text{ox}}^{\text{tot}})$ (dashed; right axis) capturing the balance between the reduction and oxidation branch fluxes extracted from the fitted decomposition. **c.** Useful cathodic output density $\Pi_{\text{dens}} = \Pi_{\text{use}}/(|\eta_{\text{eff}}| + \eta_0)$ (log scale), where $\Pi_{\text{use}} = |j_{\text{tot}}|(1 - \rho^2)$ is a cathodic throughput proxy penalized by reflection losses; circles

mark the density-optimal operating points $\eta_{\text{eff}}^*$ (one per pH) that maximize $\Pi_{\text{dens}}$ along the cathodic branch. **d.** Normalized exchange and transfer proxies derived from the same directional-flux decomposition: $\mathcal{U} = J_{\text{ox}}^{\text{tot}} + J_{\text{red}}^{\text{tot}}$ (solid) and $|\mathcal{S}| = |J_{\text{ox}}^{\text{tot}} - J_{\text{red}}^{\text{tot}}|$ (dashed), highlighting the continuous transition from an exchange-dominant regime (strong bidirectional competition) to a transfer-dominant (nearly unidirectional) regime as $\eta_{\text{eff}}$ and the pH vary.

**Conclusions**

We establish a universal mass–energy equation for finite, lossy, boundary-reflecting waveguides by expressing finite-length feedback through a boundary-composed generalized reflection coordinate and a minimal intrinsic asymmetry parameter $\xi$. The resulting invariant shell, i.e., $\mathcal{U}^2 - \mathcal{S}^2 = |\Gamma_g|^2$, provides a compact and kinematically transparent constraint linking energy-like exchanges, power-flow-like transport, and an effective standing-wave "mass" set by integrated loss and boundary reflections, thereby unifying the classic resonance and maximum-power-transfer concepts into a single geometric state space. Central to this space is the Cai–Smith chart, which renders stability margins, feedback proximity and attainable operating states directly readable and makes the point at which losses and passivity force a separation between maximum absorption and maximum useful delivery explicit. The same feedback variables yield closed-form resonance, critical coupling and ring-down predictions, i.e., $\tau = T_{\text{RT}}/(-\ln \kappa)$, tying static optima and dynamical lifetimes to a single finite-length recirculation budget.

The framework extends naturally to multi-channel systems. We establish four fundamental laws for linear, lossy, one-dimensional waveguides, providing channel-resolved absorption and emission constraints that extend the macroscopic geometry to modal power bookkeeping. We also used coherent perfect absorption at exceptional points as a stringent multiport benchmark of the same framework: the experimentally measurable condition $\sigma_{\min}(\omega_0) = 0$ provides a basis-independent CPA criterion, while contrasting a fixed coherent setting with an eigenchannel-tracking SVD-probe explains the apparent quadratic-quartic discrepancies and yields an operational classification of generic versus nongeneric CPA-EPs in terms of waveguide-invariant disk trajectories and single-loop feedback geometry. To demonstrate the framework's experimental reach, we map steady-state electrochemical polarization onto this waveguide geometry. By treating the measured current density as a power flow-like state variable, we extract a passive complex state $\Gamma_g(\eta_{\text{eff}})$ from two-channel orthogonal fits, and define phase-aware storage/transfer diagnostics and density-optimal operating points. This approach reveals a universal transition from storage dominant to transfer dominant regimes across pH, turning polarization analysis from kinetic fitting into a waveguide native design language.

Together, these results transform waveguide theory from a collection of domain-specific tool into a unified, invariant-based language for the analysis and design of dissipative, boundary-controlled wave systems. It provides a systematic path to diagnose performance limits and optimize power management across photonics, acoustics, quantum transport, and electrochemistry. The framework also opens natural extensions to cascaded non uniform structures, multimode matrix scattering, and weakly nonlinear or driven media, where deviations from the invariant shell may serve as quantitative fingerprints of new physical effects.

## Methods

### Waveguide reduction and dimensionless control parameters

A finite, linear, time-invariant, single-mode guided-wave system with a length of $L$ is modelled via an effective telegrapher equation reduction scheme, which is characterized by several per-unit-length parameters: the resistance $R'$, inductance $L'$, conductance $G'$ and capacitance $C'$. Its complex propagation constant is $\gamma(\omega)$, and terminal reflections are defined with respect to the frequency-dependent characteristic impedance $Z_c(\omega)$. This reduction serves as a domain-independent minimal model: different physical realizations (electromagnetic, acoustic, hydrodynamic, and electrochemical-transport analogues) enter only through the effective parameters $(R', L', G', C')$ and boundary conditions, while the subsequent invariant and resonance diagnostics are formulated entirely in terms of measurable guided-wave quantities (derivations and sign conventions are presented in Supplementary Methods S0-S1). Following the normalization procedure used throughout the paper, we introduce the reference scales

$$\omega_c = \frac{1}{\sqrt{L'C'}}, Z_c = \sqrt{\frac{L'}{C'}}, \qquad (M1)$$

and define the dimensionless loss budgets and frequency:

$$\delta_R = \frac{R'L}{Z_c}, \delta_G = G'LZ_c, \Omega = \frac{\omega}{\omega_c}. \qquad (M2)$$

All frequency-length dependencies are then summarized by the dimensionless complex electrical length

$$K(\Omega) = \gamma(\omega)L = \sqrt{(\delta_R + i\Omega)(\delta_G + i\Omega)} = R(\Omega) + i\Phi(\Omega), \qquad (M3)$$

whose real and imaginary parts quantify cumulative attenuation and accumulated phase advancing, respectively (Supplementary Method S1; **Extended Data Fig. 1**).

## Terminal reflections, boundary compositions and admissible Cai–Smith states

Terminal reflections are defined with respect to the characteristic impedance $Z_c(\omega)$ of the reduced single-mode port. For an impedance termination $Z_T(\omega)$,

$$\Gamma_T(\omega) = \frac{Z_T(\omega) - Z_c(\omega)}{Z_T(\omega) + Z_c(\omega)}, \tag{M4}$$

so passivity implies that $|\Gamma_T| \leq 1$ under the adopted normalization strategy (Supplementary Methods S0, S2). One round trip across the guide contributes to the propagation factor

$$E(\Omega) = e^{-2K(\Omega)}. \tag{M5}$$

Multiple reflections between the terminals reduce to the boundary-composed generalized reflection state $\bar{\Gamma}_g(\Omega)$ given in the main text (**Eq. 1**) and derived in Supplementary Method S2. Under strictly lossy propagation ($|E| < 1$) and passive terminals ($|\Gamma_S| \leq 1$, $|\Gamma_L| \leq 1$), the composition is contractive and yields the admissibility condition

$$|\bar{\Gamma}_g(\Omega)| < 1, \tag{M6}$$

which motivates the unit-disk Cai–Smith representation (Supplementary Method S2.5).

## Invariant state variables and rapidity parameterization

Energy-like and power flow-like state variables $(\mathcal{U}, \mathcal{S})$ are constructed from normalized forwards/backwards components and satisfy the invariant shell relation reported in the main text (**Eq. 2**); intrinsic asymmetry promotes the generalized reflection to an axial state via the deformation law given in the main text (**Eq. 3**), with the conventions fixed globally (Supplementary Method S3). For numerically stable regime separation and concatenation operations, we use the bounded "waveguide velocity" $\beta_w = \mathcal{S}/\mathcal{U} \in (-1,1)$ and the corresponding rapidity

$$\phi = \text{atanh}(\beta_w), \tag{M7}$$

so that

$$\mathcal{U} = |\Gamma_g|\cosh \phi, \mathcal{S} = |\Gamma_g|\sinh \phi, D_\pm = e^{\pm\phi}, \qquad (M8)$$

with the approximation thresholds and error controls fixed a priori (Supplementary Methods S4 and S5; **Extended Data Figs. 3 and 4**).

**Feedback factor, resonance proximity, decay and delay diagnostics**

Finite-length feedback is characterized by the round-trip factor $F(\Omega)$ and its polar decomposition $(\kappa, \Theta)$ defined in the main text (**Eq. 4**). We evaluate the feedback denominator consistently with the main-text convention,

$$D(\Omega) = 1 + F(\Omega), \qquad (M9)$$

and use the pole proximity functional $\Lambda(\Omega) = 1/|D(\Omega)|$ to locate resonance corridors on the $(\kappa, \Theta)$ plane and on the Cai–Smith chart (Supplementary Methods S6; **Extended Data Fig. 9**). For delay diagnostics, we compute the group delay of the feedback-controlled response using the definition given in the main text (**Eq. 12**), with phase unwrapping and finite-difference rules fixed on the frequency grid before any fitting or selection work is performed (Supplementary Methods S2.8, S4 and S6).

**Derivation pathway for the four fundamental laws**

The four fundamental laws stated in the Results section (**Eqs. 7–10**) follow from (i) the passive, single-mode, boundary-composed description given above (captured by $\bar{\Gamma}_g(\Omega)$ and $F(\Omega)$); (ii) writing absorption/emission/transfer objectives as functions of these boundary-composed quantities and the pole distance measures; and (iii) enforcing operator-level passivity/causality constraints that bound the achievable absorption and emission effects under an arbitrary (but passive) boundary control scheme. The full inequality chain, achievability conditions and extensions are given in Supplementary Method S7 to avoid duplicating the results.

**Optical CPA-EP reconstruction: fixed coherent setting versus SVD probe**

To connect the waveguide-state formulation to coherent perfect absorption (CPA) at exceptional points (EPs), we reconstruct the standard two-resonator CPA-EP within a two-mode temporal coupled-mode theory (TCMT) scattering model[41] and evaluate both frequency-domain spectra and Cai–Smith chart trajectories (**Extended Data Figs. 11-13**). At each real frequency $\omega = \omega_0 + \Delta\omega$, we compute a reciprocal $2 \times 2$ scattering matrix $\mathbf{S}(\omega)$ for two dissipative resonances (intrinsic loss rates $\gamma_{1,2}$) coupled to external waveguides (radiative leakage rates $\gamma_{c1,c2}$) and mutually coupled with rate $\kappa_{\text{TCMT}}$. Parameter sets are chosen to realize either a nongeneric CPA-EP (symmetric coupling) or a generic CPA-EP (asymmetric coupling), while satisfying the CPA-EP feasibility and defectiveness constraints in the TCMT representation (Supplementary Methods S8; **Extended Data Fig. 11**).

We compare two experimentally relevant probing protocols that separate coherent-mismatch leakage from the intrinsic absorption-eigenchannel collapse: a fixed coherent setting that freezes the CPA-tuned input at $\Delta\omega = 0$, and an eigenchannel-tracking SVD probe that re-optimizes $\mathbf{a}^*(\Delta\omega) = \mathbf{v_{min}}(\Delta\omega)$ at each detuning, thereby distinguishing protocol-induced quadratic leakage from intrinsic quartic suppression near the CPA-EP (**Extended Data Fig. 13**). (i) Fixed coherent setting: we compute the maximally absorbing input at $\Delta\omega = 0$ via the singular value decomposition $\mathbf{S}(0) = \mathbf{U\Sigma V}^\dagger$ and select the right singular vector $\mathbf{a_0} = \mathbf{v_{min}}(0)$ associated with $\sigma_{\min}(0)$. The unit-norm input $\mathbf{a_0}$ (including its relative amplitude and phase) is then held fixed while scanning detuning, yielding

$$P_{\text{out}}^{\text{fixed}}(\Delta\omega) = \|\mathbf{S}(\Delta\omega)\mathbf{a_0}\|^2, P_{\text{in}} = \|\mathbf{a_0}\|^2 = 1. \quad (M10)$$

(ii) SVD probe: at each detuning, we recompute the optimal coherent input $\mathbf{a}^*(\Delta\omega) = \mathbf{v_{min}}(\Delta\omega)$ that minimizes the output, giving the absorption-eigenchannel bound

$$P_{\text{out}}^{\text{svd}}(\Delta\omega) = \min_{\|a\|=1} \|\mathbf{S}(\Delta\omega)\mathbf{a}\|^2 = \sigma_{\min}^2(\Delta\omega). \quad (M11)$$

For two-port spectra, the outgoing amplitude is $\mathbf{b}(\Delta\omega) = \mathbf{S}(\Delta\omega)\mathbf{a}(\Delta\omega)$ and the port-resolved outputs are $P_j(\Delta\omega) = |\mathbf{b_j}(\Delta\omega)|^2$, with total output $P_{\text{tot}} = \sum_j P_j$ (**Extended Data Fig. 11d -f**). We report the channel "mass" $\rho(\Delta\omega) = \sqrt{P_{\text{out}}(\Delta\omega)/P_{\text{in}}}$ (**Extended Data Fig. 11a**) and the corresponding effective one-channel output spectrum $P_{\text{out}}/P_{\text{in}}$ (**Extended Data Fig. 11b**). Near-zero scaling exponents are obtained by linear regression of $\log P_{\text{out}}$ versus $\log |\Delta\omega|$ over a prescribed small-detuning window fixed a priori on the frequency grid (**Extended Data Fig. 11c**).

To visualize the driven channel on the Cai–Smith geometry, we compute a complex waveguide-state proxy

$$\Gamma(\Delta\omega) = \mathbf{a}^\dagger(\Delta\omega)\, \mathbf{S}(\Delta\omega)\, \mathbf{a}(\Delta\omega), \qquad (M12)$$

using $\mathbf{a}(\Delta\omega) = \mathbf{a_0}$ for the fixed coherent setting and $\mathbf{a}(\Delta\omega) = \mathbf{a}^*(\Delta\omega)$ for the SVD probe. The resulting trajectories $\Gamma(\Delta\omega)$ are plotted in the complex plane together with the unit-disk boundary to diagnose approach to the disk center and protocol-dependent drift under passivity (**Extended Data Fig. 11g, h**).

To express the same CPA-EP response in the waveguide feedback language and enable an externally measurable reconstruction, we convert $\mathbf{S}(\omega)$ to an equivalent two-port waveguide-form factorization at each frequency (**Extended Data Fig. 12**). Writing $\mathbf{r_1}(\omega) = \mathbf{S_{11}}(\omega)$, $\mathbf{r_2}(\omega) = \mathbf{S_{22}}(\omega)$ and $\mathbf{t}(\omega) = \mathbf{S_{12}}(\omega) = \mathbf{S_{21}}(\omega)$, we seek frequency-dependent effective boundaries $\Gamma_S^{\text{eff}}(\omega)$, $\Gamma_L^{\text{eff}}(\omega)$ and an effective propagation constant $K_{\text{eff}}(\omega)$ such that the waveguide-form matrix reproduces $\mathbf{S}(\omega)$ elementwise with a shared loop closure. We set the effective end-coupling amplitudes to match the TCMT channel weights, $\tau_S^{\text{eff}} = \sqrt{\gamma_{c1}}$ and $\tau_L^{\text{eff}} = \sqrt{\gamma_{c2}}$, define the round-trip factor $z(\omega) = e^{-2K_{\text{eff}}(\omega)}$ (with $\sqrt{z} = e^{-K_{\text{eff}}}$), and compute an intermediate denominator from the measured transmission,

$$D(\omega) = \frac{\sqrt{z(\omega)}\, \tau_S^{\text{eff}} \tau_L^{\text{eff}}}{t(\omega)}. \tag{M13}$$

The effective boundaries then follow by closed-form back-substitution from the reflection identities,

$$\Gamma_S^{\text{eff}}(\omega) = \frac{D(\omega)(\mathbf{r}_1(\omega) - z(\omega)\mathbf{r}_2(\omega))}{1 - z^2(\omega)}, \Gamma_L^{\text{eff}}(\omega) = \frac{D(\omega)(\mathbf{r}_2(\omega) - z(\omega)\mathbf{r}_1(\omega))}{1 - z^2(\omega)}. \tag{M14}$$

The remaining consistency condition is the single-loop closure $D(\omega) = 1 + \Gamma_S^{\text{eff}}(\omega)\Gamma_L^{\text{eff}}(\omega)z(\omega)$, which we solve frequency-by-frequency for $z(\omega)$ using a continuous parameterization $z(\omega) = e^{-u(\omega)+iv(\omega)}$ with $u(\omega) \geq 0$ and warm-start initialization across $\omega$ to avoid branch jumps; a continuous $K_{\text{eff}}(\omega) = -\frac{1}{2}\log z(\omega)$ is then obtained by phase unwrapping of $v(\omega)$. Reconstruction quality is verified by comparing singular values of the original and reconstructed matrices and by plotting the inferred $z(\omega)$, $K_{\text{eff}}(\omega)$ and $\Gamma_{S,L}^{\text{eff}}(\omega)$ over the full band (**Extended Data Fig. 12a-f**).

Finally, to isolate experimentally relevant protocol effects in a local neighborhood of the CPA frequency (**Extended Data Fig. 13**), we contrast (i) a fixed-at-$\Omega_0$ coherent excitation $\mathbf{a}_{\text{fixed}} = \mathbf{v}_{\text{min}}(\Omega_0)$ and (ii) a per-frequency optimal excitation $\mathbf{a}^*(\Omega) = \mathbf{v}_{\text{min}}(\Omega)$, reporting the singular-value ratio $\sigma_{\text{min}}/\sigma_{\text{max}}$, the effective output norm $\|S(\Omega)\mathbf{a}_{\text{fixed}}\|$, the detuning-dependent optimal amplitude ratio $|\mathbf{a}_1/\mathbf{a}_2|$ and phase difference $\Delta\psi = \arg(\mathbf{a}_1) - \arg(\mathbf{a}_2)$, and the resulting port-resolved and total outputs. All curves in **Extended Data Figs. 11-13** are generated directly from these procedures without additional fitting beyond the prescribed near-zero scaling regression.

**Electrochemical mapping and parameter estimation**

Polarization measurements provide pairs $\{(\eta_m, j_m)\}_{m=1}^M$. We include an effective series of Ohmic drops $R_\Omega \geq 0$ and define the effective interfacial overpotential $\eta_{\text{eff}} =$

$\eta - R_\Omega j$. We model the total current as a sum of two orthogonal channels $p \in \{A, B\}$, $j(\eta) = \sum_p j_p(\eta_{\text{eff}})$, where each channel follows intrinsically asymmetric Butler–Volmer kinetics with cathodic/anodic branch exponents $(b_p, a_p)$,

$$x_p(\eta_{\text{eff}}) = j_p^* \left[\exp(a_p \eta_{\text{eff}}) - \exp(b_p \eta_{\text{eff}})\right], \quad (M15)$$

together with a transport-limited saturation applied to the net channel current:

$$j_p(\eta_{\text{eff}}) = \frac{x_p(\eta_{\text{eff}})}{\sqrt{1 + \left[x_p(\eta_{\text{eff}})/j_{\text{lim},p}\right]^2}}. \quad (M16)$$

We parameterize $(a_p, b_p)$ using the intrinsic asymmetry variables $(\xi_p, \lambda_p)$ as $a_p = \xi_p \lambda_p \mathrm{F}$ and $b_p = -(1-\xi_p)\lambda_p \mathrm{F}$ with $\mathrm{F} = \mathcal{F}/(R_g T)$ (where $\mathcal{F}$ is the Faraday constant (96485 C/mol), $R_g$ is the universal gas constant (8.314 J/mol·K), and $T$ is the temperature), enforcing physical admissibility $j_p^* > 0$, $\xi_p \in (0,1)$, $\lambda_p \geq 0$, $j_{\text{lim},p} > 0$, and $R_\Omega \geq 0$ (see Supplementary Method S9).

Because $j$ appears in $\eta_{\text{eff}}$, each model evaluation executed at $\eta_m$ requires the scalar implicit equation below to be solved:

$$j = \sum_p j_p(\eta_m - R_\Omega j), \quad (M17)$$

which we compute via a safeguarded one-dimensional root finder with warm starts along the monotonic $\eta$ sweep (bracketing and backtracking; falling back to bisection if Newton steps leave the bracket), with the solver tolerances fixed across all conditions before conducting fitting (Supplementary Method S9.3).

The parameters are estimated by minimizing an asinh-transformed weighted least-squares objective to balance the low- and high-current regimes:

$$\min_{\Xi} \sum_m w_m \left[\operatorname{asinh}\left(\frac{j_{\text{model}}(\eta_m; \Xi)}{j_{\text{ref}}}\right) - \operatorname{asinh}\left(\frac{j_m}{j_{\text{ref}}}\right)\right]^2, \quad (M18)$$

where $\Xi$ collects $\{j_p^*, \xi_p, \lambda_p, j_{\lim,p}\}_p$, $R_\Omega$, $j_{\text{ref}}$ is fixed, and the $\{w_m\}$ are defined within each dataset (Supplementary Method S9.4). Fitting proceeds in two stages: Stage 1 fits the explicit $R_\Omega = 0$ approximation on a reduced grid to obtain stable initial values and a residual-informed two-channel split; Stage 2 refines all free parameters using the full implicit solver on the full grid. To avoid single-channel collapse and unrealistically large $R_\Omega$ values without imposing hard caps, we optionally include weak regularization terms as described in Supplementary Method S9.4. For each condition, we fit three candidate models (single-mode $A$, single-mode $B$, and the two-mode sum) and select the minimum-AIC model computed from the Stage 2 residual sum of squares and the number of free parameters (Supplementary Method S9.5).

To compute waveguide state diagnostics from polarization data alone, we evaluate the nonnegative directional fluxes using the unsaturated kinetics at the fitted $\eta_{\text{eff}}$, i.e.,

$$J_{\text{ox},p} = j_p^* \exp(a_p \eta_{\text{eff}}), J_{\text{red},p} = j_p^* \exp(b_p \eta_{\text{eff}}), \quad (M19)$$

and aggregate $J_{\text{ox}} = \sum_p J_{\text{ox},p}$ and $J_{\text{red}} = \sum_p J_{\text{red},p}$. We then form the exchange and transfer proxies $\mathcal{U} = J_{\text{ox}} + J_{\text{red}}$ and $\mathcal{S} = J_{\text{ox}} - J_{\text{red}}$, respectively, together with a continuous branch-balancing diagnostic

$$\beta_w = \frac{\mathcal{S}}{\mathcal{U}} = \frac{J_{\text{ox}} - J_{\text{red}}}{J_{\text{red}} + J_{\text{ox}}}, \varphi = \arccos(\beta_w) \in [0, \pi], \quad (M20)$$

as well as an amplitude ratio on the square-root (power-wave) scale:

$$\rho = \sqrt{\frac{1 - |\beta_w^2|}{1 + |\beta_w^2|}}. \quad (M21)$$

The polarization-derived generalized complex reflection factor is $\Gamma_g = \rho e^{i\varphi}$, which satisfies $|\Gamma_g| \leq 1$ based on its construction and provides a smooth, data-driven phase diagnostic without the need for impedance spectroscopy. All the Cai–Smith chart

coordinates and subsequent waveguide quantities reported in the main text are computed from $\Gamma_g(\eta_{\text{eff}})$ and evaluated along the fitted polarization curve.

In dimensionless form, when $\tilde{j}_p = j_p/j_{\text{lim},p}$ and $\tilde{x}_p = x_p/j_{\text{lim},p}$, the expression in (M16) simplifies to the following:

$$\tilde{j}_p = \frac{\tilde{x}_p}{\sqrt{1+\tilde{x}_p^2}}, \qquad (M22)$$

with $\tilde{x}_p = \left(\frac{j_{*,p}}{j_{\text{lim},p}}\right)\left[\exp(\xi_p\tilde{\eta}) - \exp(-(1-\xi_p)\tilde{\eta})\right]$ and $\tilde{\eta} = \lambda_p f \eta_{\text{eff}}$. This hyperbolic mapping ensures passivity and makes the transition from exponential growth to saturation explicit in a single closed form. In practice, the relevant regimes are most cleanly demarcated by the unsaturated power-flow-like channel $\mathcal{S}(\tilde{\eta})$ (equivalently $\tilde{x}$ in **Eq. M22**) rather than by fixed $\tilde{\eta}$ cutoffs. Accordingly, in **Extended Data Fig. 25** we visualize $\tilde{j}(\tilde{\eta})$ for representative asymmetries $\xi = \{0.2, 0.5, 0.8\}$ using the flux-based scheme construction ($J_{\text{ox}}, J_{\text{red}} \to \mathcal{U}, \mathcal{S}$) and mark per-curve, per-branch regime boundaries by solving $\mathcal{S}(\tilde{\eta}) = \pm \mathcal{S}_{\text{thr}}$ with $\mathcal{S}_{\text{thr}} \in \{1, 3\}$ (dotted: $|\mathcal{S}| = 1$; dashed: $|\mathcal{S}| = 3$). Because $\mathcal{S}(\tilde{\eta})$ is $\xi$-biased, the resulting boundaries occur at different $|\tilde{\eta}|$ on the positive and negative branches, making the asymmetry-shifted curvature and saturation onset directly readable. (1) Linear (exponential) regime ($|\mathcal{S}| \ll 1$): $\tilde{j} \approx \mathcal{S}$, recovering Butler-Volmer-like exponential kinetics; to first order about $\tilde{\eta} = 0$, $\mathcal{S} \approx j_p^* \lambda \tilde{\eta}$ (independent of $\xi$ at leading order), and energy is primarily directed towards transmission (power-flow-dominant, low effective mass). (2) Transition regime ($|\mathcal{S}| \approx 1$): saturation begins to limit growth, producing a pronounced curvature in $\tilde{j}(\tilde{\eta})$; intrinsic asymmetry shifts the curvature onset so that $\xi < 0.5$ advances the negative-branch departure from linearity whereas $\xi > 0.5$ advances the positive-branch departure. (3) Saturation regime ($|\mathcal{S}| \gg 1$): $\tilde{j} \to$

$\pm 1$ while the retention proxy derived from the same flux decomposition remains $\rho(\tilde{\eta}) = \sqrt{(1-|\beta_w|)/(1+|\beta_w|)} = \exp(-|\lambda\eta|/2)$ with $\beta_w = \mathcal{S}/\mathcal{U}$, highlighting how the bounded output coexists with a state-dependent storage/feedback coordinate.

**Data availability**

All data supporting the findings of this work are available in the paper and its Supplementary Information. Specifically, the external electrochemical dataset analyzed in this study was originally published in ref.[46] (Copyright 2013 Springer Nature), and was directly accessed via the Source Data files provided with ref.[47] (Copyright 2026 Springer Nature). The MATLAB scripts used for reproducing all figures are openly released at https://github.com/Huayangcai/universal-waveguide-mass-energy.

**Author contributions**

H.C. conceived the research direction and supervised the project. B.C. applied the model to fit the electrochemical data. All authors contributed to revising the manuscript and approved the final version.

**Acknowledgements**

This research was supported by the Guangdong Basic and Applied Basic Research Foundation (Grant No. 2023B1515040028), the National Natural Science Foundation of China (Grant No. 52279080, 42376097), and the Guangdong Provincial Department of Science and Technology (Grant No. 2019ZT08G090).

**Extended Data Figures**

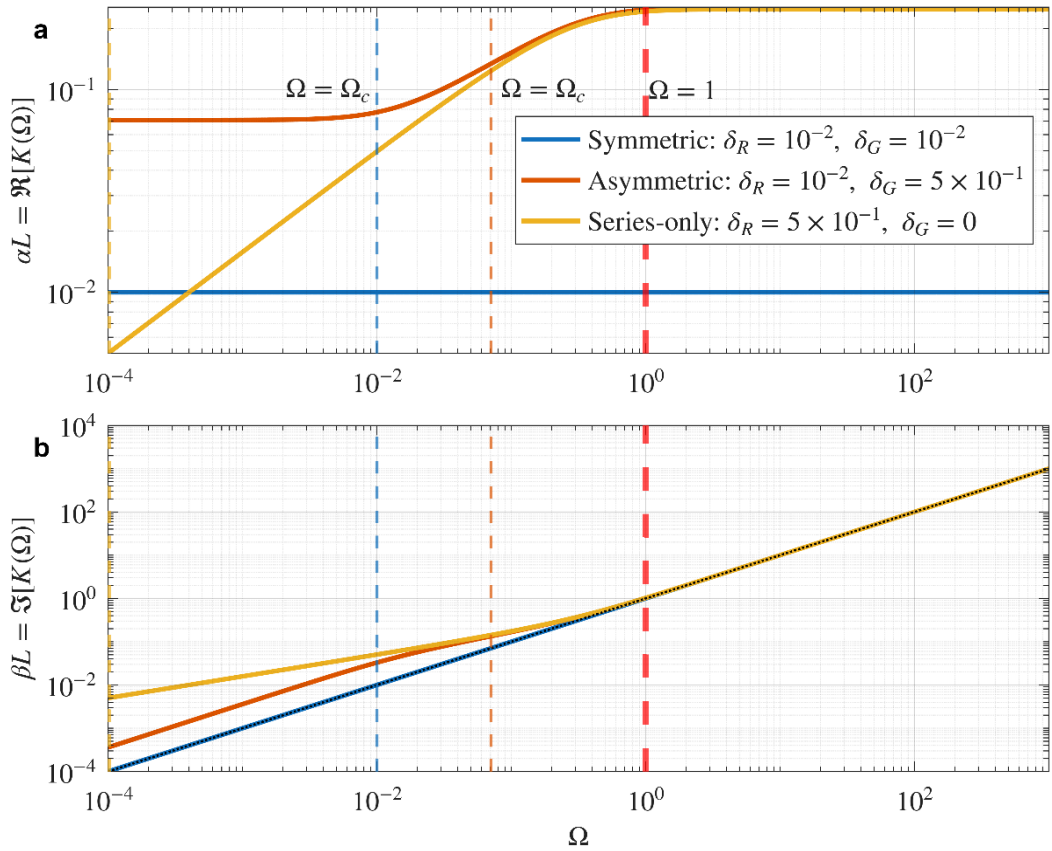

**Extended Data Fig. 1 | Dispersion classification based on the dimensionless RLGC electrical length. a.** $\alpha L = \Re[K(\Omega)]$ ; **b.** $\beta L = \Im[K(\Omega)]$ , with $K(\Omega) = \sqrt{(\delta_R + i\Omega)(\delta_G + i\Omega)}$. The solid curves show representative loss budgets: symmetric $(\delta_R = \delta_G = 10^{-2})$ , asymmetric $(\delta_R = 10^{-2}, \delta_G = 5 \times 10^{-1})$ , and series-only budgets $(\delta_R = 5 \times 10^{-1}, \delta_G = 0)$. The coloured dashed vertical lines mark the crossover point $\Omega_c = \sqrt{\delta_R \delta_G}$ at which $\Re[K(\Omega_c)] = \Im[K(\Omega_c)]$; the thick red dashed line indicates $\Omega = 1$.

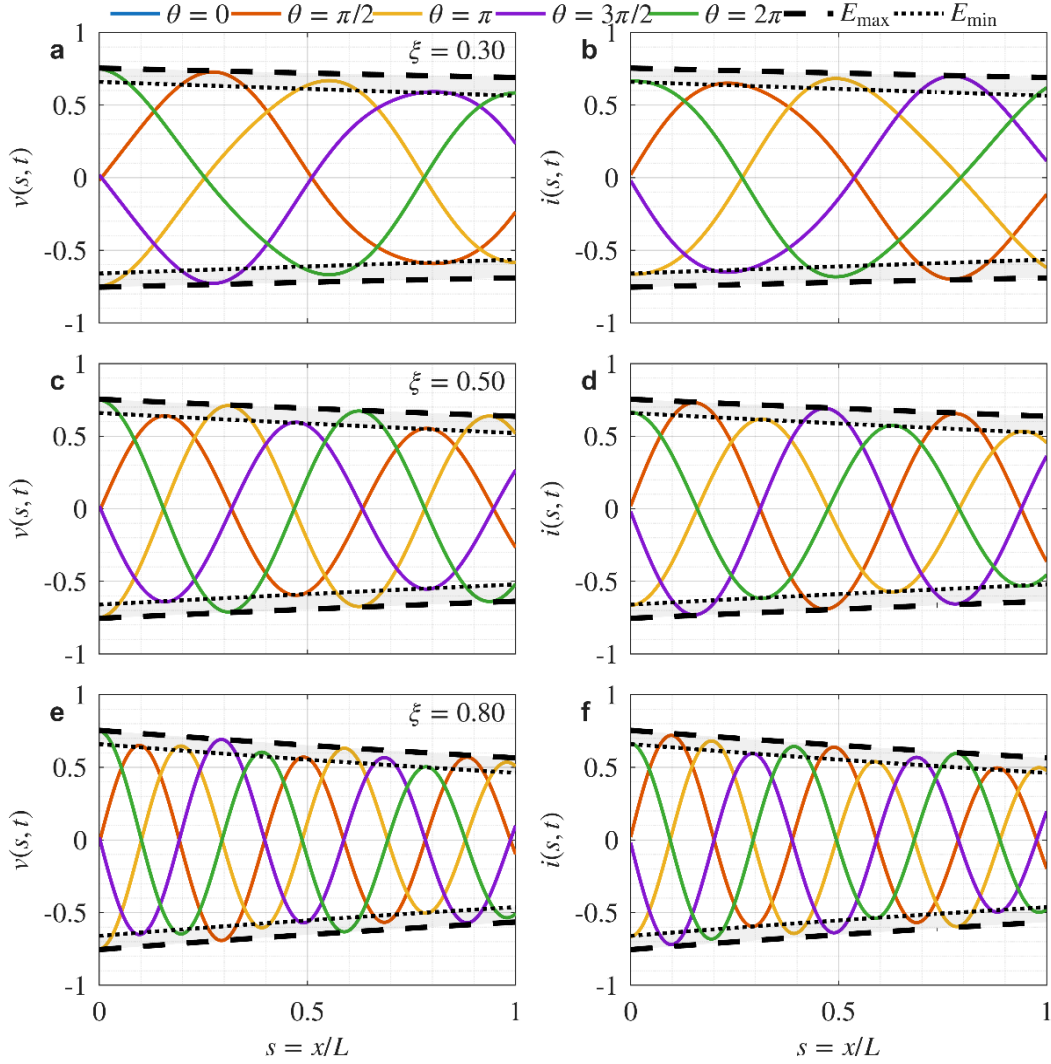

**Extended Data Fig. 2 | Asymmetry-controlled snapshots and extreme envelopes in a lossy waveguide.** Instantaneous profiles of the voltage $v(s,t)$ (left column) and current $i(s,t)$ (right column) are plotted over the normalized coordinate $s = x/L$ for phase snapshots $\theta \in \{0, \pi/2, \pi, 3\pi/2, 2\pi\}$ (coloured curves). The rows correspond to $\xi = 0.30$ (**a, b**), $\xi = 0.50$ (**c, d**), and $\xi = 0.80$ (**e, f**). The thick dashed and dotted black curves denote the phase-independent reachable envelopes $E_{\max}(s) = |A|(e^{-2\xi x} + \rho e^{2(1-\xi)x})$ and $E_{\min}(s) = |A||e^{-2\xi x} - \rho e^{2(1-\xi)x}|$ (shown as $\pm E_{\max}$ and $\pm E_{\min}$), respectively, where $x = s\,\Re[K(\Omega)]$ and $\rho = |\Gamma_g|$. Parameters: $\delta_R = 10^{-1}$, $\delta_G = 3 \times 10^{-1}$, $\Omega = 10$, $\Gamma_L = 0.1 e^{i\pi/2}$, and $\Gamma_S = 0$.

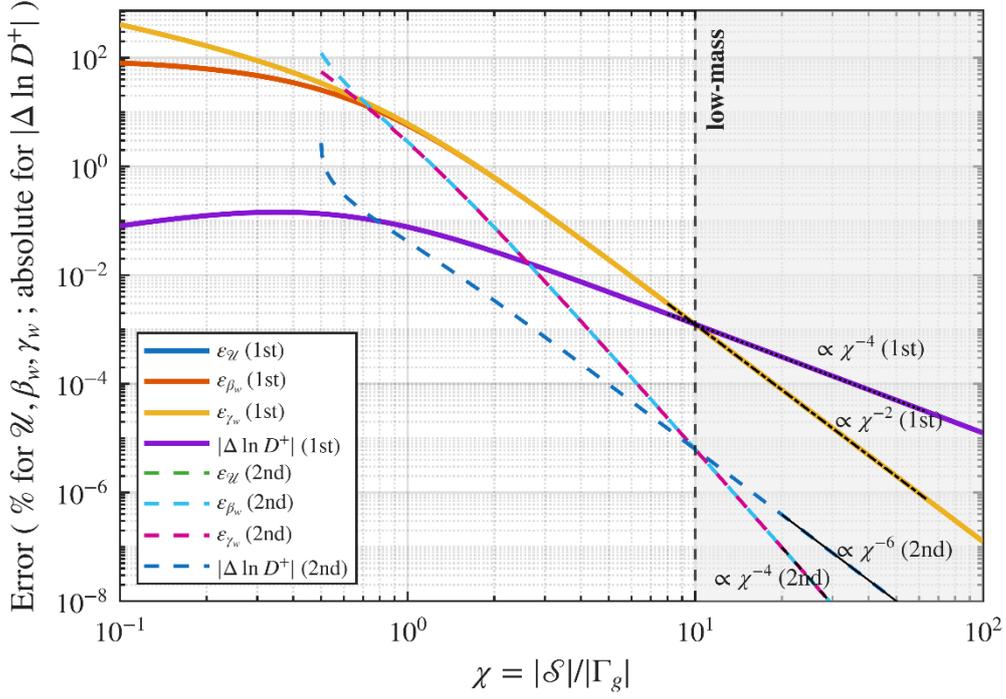

**Extended Data Fig. 3 | Universal error scaling of the low-mass approximation.** The errors induced by the low-mass expansion of the mass–energy identity $\mathcal{U}^2 = \mathcal{S}^2 + |\Gamma_g|^2$ are reported versus $\chi = |\mathcal{S}|/|\Gamma_g|$. The solid curves represent the first-order truncation results, and the dashed curves represent the second-order truncation results. We report the relative errors (in %) for $\mathcal{U}$, $\beta_w = \mathcal{S}/\mathcal{U}$, and $\gamma_w = \mathcal{U}/|\Gamma_g|$ and the absolute rapidity (log-Doppler) error $|\Delta \ln D_+|$ with $\ln D_+ = \mathrm{atanh}(\beta_w)$, which is evaluated only where $|\beta_w| < 1$. The vertical dashed line indicates the low-mass threshold $\chi_{\mathrm{th}} = 10$. The guide lines indicate the asymptotic power laws: $\propto \chi^{-4}$ for $\epsilon_\mathcal{U}, \epsilon_\beta, \epsilon_\gamma$ and $\propto \chi^{-2}$ for $|\Delta \ln D_+|$ at the first order, which improve to $\propto \chi^{-6}$ and $\propto \chi^{-4}$, respectively, at the second order.

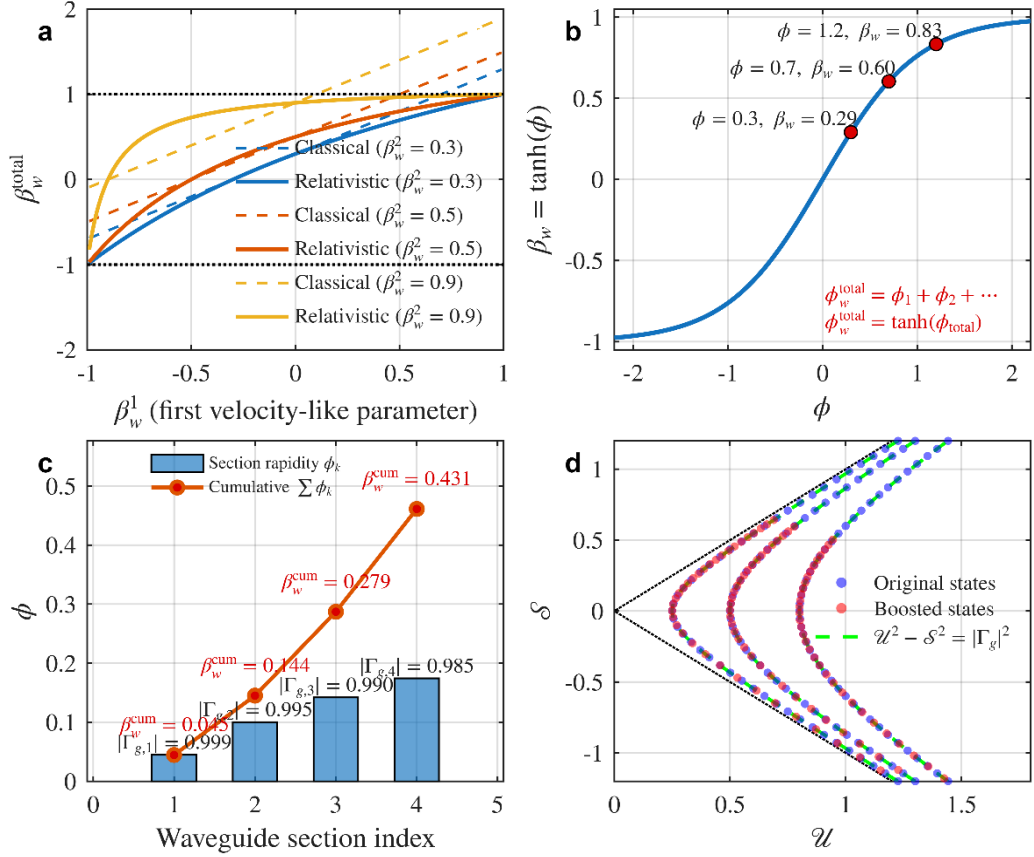

**Extended Data Fig. 4 | Waveguide relativity: velocity composition, rapidity linearization, and Lorentz boosts in $\mathcal{U}-\mathcal{S}$ space. a.** Classic (Galilean) versus Lorentz velocity additions for a fixed $\beta_w^2$; the Lorentz rule $\beta_w^{\text{tot}} = (\beta_w^1 + \beta_w^2)/(1 + \beta_w^1 \beta_w^2)$ enforces the bound $|\beta_w| \leq 1$ implied by $\beta_w = \mathcal{S}/\mathcal{U}$. **b.** Rapidity mapping $\beta_w = \tanh \phi$ with linear additivity $\phi_{\text{tot}} = \phi_1 + \phi_2 + \cdots$ and $\beta_w^{\text{tot}} = \tanh(\phi_{\text{tot}})$. **c.** Multisection composition at $\mathcal{U} = 1$: the section masses $|\Gamma_{g,k}|$ determine $\beta_w^k = \sqrt{1 - |\Gamma_{g,k}|^2}$ and $\phi_k = \operatorname{atanh}(\beta_w^k)$, whose sum yields $\beta_w^{\text{cum}}$. **d.** Lorentz boosting as a hyperbolic rotation in the $(\mathcal{U}, \mathcal{S})$ plane for preserving $\mathcal{U}^2 - \mathcal{S}^2 = |\Gamma_g|^2$ (green dashed), bounded by the light cone $\mathcal{U} = |\mathcal{S}|$ (black dotted).

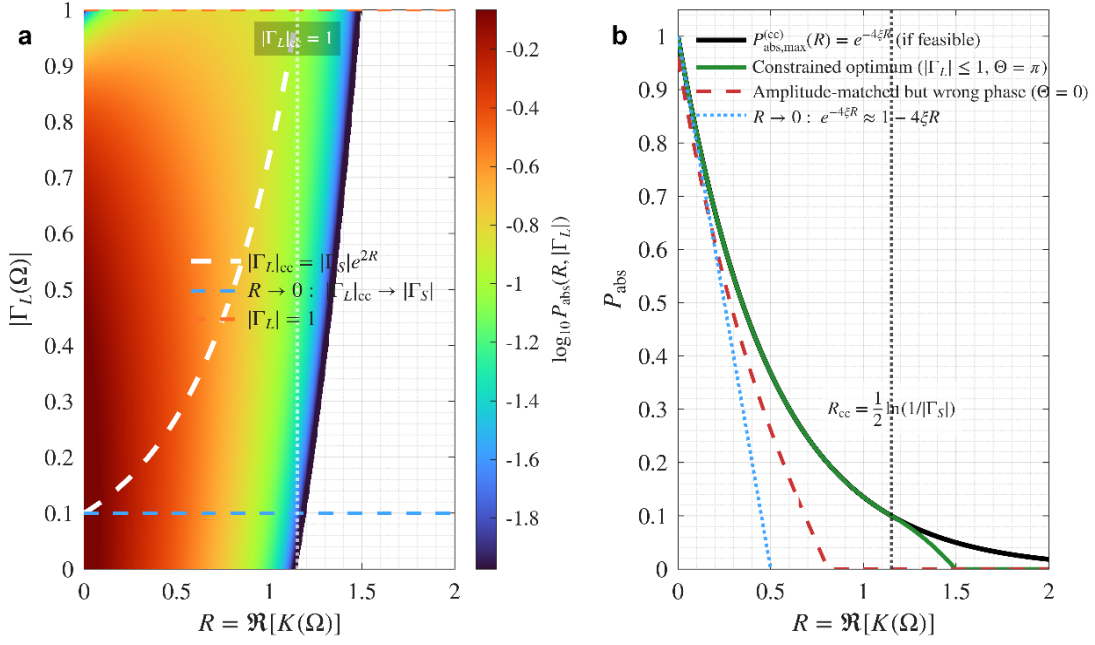

**Extended Data Fig. 5 | Resonance-aware absorbed power landscape and feasibility-limited critical coupling optimum. a.** Heatmap of the normalized absorbed power $P_{\text{abs}}(R, |\Gamma_L|)$ at $\xi = 0.5$ and $|\Gamma_S| = 0.1$, evaluated on the constructive (odd-$\pi$) interference class $\Theta = \pi$ (set here by $\phi_S = 0°$ and $\phi_L = 180°$). The colours encode $\log_{10} P_{\text{abs}}$ over the passive load domain $0 \leq |\Gamma_L| \leq 1$. The white dashed curve indicates the critical coupling condition $|\Gamma_L|_{\text{cc}} = |\Gamma_S| e^{2R}$ for which $\bar{\Gamma}_g = 0$ and $P_{\text{abs}} = e^{-4\xi R}$; the blue dashed horizontal line indicates the small-loss limit $R \to 0$ where $|\Gamma_L|_{\text{cc}} \to |\Gamma_S|$. The vertical dotted line at $R_{\text{cc}} = \frac{1}{2}\ln(1/|\Gamma_S|)$ indicates where $|\Gamma_L|_{\text{cc}} = 1$, beyond which critical coupling is infeasible under passivity. **b.** Absorbed power optimum versus $R$: the black curve shows the feasible critical coupling bound $P_{\text{abs,max}}^{(\text{cc})}(R) = e^{-4\xi R}$; the green curve shows the constrained optimum under $|\Gamma_L| \leq 1$ at $\Theta = \pi$, which follows the bound for $R \leq R_{\text{cc}}$ and rolls off once $|\Gamma_L|$ saturates at the passive boundary. The red dashed curve shows an amplitude-matched but phase-mismatched case ($\Theta = 0$); the blue dotted line gives the small-$R$ expansion $e^{-4\xi R} \approx 1 - 4\xi R$.

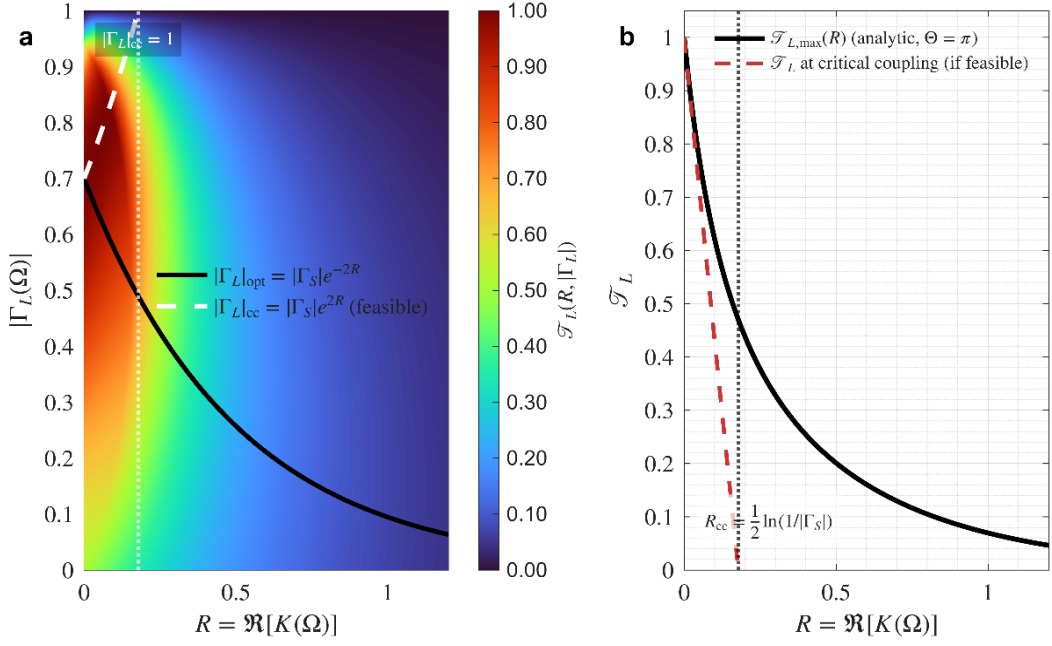

**Extended Data Fig. 6 | Resonance-aware useful load power landscape and its analytic maximum under fixed attenuation conditions. a.** Useful load power fraction $\mathcal{T}_L(R, |\Gamma_L|)$ on the odd-$\pi$ class ($\Theta = \pi$), computed from $F = \Gamma_S \Gamma_L e^{-2K}$ with $K = R + i\Phi$ (here, $\Phi = \Phi_0$ and is fixed), $\bar{\Gamma}_g = (\Gamma_L e^{-2K} + \Gamma_S)/(1 + F)$, $\mathcal{A} = 1 - |\bar{\Gamma}_g|^2$, and $\mathcal{T}_L = \mathcal{A}(1 - |\Gamma_L|^2)e^{-2R}/|1 + F|^2$. The black curve represents the useful power optimum $|\Gamma_L|_{\text{opt}} = |\Gamma_S| e^{-2R}$; the white dashed curve represents the critical coupling branch $|\Gamma_L|_{\text{cc}} = |\Gamma_S| e^{2R}$ (where feasible); and the vertical dotted line represents $|\Gamma_L|_{\text{cc}} = 1$ at $R_{\text{cc}} = \frac{1}{2}\ln(1/|\Gamma_S|)$. **b.** Analytic maximum $\mathcal{T}_{L,\max}(R) = \frac{(1-|\Gamma_S|^2)e^{-2R}}{1-|\Gamma_S|^2 e^{-4R}}$ (black) compared with the $\mathcal{T}_L$ evaluated on the critical coupling branch (red dashed line) over $R \leq R_{\text{cc}}$. Parameters: $|\Gamma_S| = 0.7$, $\arg \Gamma_S = 0$, and $\Phi_0 = 0$ (with $\Theta = \pi$ enforced by fixing $\arg \Gamma_L$).

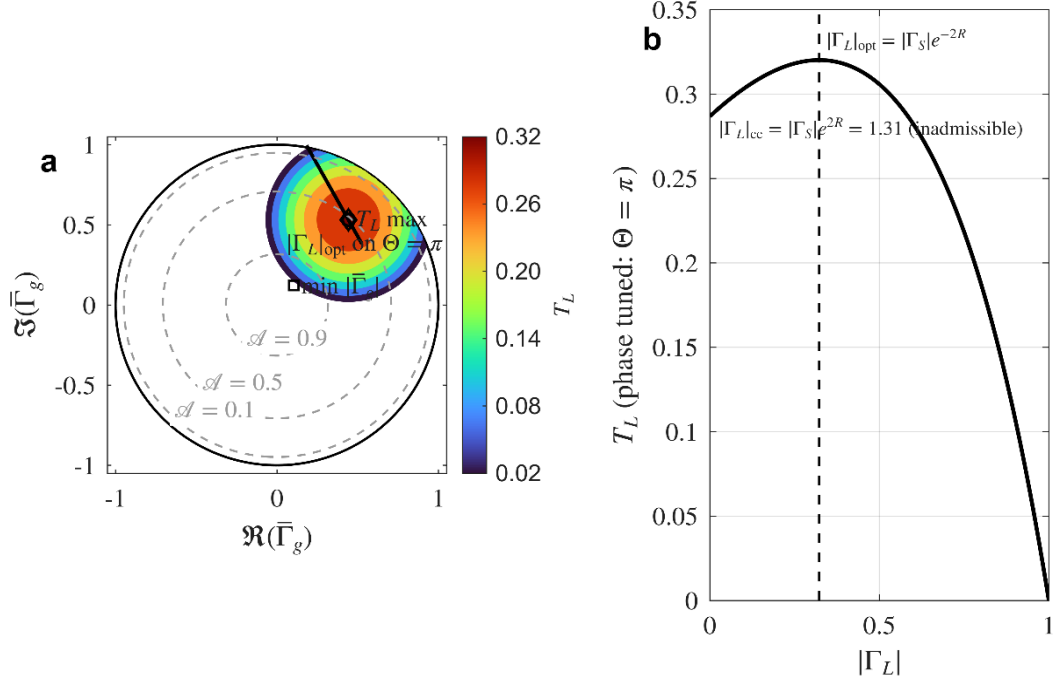

**Extended Data Fig. 7 | Useful load power versus one-port absorptivity on the Cai–Smith chart. a.** Cai–Smith representation of the boundary-composed reflection $\bar{\Gamma}_g(\Omega)$ for a finite lossy guide with $K(\Omega) = R + i\Phi$ and a round-trip factor $E(\Omega) = e^{-2K(\Omega)} = e^{-2R}e^{-2i\Phi}$. The source reflection is fixed to $\Gamma_S = |\Gamma_S|e^{i\arg \Gamma_S}$ (here, $|\Gamma_S| = 0.65$, and $\arg \Gamma_S = 0.2\pi$), and $R = 0.35$ and $\Phi = 0.9\pi$. Scanning the passive load $\Gamma_L$ over $|\Gamma_L| \leq 1$ generates the reachable set of $\bar{\Gamma}_g$. The colourmap reports the load-delivered power fraction $T_L = \frac{(1-|\Gamma_S|^2)(1-|\Gamma_L|^2)e^{-2R}}{|1+\Gamma_S\Gamma_L E|^2}$ (with the colour scale matched to the reachable range). The grey dashed circles denote the absorptivity contours $\mathcal{A} = 1 - |\bar{\Gamma}_g|^2$ produced at $\mathcal{A} = 0.1, 0.5, 0.9$. The black curve indicates the resonant phase-tuned path $\Theta = \arg(\Gamma_S\Gamma_L E) = \pi$, along which the resonance-constrained optimum occurs at $|\Gamma_L|_{\text{opt}} = |\Gamma_S|e^{-2R}$ (diamond). **b.** Under $\Theta = \pi$, $T_L$ versus $|\Gamma_L|$ exhibits a unique interior maximum at $|\Gamma_L|_{\text{opt}} = |\Gamma_S|e^{-2R}$, whereas one-port critical coupling ($\bar{\tilde{\Gamma}}_g = 0$) would require $|\Gamma_L|_{\text{cc}} = |\Gamma_S|e^{2R} > 1$ (inadmissible),

demonstrating that maximizing the useful delivered power is generally distinct from maximizing the total absorption rate.

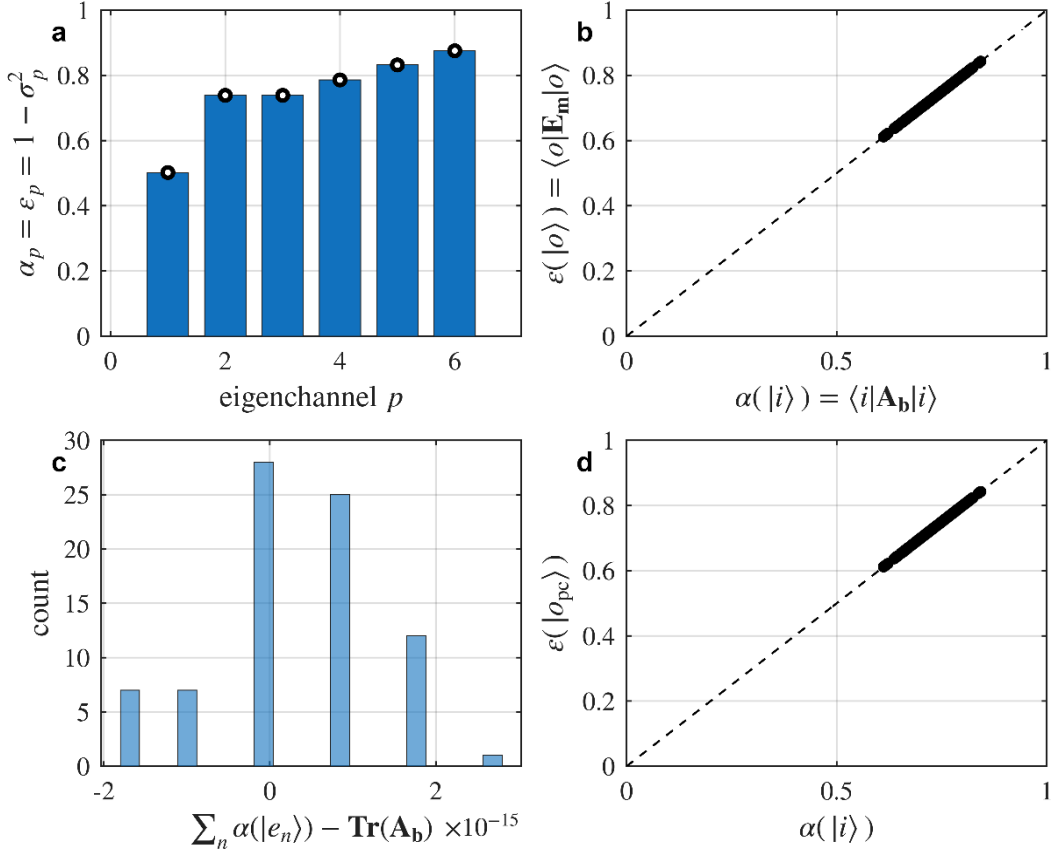

**Extended Data Fig. 8 | Four fundamental laws for power absorption and emissions in finite, linear, passive waveguides. a.** Law 1 (eigenchannel equivalence): on the eigenchannel basis, $\alpha_p = \epsilon_p = 1 - \sigma_p^2$ for each channel $p$. **b.** Law 2 (same modal power splitting scheme): for trials where the input $|i\rangle$ and output $|o\rangle$ share identical eigenchannel intensity weights $\{|c_p|^2\}$, the emissions equal the absorption level, i.e., $\epsilon(|o\rangle) = \alpha(|i\rangle)$. **c.** Law 3 (basis-independent sum rule): $\sum_n \alpha(|e_n\rangle) = \mathbf{Tr}(\mathbf{A_b})$ for any complete orthonormal input basis $\{|e_n\rangle\}$. **d.** Law 4 (reciprocal phase-conjugate pairing): for reciprocal systems, $\epsilon(|o_{\mathrm{pc}}\rangle) = \alpha(|i\rangle)$ for the phase-conjugate paired output. All panels are evaluated at a fixed $\Omega^*$ for a reciprocal multiport scattering

operator $\mathbf{S}(\Omega^*)$ with $\mathbf{S} = \mathbf{U\Sigma V^\dagger}$ and singular values $\{\sigma_p\}$. Absorption and emissions are quantified by $\mathbf{A_b} = \mathbf{I} - \mathbf{S^\dagger S}$ and $\mathbf{E_m} = \mathbf{I} - \mathbf{SS^\dagger}$, respectively, yielding $\alpha(|i\rangle) = \langle i|\mathbf{A_b}|i\rangle$ and $\epsilon(|o\rangle) = \langle o|\mathbf{E_m}|o\rangle$.

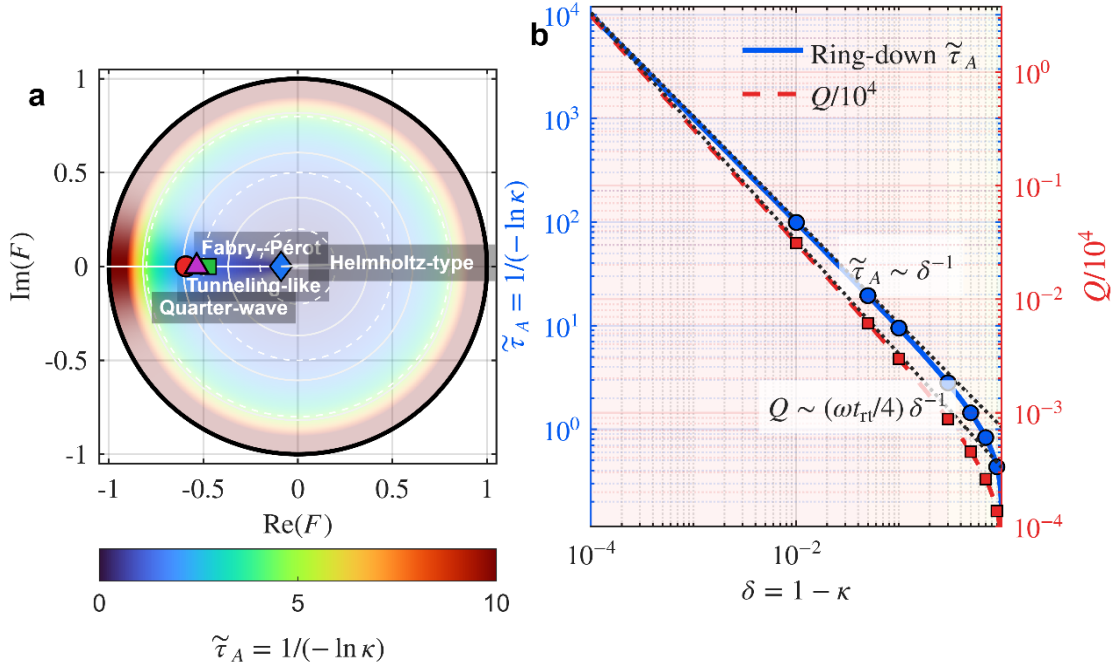

**Extended Data Fig. 9 | Cai–Smith map of the ring-down time and quality factor. a.** Cai–Smith chart of the complex round-trip feedback factor $F = \kappa e^{i\Theta}$ ($|F| \leq 1$). The colours encode the dimensionless amplitude ring-down time $\tilde{\tau}_A = \tau_A/T_{RT} = 1/(-\ln\kappa)$ (capped at 10 for display purposes), which diverges as $\kappa \to 1^-$. The dashed circles indicate constant $\kappa$ levels, and the white contours indicate the representative $\tilde{\tau}_A$ isolines. Transparency is enhanced within a narrow detuning corridor around the odd-$\pi$ direction using $\exp[-(\Delta\Theta/\Theta_\Sigma)^2]$ with $\Delta\Theta = \text{wrap}(\Theta - \pi)$. Markers indicate canonical archetypes (quarter-wave, Fabry–Pérot, Helmholtz-type, and tunnelling-like archetypes). **b.** Universal $\delta$-domain scaling with $\delta = 1 - \kappa$ (log-log): the exact relation $\tilde{\tau}_A = 1/\{-\ln(1-\delta)\}$ follows $\tilde{\tau}_A \sim \delta^{-1}$ for $\delta \ll 1$. The quality factor is computed as $Q = (\omega_{\text{trt}}/4)\tilde{\tau}_A$ (here, $\omega_{\text{trt}}/4 = \pi$) and is shown as $Q/10^4$, yielding $Q \sim (\omega_{\text{trt}}/4)\delta^{-1}$.

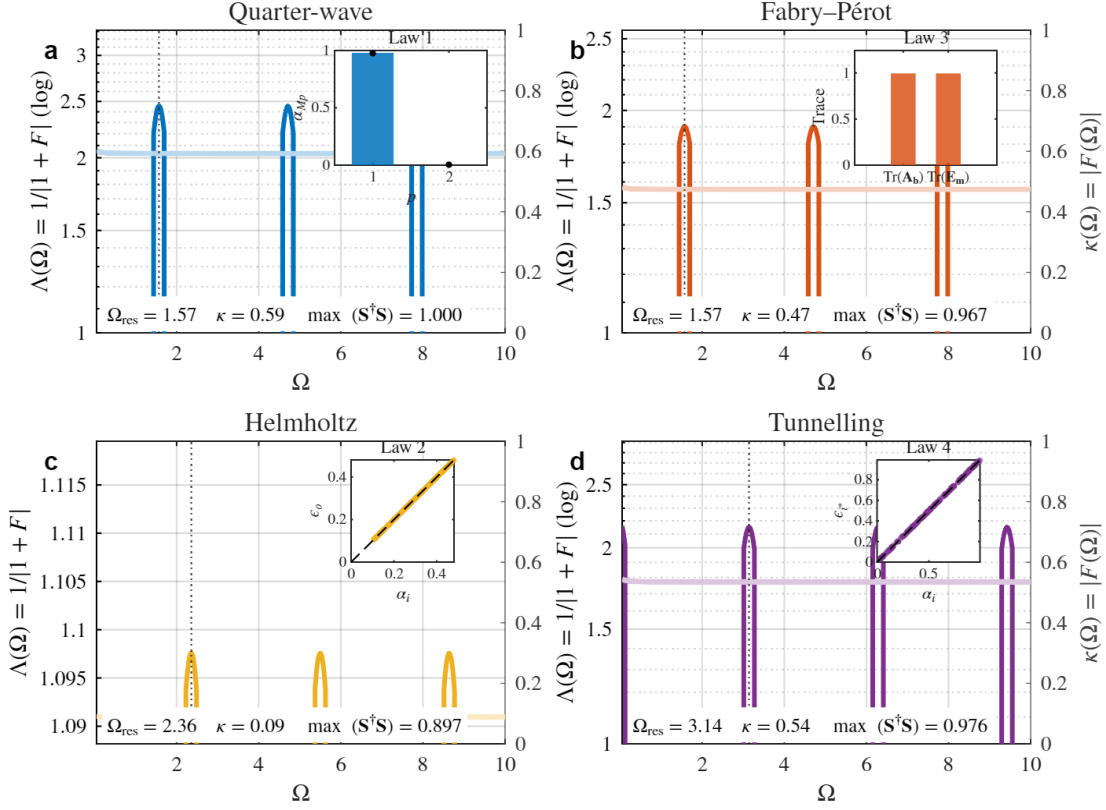

**Extended Data Fig. 10 | Selected waveguide law diagnostics for four canonical resonator archetypes.** For each structure (quarter-wave, Fabry–Pérot, Helmholtz, and tunnelling archetypes, **a-d**), the feedback factor $F(\Omega) = \Gamma_S \Gamma_L e^{-2K(\Omega)}$ defines the denominator $D(\Omega) = 1 + F(\Omega)$ and enhancement effect $\Lambda(\Omega) = 1/|D(\Omega)|$ (left axis; log scale except for the Helmholtz archetype, where a linear scale resolves the weak peak), together with the round-trip survival rate $\kappa(\Omega) = |F(\Omega)|$ (right axis). The vertical dotted line indicates the corridor-selected resonance $\Omega_{\text{res}}$ (the maximum $\Lambda$ under $|\arg F - \pi| < \Theta_\Sigma$), and the annotations report $(\Omega_{\text{res}}, \kappa(\Omega_{\text{res}}), \max(\mathbf{S}^\dagger \mathbf{S}))$ as a passivity check for a passive reciprocal two-port realization at $\Omega_{\text{res}}$. The insets validate a resonator-appropriate subset of Laws 1-4 at $\Omega_{\text{res}}$, using $\mathbf{A_b} = \mathbf{I} - \mathbf{S}^\dagger \mathbf{S}$ and $\mathbf{E_m} = \mathbf{I} - \mathbf{S}\mathbf{S}^\dagger$.

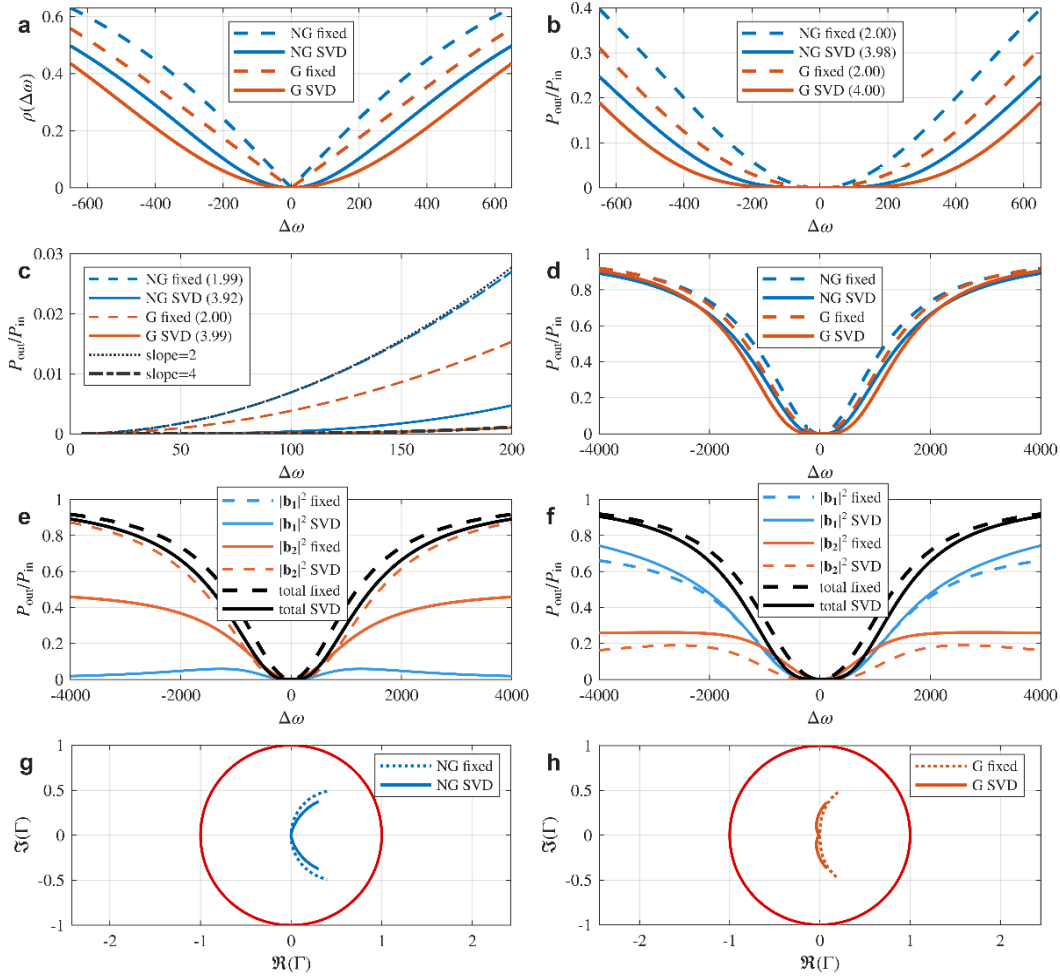

**Extended Data Fig. 11 | Waveguide-theoretic reconstruction of optical CPA-EPs under fixed coherent driving versus SVD eigenchannel probing.** A two-mode coupled-resonator scattering model is evaluated at real detuning $\Delta\omega = \omega - \omega_0$ for nongeneric (NG; symmetric coupling) and generic (G; asymmetric coupling) CPA exceptional points. Solid curves denote a fixed coherent setting in which the input state is tuned at $\Delta\omega = 0$ to maximize absorption and then held constant during the detuning scan; dashed curves denote an SVD probe in which the input is re-optimized at each $\Delta\omega$ as the minimum-singular-vector channel, yielding the minimum achievable output. **a.** Channel "mass" $\rho(\Delta\omega) = \sqrt{P_{\text{out}}/P_{\text{in}}}$ for NG and G under the two probing protocols. **b.** Effective one-channel output spectrum $P_{\text{out}}/P_{\text{in}}$ highlighting quadratic versus quartic suppression near $\Delta\omega = 0$ depending on CPA-EP class and probing

protocol; numbers in parentheses indicate fitted near-zero exponents. **c.** Near-zero scaling of the total two-port output $P_{\text{out}}$ versus $|\Delta\omega|$ with slope-2 and slope-4 guides, demonstrating that NG exhibits quadratic behavior under fixed driving but approaches quartic suppression under SVD optimization, whereas G remains quartic. **d.** Total two-port output spectra versus $\Delta\omega$ comparing fixed and SVD protocols for NG and G. **e, f.** Port-resolved outputs $|\mathbf{b_1}|^2$ and $|\mathbf{b_2}|^2$ (normalized by $P_{\text{in}}$) and their sum (black) for NG (**e**) and G (**f**), showing how optimal eigenchannel probing reshapes the detuning-dependent power partition while reducing the total output. **g, h.** Cai–Smith chart trajectories of the driven channel's complex waveguide state proxy $\Gamma\Delta\omega) = \mathbf{a}^{\dagger}\mathbf{S}(\Delta\omega)\mathbf{a}$ for NG (**g**) and G (**h**); the unit-circle boundary (red) indicates the passivity bound, and the distinct fixed-versus-SVD trajectories provide a geometric diagnostic of the CPA-EP class and probing protocol.

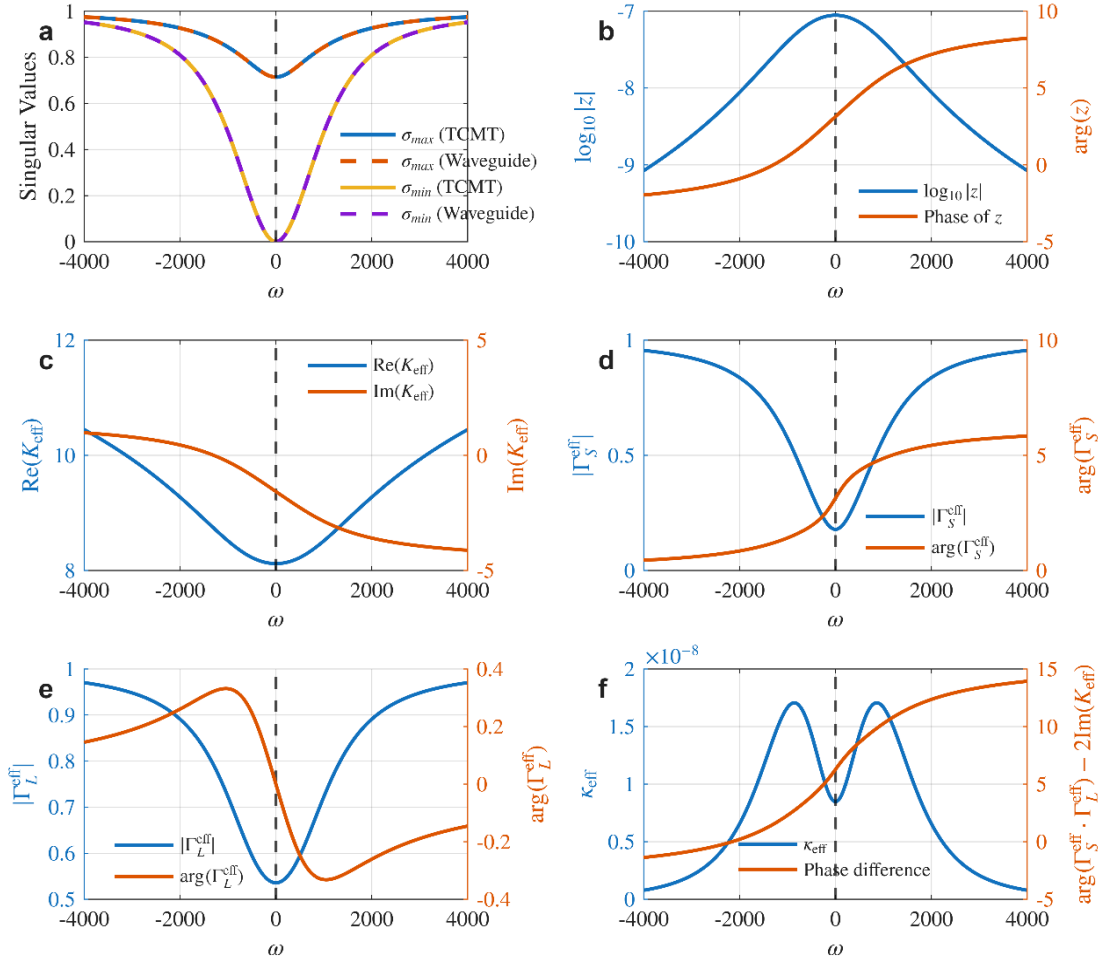

**Extended Data Fig. 12 | Frequency-domain mapping of a two-resonator temporal coupled-mode theory (TCMT) model to an equivalent waveguide representation.** At each frequency $\omega$, the scattering coefficients $(\mathbf{r}_1, \mathbf{r}_2, \mathbf{t})$ of the TCMT model are mapped to a waveguide-form scattering matrix $\hat{\mathbf{S}}_{\text{WG}}(\omega)$ by solving for the complex round-trip factor $z(\omega) = e^{-u(\omega)+iv(\omega)}$ and the effective boundary reflectivities $\Gamma_S^{\text{eff}}(\omega)$ and $\Gamma_L^{\text{eff}}(\omega)$. The waveguide closure relation $D(\omega) = 1 + \Gamma_S^{\text{eff}}(\omega)\Gamma_L^{\text{eff}}(\omega)z(\omega)$ is enforced, with $\mathbf{t}(\omega) = \sqrt{z(\omega)}\,\tau_S\tau_L/D(\omega)$. **a.** Singular values $\sigma_{\max}(\omega)$ (red) and $\sigma_{\min}(\omega)$ (blue) of the original $\mathbf{S}_{\text{TCMT}}$ (solid lines) and the reconstructed $\hat{\mathbf{S}}_{\text{WG}}$ (dashed lines). The two sets of curves coincide within machine precision. The condition $\sigma_{\min} \to 0$ at $\omega = 0$ indicates coherent perfect absorption (CPA). **b.** Magnitude $|z(\omega)|$ (left axis, log scale) and phase $\arg z(\omega)$ (right axis) of

the extracted round-trip factor. **c.** Real and imaginary parts of the effective propagation constant $K_{\text{eff}}(\omega) = -\frac{1}{2}\log z(\omega)$, obtained via phase unwrapping. **d, e.** Magnitude (left axes) and phase (right axes) of the frequency-dependent effective source and load boundaries, $\Gamma_S^{\text{eff}}(\omega)$ and $\Gamma_L^{\text{eff}}(\omega)$. **f.** Inferred feedback magnitude $\kappa_{\text{eff}}(\omega) = |\Gamma_S^{\text{eff}}(\omega)\Gamma_L^{\text{eff}}(\omega)|\, e^{-2\text{Re}(K_{\text{eff}}(\omega))}$ (left axis) and the corresponding phase-closure condition $\arg(\Gamma_S^{\text{eff}}\Gamma_L^{\text{eff}}) - 2\,\text{Im}(K_{\text{eff}})$ (right axis). The vertical dashed line at $\omega = 0$ marks the CPA exceptional point (CPA-EP) condition, where the TCMT parameters satisfy the CPA balance and the waveguide inversion yields the corresponding effective parameters.

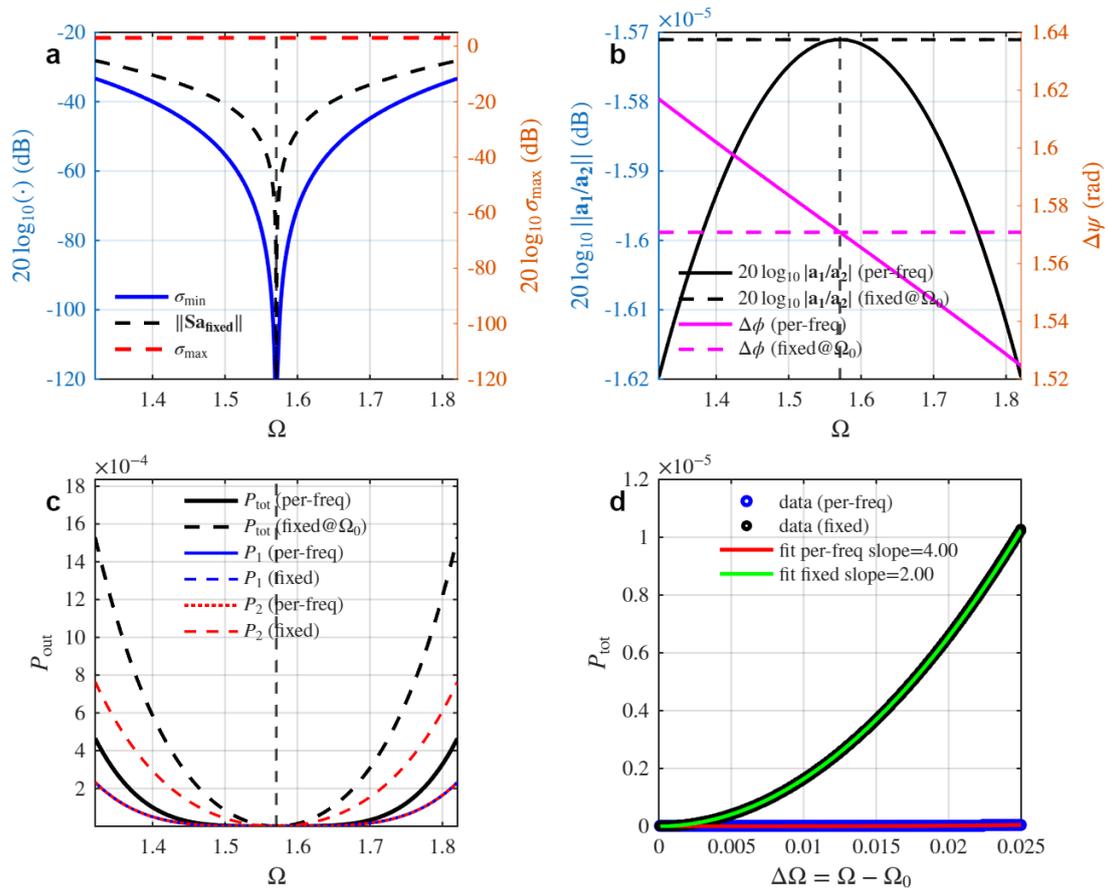

**Extended Data Fig. 13 | Comparison of coherent input strategies in a two-port waveguide system with CPA-EP at given frequency $\Omega_0 = \pi/2$.** Model and parameters: We use the reciprocal two-port waveguide form with effective propagation

constant $K(\Omega) = \sqrt{(i\Omega + \delta_R)(i\Omega + \delta_G)}$. The round-trip factor is $E(\Omega) = e^{-2K(\Omega)}$ and the common denominator is $D(\Omega) = 1 + \Gamma_S \Gamma_L E(\Omega)$, giving $\mathbf{r_1} = (\Gamma_S + \Gamma_L E)/D$, $\mathbf{r_2} = (\Gamma_L + \Gamma_S E)/D$, and $\mathbf{t} = (e^{-K} \tau_S \tau_L)/D$ with $\tau_{S,L} = \sqrt{1 - |\Gamma_{S,L}|^2}$. In the representative low-loss setting shown here, $\delta_R = \delta_G = 10^{-5}$, $\Gamma_S = 0.486 e^{i0}$, and $\Gamma_L = |\Gamma_L| e^{i\pi}$ where the CPA condition is obtained by solving $N(\Omega_0) = 0$, yielding $|\Gamma_L| = 0.337$ and $\Omega_0 = \pi/2$ (vertical dashed line). **a.** Singular values and output norms. The blue curve shows the minimum singular value $\sigma_{\min}(\Omega)$ of the scattering matrix, while the black dashed curve indicates the effective output norm $\|\mathbf{S}(\Omega)\mathbf{a}_{\text{fixed}}\|$ obtained using the input vector optimized at the CPA frequency $\Omega_0$ (vertical dashed line). $\sigma_{\max}(\Omega)$ is shn in red dashed for comparison. **b.** Optimal coherent input parameters. Black curves (left axis) show the amplitude ratio $20\log_{10} |\mathbf{a_1}/\mathbf{a_2}|$ in dB; magenta curves (right axis) show the phase difference $\Delta\psi = \arg(\mathbf{a_1}/\mathbf{a_2})$. Solid lines correspond to the per-frequency optimal input, dashed lines to the fixed input optimized at $\Omega_0$. **c.** Output powers. Total output power $P_{\text{tot}}$ (black), port 1 power $P_1$ (blue) and port 2 power $P_2$ (red) are shown for both strategies (solid: per-frequency; dashed: fixed at $\Omega_0$). **d.** Scaling of total output power with frequency detuning. Log-log plot of $P_{\text{tot}}$ versus $\Delta\Omega = \Omega - \Omega_0$ for the frequency range to the right of $\Omega_0$. Markers represent data for per-frequency and fixed strategies with corresponding linear fits, giving slopes $\simeq 4.00$ (per-frequency) and $\simeq 2.00$ (fixed@$\Omega_0$) for this parameter setting.

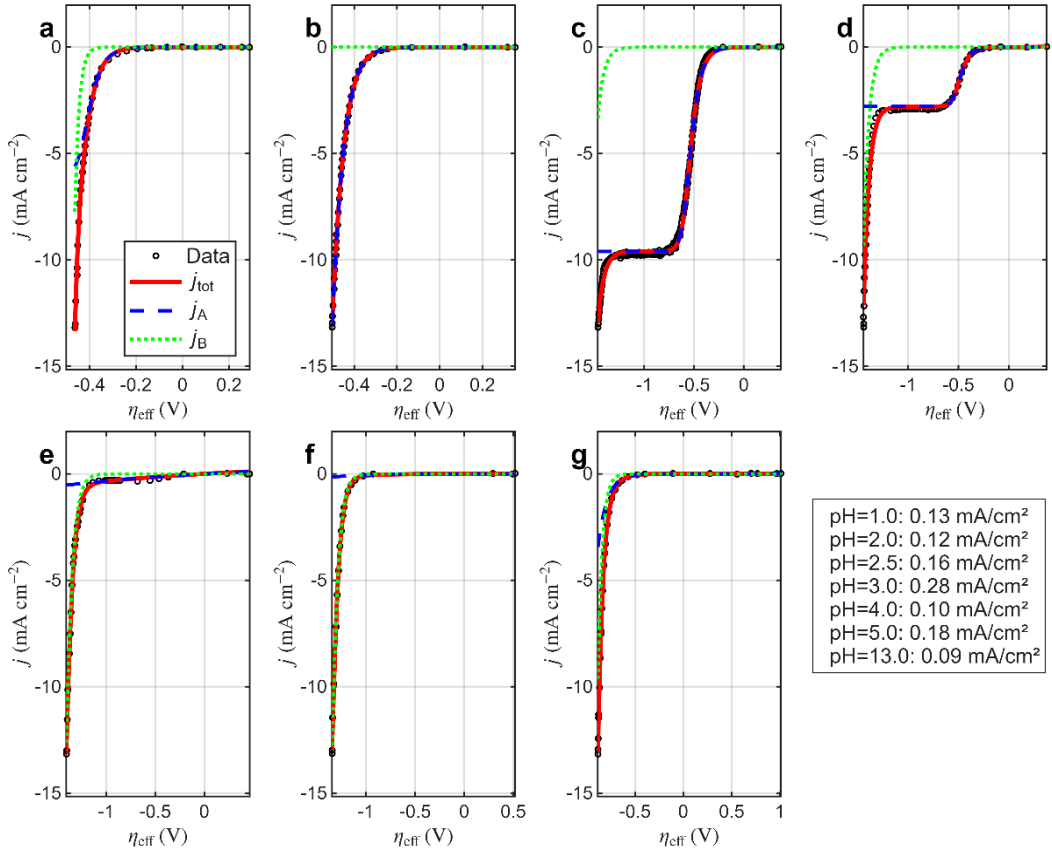

**Extended Data Fig. 14 | Polarization curves and waveguide-invariant two-mode fitting results obtained on Au(111) across different pH values.** Measured[46,47] (Copyright 2013 and 2026 Springer Nature) polarization curves $j(\eta)$ (open circles) and best-fitting two-mode model responses (solid) for electrolytes at pH =1, 2, 2.5, 3, 4, 5, and 13 (**a–g**). The current density $j$ versus the effective interfacial overpotential $\eta_{\text{eff}}$ (including ohmic correction) is shown. The fitted total current $j_{\text{tot}}$ is decomposed into two orthogonal modal contributions $j_A$ (dashed) and $j_B$ (dotted), revealing a systematic, pH-dependent rebalancing scheme while capturing both the near-equilibrium and strongly cathodic regimes.

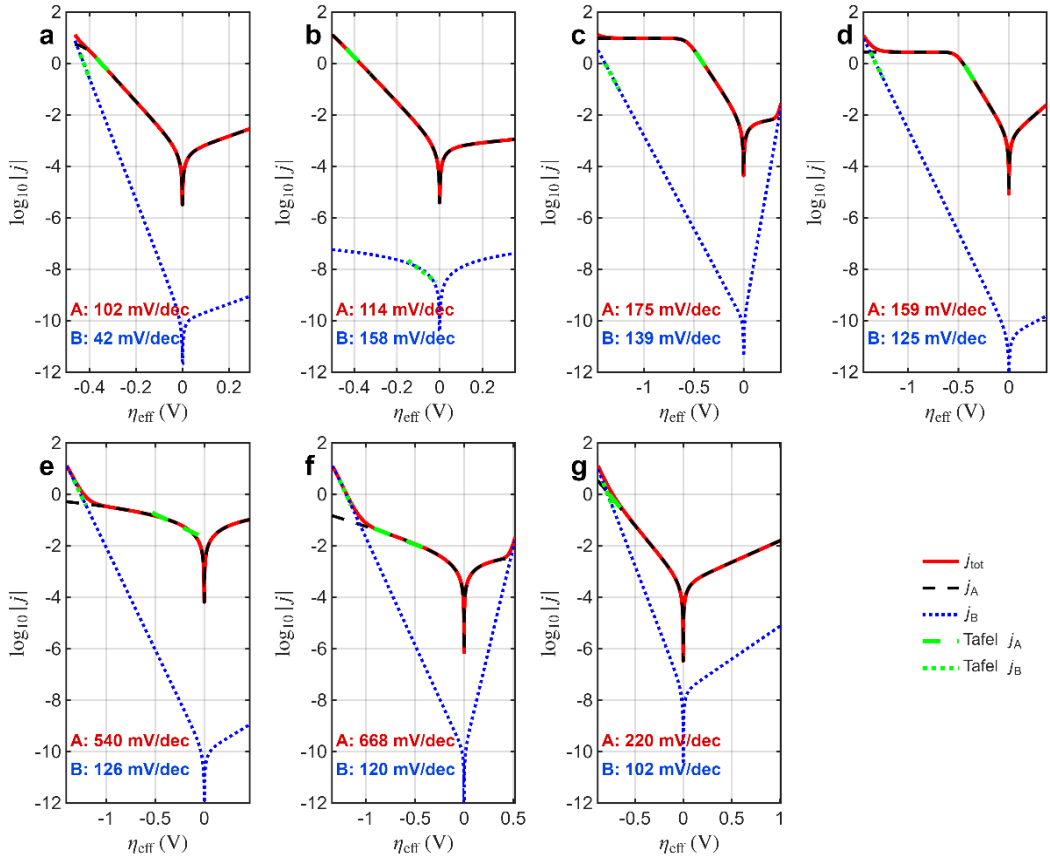

**Extended Data Fig. 15 | Modal Tafel analysis of the waveguide-invariant decomposition process on Au(111) across different pH values.** Semilogarithmic Tafel representations of $\log_{10} |j|$ versus $\eta_{\text{eff}}$ for pH =1, 2, 2.5, 3, 4, 5, and 13 (**a–g**). The fitted total current $j_{\text{tot}}$ (solid) is decomposed into orthogonal modal contributions $j_A$ (dashed) and $j_B$ (dotted). The green segments indicate the best linear Tafel regimes for each mode, with the modal Tafel slopes (mV dec$^{-1}$) annotated.

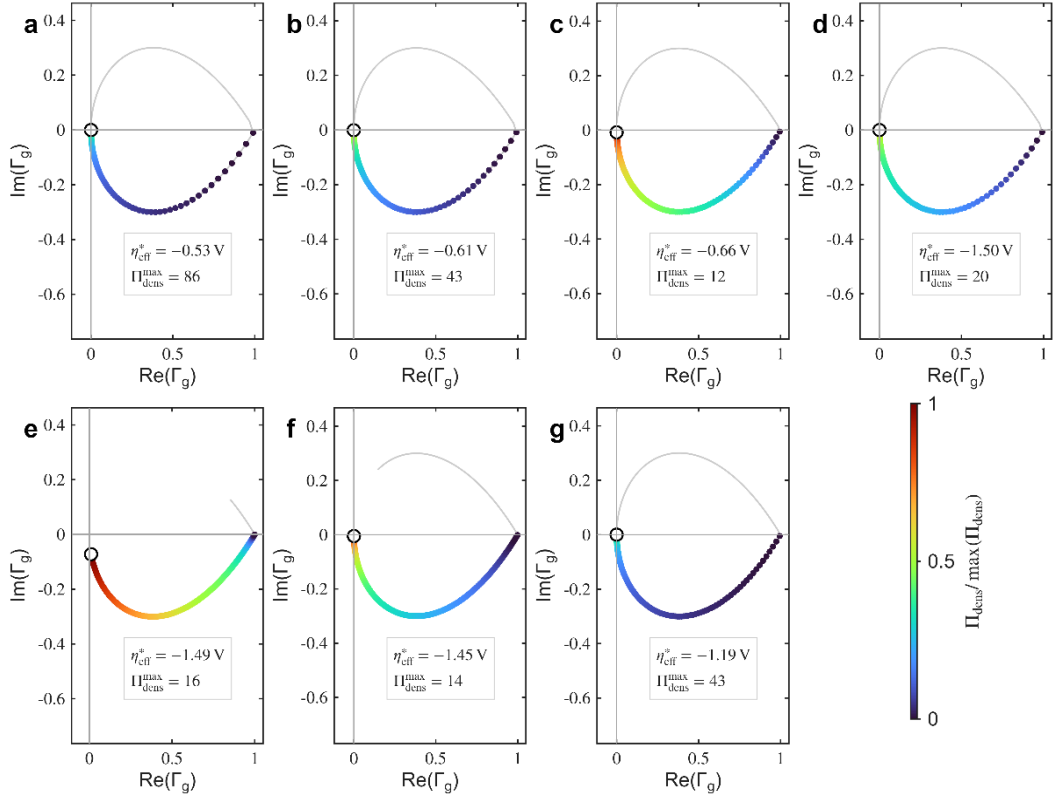

**Extended Data Fig. 16 | Cai–Smith plots coloured according to the normalized cathodic output densities obtained for Au(111) across different pH values.** For each pH, the fitted two-mode polarization model is mapped to the generalized reflection factor $\Gamma_g(\eta_{\text{eff}}) = \rho e^{i\varphi}$ on the Cai–Smith chart and coloured according to the per-pH normalized output density $\Pi_{\text{dens}}(\eta_{\text{eff}})/\max_{\eta_{\text{eff}}<0} \Pi_{\text{dens}}$ (**a–g**). Here, $\Pi_{\text{dens}} = \Pi_{\text{use}}/(|\eta_{\text{eff}}| + \eta_0)$ with $\Pi_{\text{use}} = \max(0, -j_{\text{tot}})(1-\rho^2)$ on the cathodic (HER) branch, and $\eta_0 > 0$ regularizes the density near $\eta_{\text{eff}} = 0$. The open circle indicates the origin; the faint arc indicates the unit-circle reference. The annotated marker highlights the density-optimal operating point $\eta_{\text{eff}}^*$ together with $\Pi_{\text{dens}}^{\max} = \max_{\eta_{\text{eff}}<0} \Pi_{\text{dens}}$ for each pH.

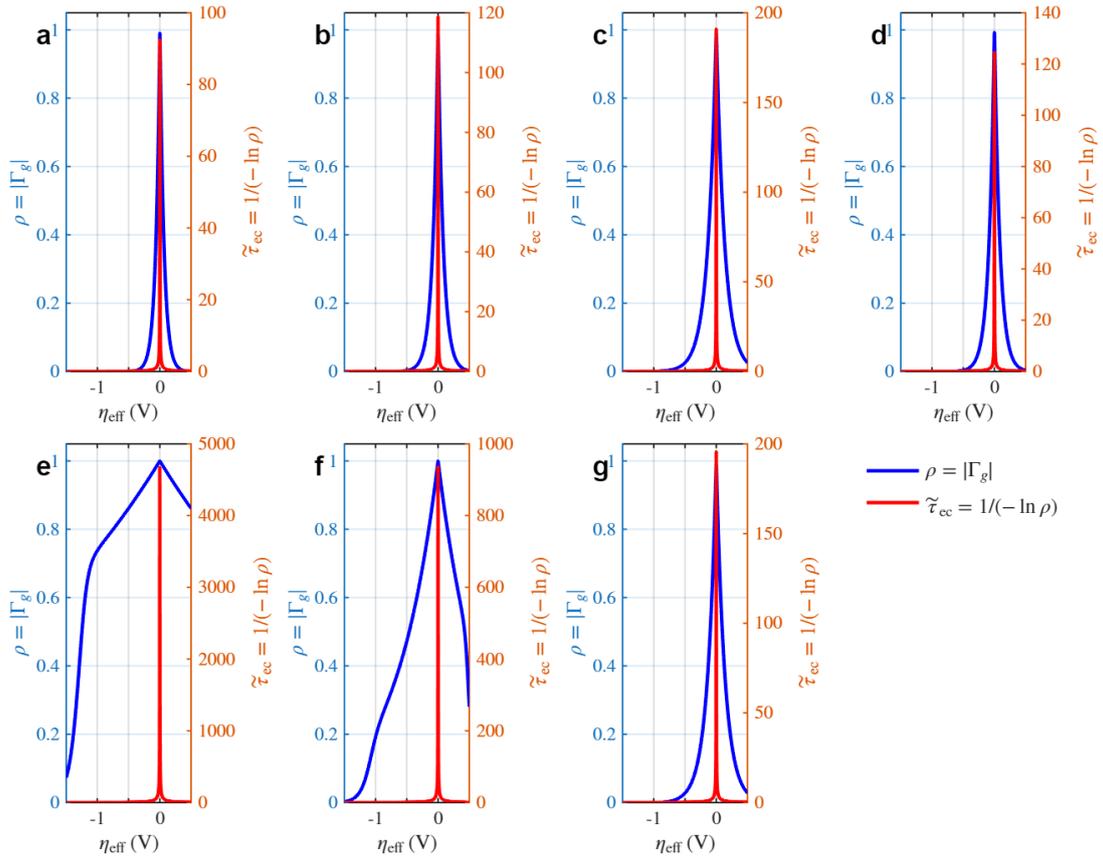

**Extended Data Fig. 17 | Feedback strength and retention proxy versus the effective interfacial overpotential on Au(111) across different pH values.** For each pH, the feedback magnitude $\rho(\eta_{\text{eff}}) = |\Gamma_g|$ (left axis) and the retention-like proxy $\tilde{\tau}_{ec}(\eta_{\text{eff}}) = 1/[-\ln \rho]$ (right axis) are plotted versus the $\eta_{\text{eff}}$ computed with the fitted ohmic correction (**a–g**). The dashed vertical line indicates the cathodic density optimum $\eta_{\text{eff}}^*$ defined by the maximization of $\Pi_{\text{dens}}$ over $\eta_{\text{eff}} < 0$.

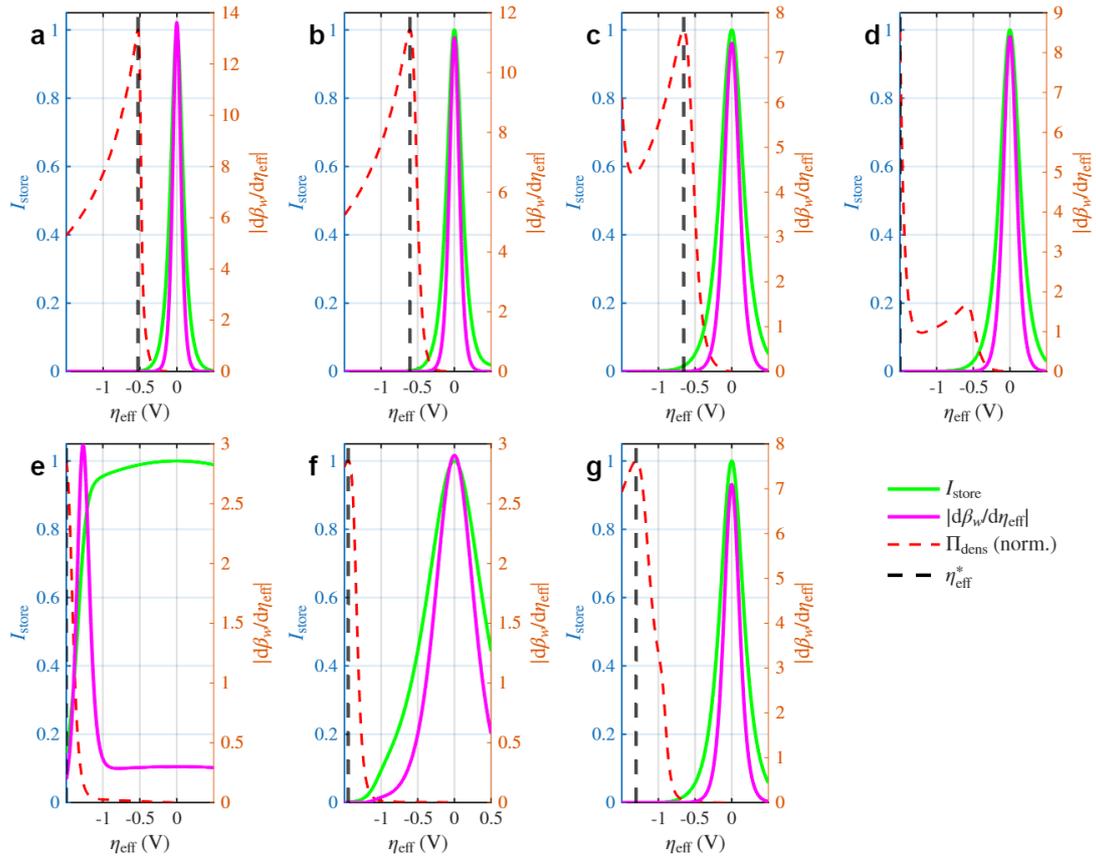

**Extended Data Fig. 18 | Storage intensity, susceptibility, and density optimum on Au(111) across different pH values.** For each pH, the standing-wave/storage intensity $I_{\text{store}}(\eta_{\text{eff}}) = 2\sqrt{J_{\text{ox}}J_{\text{red}}}/(J_{\text{ox}} + J_{\text{red}})$ (left axis) is shown together with a susceptibility diagnostic $|d\beta_w/d\eta_{\text{eff}}|$ (right axis), where $\beta_w = (J_{\text{ox}} - J_{\text{red}})/(J_{\text{red}} + J_{\text{ox}})$ is the branch-balancing (coherence) metric computed from the unsaturated directional fluxes $J_{\text{ox}}, J_{\text{red}}$ (**a–g**). The normalized output density $\Pi_{\text{dens}}/\max_{\eta_{\text{eff}}<0} \Pi_{\text{dens}}$ is overlaid (dashed), and the dashed vertical line indicates $\eta_{\text{eff}}^*$.

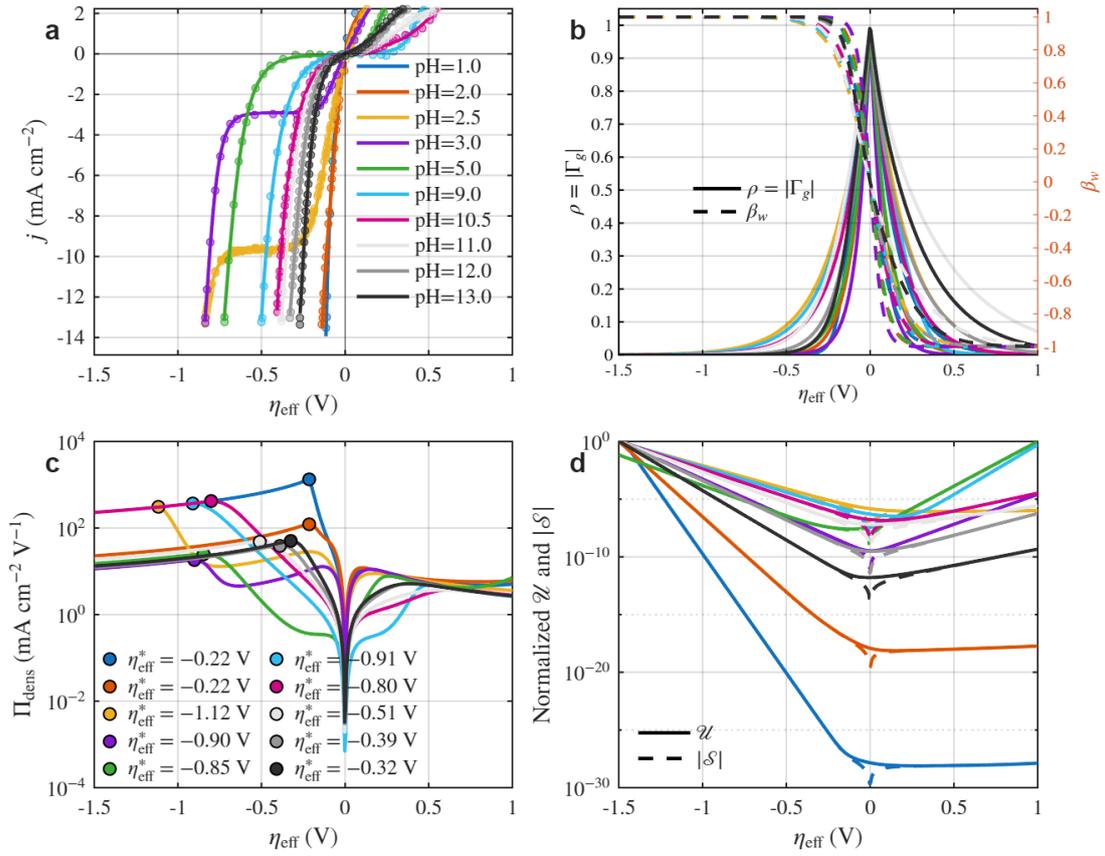

**Extended Data Fig. 19 | Waveguide-invariant mapping of the electrochemical polarization kinetics observed for Pt(111) across different pH values. a.** Measured[46,47] polarization curves $j(\eta)$ (markers) and fitted two-channel model responses (solid) obtained for Pt(111) across pH =1, 2, 2.5, 3, 5, 9, 10.5, 11, 12, and 13, plotted versus $\eta_{\text{eff}}$ (including ohmic correction). **b.** Polarization-derived state diagnostics versus $\eta_{\text{eff}}$: the reflection amplitude $\rho = |\Gamma_g|$ (solid; left axis) and the coherence metric $\beta_w = (J_{\text{ox}}^{\text{tot}} - J_{\text{red}}^{\text{tot}})/(J_{\text{red}}^{\text{tot}} + J_{\text{ox}}^{\text{tot}})$ (dashed; right axis) are computed from the fitted directional flux decomposition results. **c.** Useful cathodic output density $\Pi_{\text{dens}} = \Pi_{\text{use}}/(|\eta_{\text{eff}}| + \eta_0)$ (log scale), where $\Pi_{\text{use}} = |j_{\text{tot}}|(1 - \rho^2)$; circles mark the $\eta_{\text{eff}}^*$ values that maximize $\Pi_{\text{dens}}$ along the cathodic branch. **d.** Normalized exchange and transfer proxies derived from the same decomposition scheme: $\mathcal{U} = J_{\text{ox}}^{\text{tot}} + J_{\text{red}}^{\text{tot}}$ (solid) and $|\mathcal{S}| = |J_{\text{ox}}^{\text{tot}} - J_{\text{red}}^{\text{tot}}|$ (dashed).

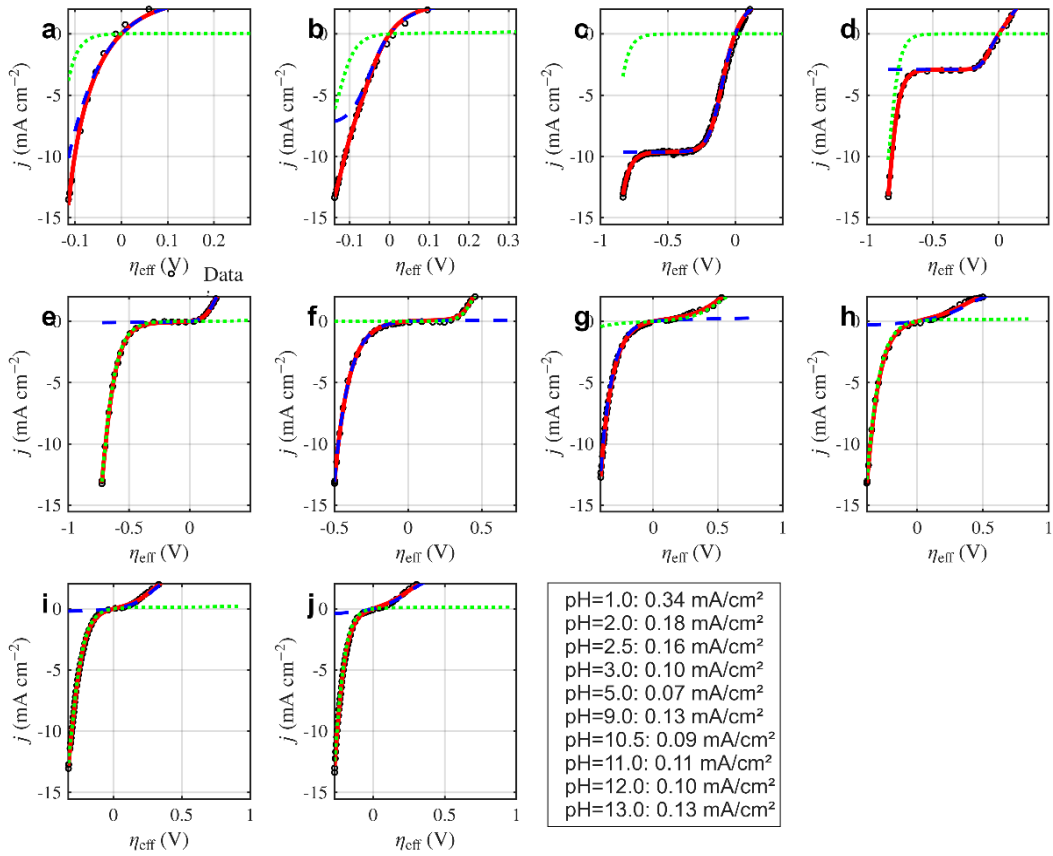

**Extended Data Fig. 20 | Polarization curves and waveguide-invariant two-mode fits obtained for Pt(111) across different pH values.** Measured[46,47] polarization curves $j(\eta)$ (open circles) and best-fitting two-mode model responses (solid) obtained for electrolytes at pH =1, 2, 2.5, 3, 5, 9, 10.5, 11, 12, and 13 (**a–j**). The current density $j$ is shown versus $\eta_{\text{eff}}$. The fitted $j_{\text{tot}}$ is decomposed into orthogonal modal contributions $j_A$ (dashed) and $j_B$ (dotted), capturing both the near-equilibrium and strongly cathodic regimes.

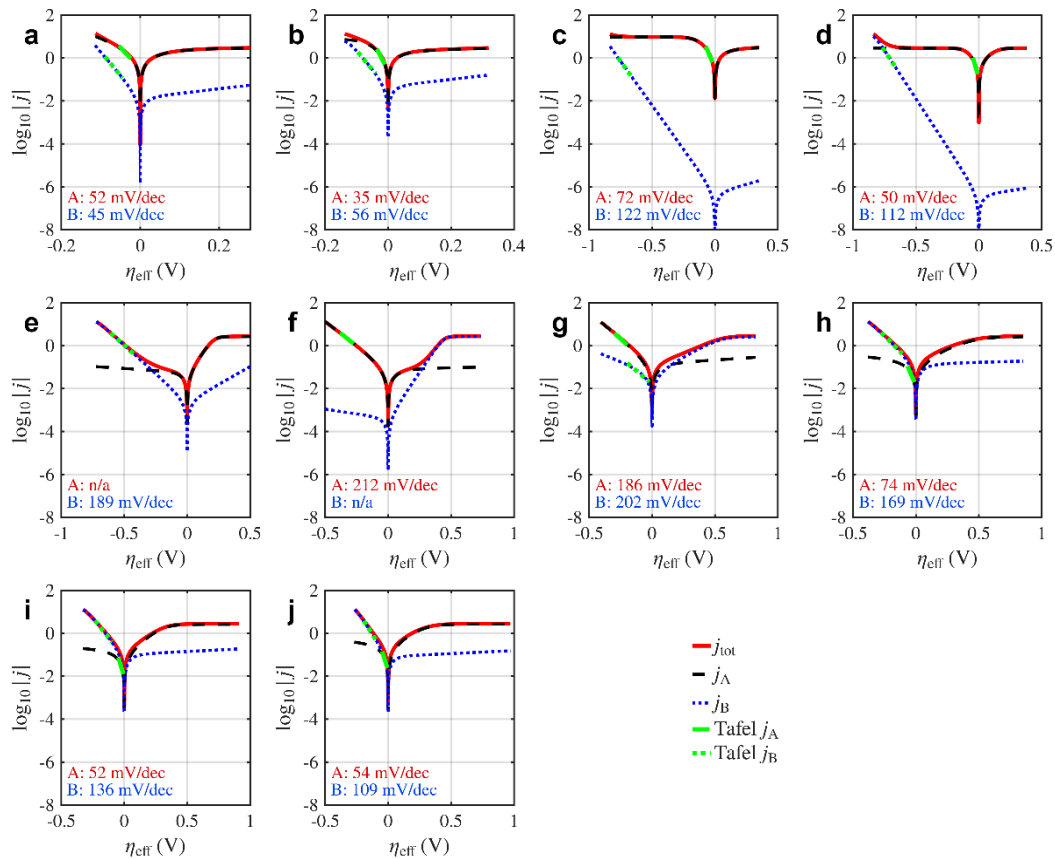

**Extended Data Fig. 21 | Modal Tafel analysis of the waveguide-invariant decomposition scheme on Pt(111) across different pH values.** $\log_{10}|j|$ versus $\eta_{\text{eff}}$ for pH values of 1, 2, 2.5, 3, 5, 9, 10.5, 11, 12, and 13 (**a–j**). The fitted $j_{\text{tot}}$ (solid) is decomposed into $j_A$ (dashed) and $j_B$ (dotted). The best linear Tafel regimes are indicated (segments), and the corresponding modal Tafel slopes (mV dec$^{-1}$) are annotated.

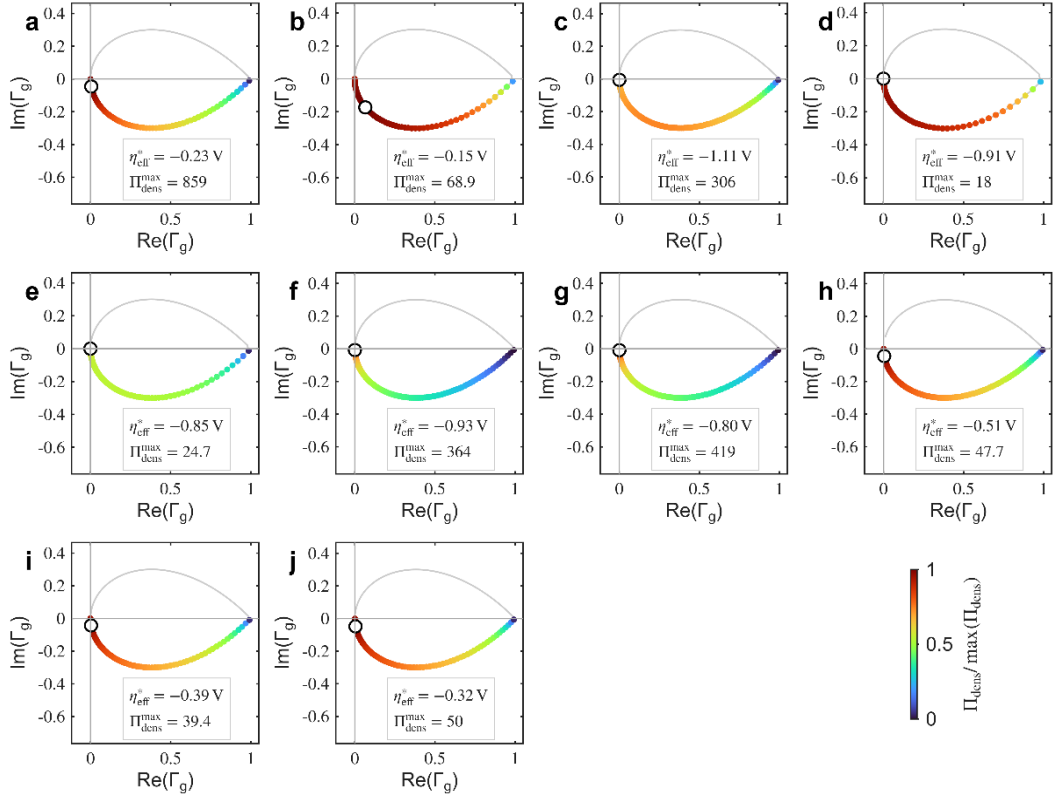

**Extended Data Fig. 22 | Cai–Smith plots coloured according to the normalized cathodic output densities achieved on Pt(111) across different pH values.** For each pH, the fitted model is mapped to $\Gamma_g(\eta_{\text{eff}}) = \rho e^{i\varphi}$ and coloured by $\Pi_{\text{dens}}(\eta_{\text{eff}})/\max_{\eta_{\text{eff}}<0} \Pi_{\text{dens}}$, with $\Pi_{\text{dens}} = \Pi_{\text{use}}/(|\eta_{\text{eff}}| + \eta_0)$ and $\Pi_{\text{use}} = \max(0, -j_{\text{tot}})(1-\rho^2)$ on the cathodic branch (**a–j**). The origin and unit-circle reference are indicated; the annotated markers highlight $\eta_{\text{eff}}^*$ and $\Pi_{\text{dens}}^{\max}$ for each pH.

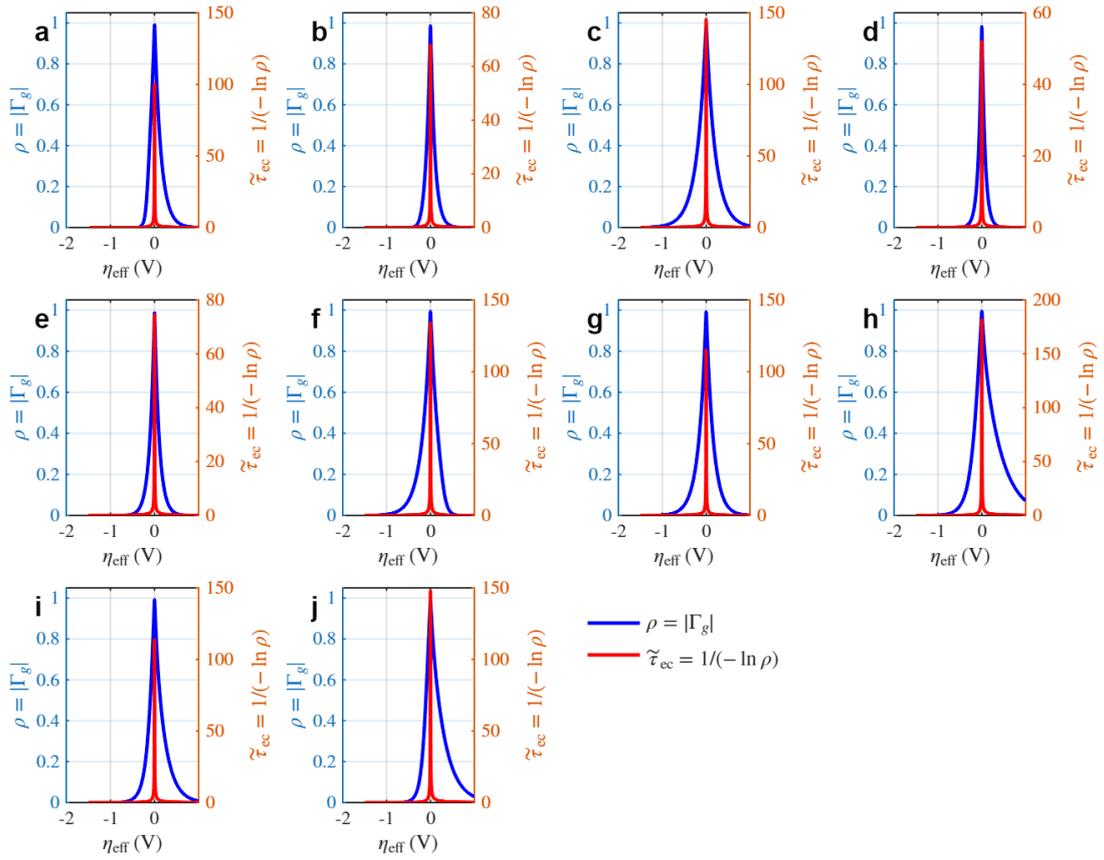

**Extended Data Fig. 23 | Feedback strength and retention proxy versus the effective interfacial overpotential for Pt(111) across different pH values.** For each pH, $\rho(\eta_{\text{eff}}) = |\Gamma_g|$ and $\tilde{\tau}_{\text{ec}}(\eta_{\text{eff}}) = 1/[-\ln \rho]$ are plotted versus $\eta_{\text{eff}}$ with a fitted ohmic correction. The dashed vertical lines mark the $\eta_{\text{eff}}^*$ values that maximize $\Pi_{\text{dens}}$ over $\eta_{\text{eff}} < 0$ (**a-j**).

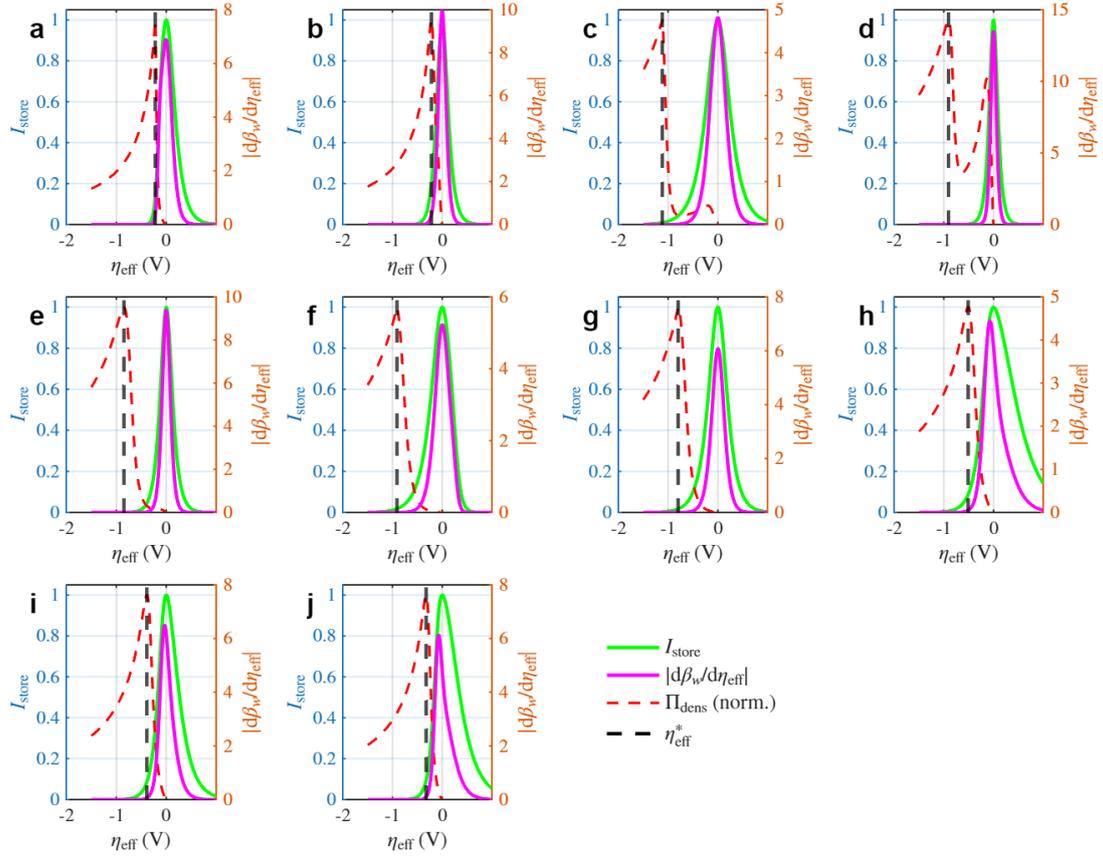

**Extended Data Fig. 24 | Storage intensity, susceptibility, and density optimum obtained for Pt(111) across different pH values.** For each pH, $I_{\text{store}}(\eta_{\text{eff}}) = 2\sqrt{J_{\text{ox}}J_{\text{red}}}/(J_{\text{ox}}+J_{\text{red}})$ and $|d\beta_w/d\eta_{\text{eff}}|$ are shown versus $\eta_{\text{eff}}$, where $\beta_w = (J_{\text{ox}} - J_{\text{red}})/(J_{\text{red}}+J_{\text{ox}})$ is computed from the unsaturated $J_{\text{ox}}, J_{\text{red}}$. The normalized density $\Pi_{\text{dens}}/\max\limits_{\eta_{\text{eff}}<0} \Pi_{\text{dens}}$ is overlaid, and $\eta_{\text{eff}}^*$ is indicated (**a–j**).

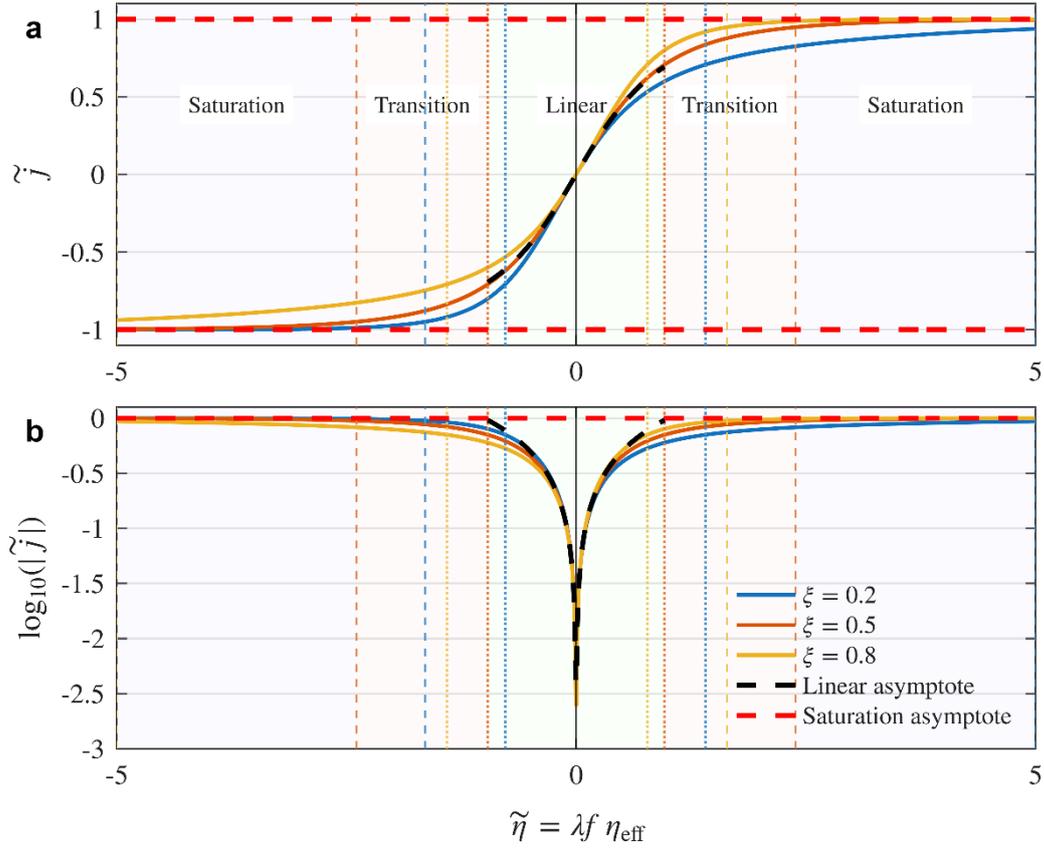

**Extended Data Fig. 25 | Dimensionless current density versus the dimensionless overpotential across various regimes and asymmetry conditions. a.** Single-mode, flux-based dimensionless polarization curves $\tilde{j}(\tilde{\eta})$ for $\xi = \{0.2, 0.5, 0.8\}$, where the unsaturated directional fluxes are $J_{ox} = j_p^* \exp(\xi \lambda \eta)$ and $J_{red} = j_p^* \exp(-(1-\xi)\lambda \eta)$, giving $\mathcal{U} = J_{ox} + J_{red}$ and $\mathcal{S} = J_{ox} - J_{red}$, and the displayed bounded power-flow proxy is $\tilde{j} = \mathcal{S}/\sqrt{1+\mathcal{S}^2}$ (here $j_p^* = 1$ and $\lambda = 1$ so that $\tilde{\eta} = \lambda f \eta_{eff}$). The red dashed lines indicate the saturation asymptotes $\tilde{j} = \pm 1$, and the black dashed line indicates the first-order linear asymptote $\tilde{j} \approx j_p^* \lambda \tilde{\eta}$ near $\tilde{\eta} = 0$ (equivalently $\tilde{j} \approx \mathcal{S}$ for $|\mathcal{S}| \ll 1$). Regime boundaries are computed \emph{for each} $\xi$ and \emph{for each} branch by solving $\mathcal{S}(\tilde{\eta}) = \pm \mathcal{S}_{thr}$ with $\mathcal{S}_{thr} \in \{1, 3\}$; faint coloured vertical lines mark the resulting asymmetric boundaries (dotted: $|\mathcal{S}| = 1$, dashed: $|\mathcal{S}| = 3$). Background shading provides a symmetric reference partition using the $\xi = 0.5$ case.

**b.** Semilog representation $\log_{10} |\tilde{\jmath}|$ versus $\tilde{\eta}$ for the same cases, with the same per-branch boundaries and asymptotes, highlighting the asymmetry-shifted curvature and the onset of saturation.

**Extended Data Tables**

**Extended Data Table 1 | Optimized two-mode waveguide parameters for the HER on Au(111) across different pH values.** Best-fitting parameters $\{j_A^*, \xi_A, \lambda_A, j_{\lim,A}^*, j_B^*, \xi_B, \lambda_B, j_{\lim,B}^*, R_\Omega\}$ for each pH, obtained from the least-squares fitting results produced for $j(\eta)$ using $\eta_{\text{eff}} = \eta - R_\Omega j$ and two-mode waveguide-invariant kinetics (Methods).

| pH | $j_A^*$ | $\xi_A$ | $\lambda_A$ | $j_{lim,A}^*$ | $j_B^*$ | $\xi_B$ | $\lambda_B$ | $j_{lim,B}^*$ | $R_\Omega$ |
|---|---|---|---|---|---|---|---|---|---|
| 1 | 3.41E-04 | 0.24 | 0.78 | 6.16 | 9.96E-11 | 0.12 | 1.59 | 48.38 | 4.75E-07 |
| 2 | 5.69E-04 | 0.09 | 0.57 | 31.55 | 1.60E-05 | 0.91 | 0.00 | 93.90 | 5.46E-02 |
| 2.5 | 4.57E-03 | 0.09 | 0.38 | 9.60 | 9.42E-11 | 0.29 | 0.60 | 320.12 | 3.71E-02 |
| 3 | 1.37E-03 | 0.35 | 0.58 | 2.79 | 2.70E-11 | 0.21 | 0.60 | 373.95 | 2.82E-11 |
| 4 | 4.31E-01 | 0.06 | 0.02 | 2.63 | 9.72E-11 | 0.24 | 0.62 | 26.98 | 2.22E-01 |
| 5 | 4.00E-03 | 0.09 | 0.08 | 360.46 | 9.59E-11 | 0.66 | 1.43 | 21.49 | 1.05E-06 |
| 13 | 3.46E-04 | 0.27 | 0.37 | 22.61 | 2.22E-08 | 0.27 | 0.80 | 24.38 | 1.89E-02 |

**Extended Data Table 2 | Optimized two-mode waveguide parameters for the HER on Pt(111) across different pH values.** Best-fitting parameters $\{j_A^*, \xi_A, \lambda_A, j_{\lim,A}^*, j_B^*, \xi_B, \lambda_B, j_{\lim,B}^*, R_\Omega\}$ for each pH, obtained using the same fitting protocol as that employed in Extended Data Table 1.

| pH | $j_A^*$ | $\xi_A$ | $\lambda_A$ | $j_{lim,A}^*$ | $j_B^*$ | $\xi_B$ | $\lambda_B$ | $j_{lim,B}^*$ | $R_\Omega$ |
|---|---|---|---|---|---|---|---|---|---|
| 1 | 2.81E+00 | 0.02 | 0.35 | 162.40 | 1.48E-02 | 0.09 | 1.36 | 329.12 | 7.07E-03 |
| 2 | 2.14E+00 | 0.05 | 0.51 | 7.02 | 1.01E-01 | 0.09 | 0.92 | 9.32 | 1.78E-02 |
| 2.5 | 2.96E+00 | 0.03 | 0.25 | 9.64 | 4.93E-07 | 0.25 | 0.65 | 395.32 | 2.88E-02 |
| 3 | 6.09E-01 | 0.44 | 0.69 | 2.91 | 3.91E-07 | 0.09 | 0.59 | 14.79 | 2.62E-02 |
| 5 | 5.62E-02 | 0.95 | 0.48 | 2.65 | 2.31E-03 | 0.39 | 0.51 | 23.15 | 8.10E-02 |
| 9 | 5.88E-03 | 0.39 | 0.88 | 2.72 | 4.66E-05 | 0.09 | 0.73 | 12.14 | 1.59E-01 |
| 10.5 | 1.38E-01 | 0.07 | 0.31 | 381.72 | 4.76E-02 | 0.58 | 0.33 | 2.60 | 3.35E-12 |
| 11 | 2.81E-01 | 0.91 | 0.14 | 2.64 | 1.38E-01 | 0.03 | 0.32 | 28.76 | 8.28E-02 |
| 12 | 1.53E-01 | 0.91 | 0.25 | 2.61 | 1.02E-01 | 0.04 | 0.42 | 18.44 | 1.82E-01 |
| 13 | 3.87E-01 | 0.91 | 0.18 | 2.67 | 8.26E-02 | 0.03 | 0.52 | 19.77 | 2.18E-01 |

**Extended Data Table 3 | Density-optimal operating points for the HER on Au(111) across different pH values.** The density-optimal driving $\eta^*_{eff} = \arg\max_{\eta_{eff}<0} \Pi_{dens}(\eta_{eff})$ and the corresponding $\Pi^*_{dens} = \Pi_{dens}(\eta^*_{eff})$, computed from the fitted Au(111) polarization mapping, are reported.

| pH | $\eta^*_{eff}$ (V) | $\Pi^*_{dens}$ (mA cm$^{-2}$ V$^{-1}$) |
|---|---|---|
| 1.0 | −0.53 | 92.9 |
| 2.0 | −0.61 | 45.9 |
| 2.5 | −0.65 | 12.9 |
| 3.0 | −1.50 | 20.3 |
| 4.0 | −1.49 | 16.7 |
| 5.0 | −1.45 | 14.2 |
| 13.0 | −1.14 | 37.8 |

**Extended Data Table 4 | Density-optimal operating points for the HER on Pt(111) across different pH values.** The same quantities as those presented in Extended Data Table 3 are computed from the fitted Pt(111) polarization mapping.

| pH | $\eta^*_{\text{eff}}$ (V) | $\Pi^*_{\text{dens}}$ (mA cm$^{-2}$ V$^{-1}$) |
|---|---|---|
| 1.0 | -0.33 | 930 |
| 2.0 | -0.17 | 73.1 |
| 2.5 | -1.12 | 305 |
| 3.0 | -0.90 | 18 |
| 5.0 | -0.85 | 24.7 |
| 9.0 | -0.90 | 373 |
| 10.5 | -0.80 | 419 |
| 11.0 | -0.51 | 48.1 |
| 12.0 | -0.39 | 39.4 |
| 13.0 | -0.32 | 50.3 |

# A universal waveguide mass–energy relation for lossy one-dimensional waves in nature

Huayang Cai[1,2,3] and Bishuang Chen[4]*

[1] Institute of Estuarine and Coastal Research, School of Ocean Engineering and Technology, Sun Yat-Sen University / Southern Marine Science and Engineering Guangdong Laboratory (Zhuhai), Zhuhai, Guangdong 519082, China.

[2] State and Local Joint Engineering Laboratory of Estuarine Hydraulic Technology / Guangdong Provincial Engineering Research Center of Coasts, Islands and Reefs / Guangdong Provincial Key Laboratory of Marine Resources and Coastal Engineering / Guangdong Provincial Key Laboratory of Information Technology for Deep Water Acoustics / Key Laboratory of Comprehensive Observation of Polar Environment (Sun Yat-Sen University), Ministry of Education, Zhuhai, Guangdong 519082, China.

[3] Zhuhai Research Center, Hanjiang National Laboratory, Zhuhai, Guangdong 519082, China.

[4] School of Marine Sciences, Sun Yat-Sen University, Zhuhai 519080, China.

*Corresponding Author

Bishuang Chen: chenbsh23@mail.sysu.edu.cn

**Supplementary Methods S0 | Reader's guide, notation and sign conventions**

**S0.0 Scope and usage of this supplementary methods document**

This Supplementary Methods document provides a structured and reproducible specification of the modeling reduction, state-variable definitions, asymmetry parameterization, approximations, proofs, and data-mapping procedures used throughout the main text and Extended Data. The document is organized with an index-first approach: each section concludes with an explicit statement of which Extended Data Figures it reproduces and which main-text claims it supports. Unless noted otherwise, all definitions and sign conventions in S0 apply entirely to the entire manuscript (main text, Extended Data, and Supplementary Information).

**S0.1 Symbols, normalization, and dimensional bookkeeping**

**Time-harmonic convention**

We adopt the complex time dependence $e^{+i\omega t}$ throughout. Spatial dependence for a forward-travelling guided wave is written as $\exp(-\gamma x)$ with the complex propagation constant $\gamma(\omega) = \alpha(\omega) + i\beta(\omega)$, where $\alpha \geq 0$ is the attenuation constant and $\beta$ is the phase constant. With this convention, a one-way traversal over length $L$ contributes a factor $\exp(-\gamma L)$ and a round trip contributes $\exp(-2\gamma L)$. When convenient, we write the dimensionless electrical length as $K(\Omega)$ and use $\exp(-2K)$ to denote the corresponding (dimensionless) round-trip factor; the precise definition of $K$ under the RLGC reduction is given in Supplementary Methods S1.

**Telegrapher-type reduction**

A finite, linear, time-invariant, single-mode guided system of length $L$ is represented by effective per-unit-length parameters $(R', L', G', C')$, a characteristic impedance $Z_c(\omega)$, and the complex propagation constant $\gamma(\omega)$. Different physical realizations (electromagnetic, acoustic, photonic, electrochemical transport analogues) enter only through these effective parameters and boundary conditions; all subsequent invariants and resonance/transfer diagnostics are formulated in measurable guided-wave quantities.

**Reference scales and dimensionless parameters**

We introduce reference scales $\omega_c$ and $Z_c$ (defined in the main-text Methods and Supplementary Methods S1) and define the dimensionless angular frequency $\Omega = \omega/\omega_c$. Distributed loss budgets are represented by non-negative dimensionless parameters (denoted generically by $\delta_R, \delta_G$) that quantify resistive-type and leakage-type losses, respectively, under the chosen normalization. The crossover frequency $\Omega_c$ separating loss-dominated and phase-dominated regimes, and the resulting behaviours of $\alpha L$ and $\beta L$, are derived and visualized in Supplementary Methods S1 and Extended Data Fig. 1.

**State variables and invariant**

From normalized forward/backward modal amplitudes we construct a dimensionless energy-like density $\mathcal{U}(\zeta)$ and a power-flow-like density $\mathcal{S}(\zeta)$ (with $\zeta = x/L \in [0,1]$) such that the exact single-mode invariant takes the form

$$\mathcal{U}^2(\zeta) - \mathcal{S}^2(\zeta) = |\Gamma_g(\zeta)|^2. \qquad (S0-1)$$

Equivalently, $\mathcal{U}^2 = \mathcal{S}^2 + |\Gamma_g|^2$, which organizes all reachable states as "mass shells" in the $(\mathcal{S}, \mathcal{U})$ plane with the effective standing-wave "mass" given by $|\Gamma_g|$. The construction of $(\mathcal{U}, \mathcal{S})$ and the proof of Eq. (S0-1) within the reduced model are detailed in Supplementary Methods S2; low-mass/high-speed expansions and quantitative error bounds are given in Supplementary Methods S4.

**Rapidity and doppler factors**

We use the bounded "waveguide velocity" $\beta_w = \mathcal{S}/\mathcal{U} \in (-1,1)$ and the corresponding rapidity $\phi = \text{atanh}(\beta_w)$ so that $\mathcal{U} = |\Gamma_g|\cosh\phi$ and $\mathcal{S} = |\Gamma_g|\sinh\phi$. We further define the bidirectional Doppler factors $D_\pm = \exp(\pm\phi)$, which provide a numerically stable parameterization for regime separation, composition under cascaded segments, and visualization of "light-cone" constraints. These definitions and their composition rules are developed in Supplementary Methods S5 and visualized in Extended Data Fig. 4.

**Boundary reflection coefficients**

Terminal reflections are defined with respect to $Z_c(\omega)$. For an impedance termination $Z_T(\omega)$, the corresponding reflection coefficient is

$$\Gamma_T(\omega) = \frac{Z_T(\omega) - Z_c(\omega)}{Z_T(\omega) + Z_c(\omega)}. \qquad (S0-2)$$

We denote the source-side and load-side terminal reflection coefficients by $\Gamma_S$ and $\Gamma_L$ (both may be frequency dependent). Under passive terminations, $|\Gamma_S| \leq 1$ and $|\Gamma_L| \leq 1$. The boundary-composed generalized reflection coefficient (the Cai–Smith state variable) is written as a Möbius map of $(\Gamma_S, \Gamma_L, \exp(-2K))$ and is defined explicitly in Supplementary Methods S2; it generates trajectories on (or within) the unit disk when scanned in frequency or operating condition.

**Intrinsic asymmetry**

Intrinsic asymmetry is encoded by a single scalar parameter $\xi \in (0,1)$, with $\xi = 1/2$ denoting symmetry and $\xi \neq 1/2$ denoting directional asymmetry. Operationally, $\xi$ modulates the radial deformation of the generalized reflection magnitude $|\Gamma_g|$ (and therefore the effective "mass" $|\Gamma_g|$) relative to the symmetric baseline at fixed loss budget. Under the global convention adopted here, increasing $\xi$ above 1/2 shifts the reachable envelope radially inward (smaller $|\Gamma_g|$) for fixed distributed loss, whereas decreasing $\xi$ below 1/2 shifts it outward (larger $|\Gamma_g|$). The precise deformation law, the associated phase snapshot construction, and the envelope bounds are given in Supplementary Methods S3 and visualized in Extended Data Fig. 2.

**Feedback factor and ring-down metrics**

For the unified resonance description, we use the complex feedback factor

$$F(\Omega) = \kappa(\Omega)\, e^{i\Theta(\Omega)}, \qquad (S0-3)$$

where $\kappa \in [0,1)$ is the round-trip magnitude (net feedback strength) and $\Theta$ is the round-trip phase detuning. Ring-down time and quality-factor proxies are expressed in terms of $\kappa$ (or equivalently $\delta = 1 - \kappa$ in the high-$Q$ regime). The whole definitions and asymptotics are specified in Supplementary Methods S6; the corresponding maps and bands are shown in Extended Data Fig. 9 and related panels.

**S0.2 Sign conventions and orientation on the Cai–Smith disk**

**Forward direction and $\mathcal{S}$ sign**

We define the forward direction as from source to load (increasing $x$). The power-flow-like state variable $\mathcal{S}$ is defined so that $\mathcal{S} > 0$ corresponds to net forward power flow and $\mathcal{S} < 0$ corresponds to net backward power flow. With this convention, $\beta_w = \mathcal{S}/\mathcal{U}$ is positive for forward-dominated states and negative for backward-dominated states. The absolute energy-like density satisfies $\mathcal{U} \geq |\mathcal{S}|$ by construction.

**Generalized reflection $\Gamma_g$ and disk geometry**

The Cai–Smith representation uses the complex plane of the generalized reflection coefficient $\Gamma_g$ restricted to the unit disk $|\Gamma_g| \leq 1$ under passive boundaries. Radial position encodes the effective standing-wave "mass" $|\Gamma_g|$, while the polar angle encodes the generalized reflection phase. Frequency or operating-condition sweeps generate trajectories $\Gamma_g(\Omega)$ within the disk; "critical coupling" and "maximum power transfer" conditions correspond to trajectories approaching the disk center (small $|\Gamma_g|$) under the appropriate objective, as detailed in Supplementary Methods S8 and S9.

**Asymmetry parameter $\xi$ sign meaning**

Throughout this work $\xi = 1/2$ denotes intrinsic symmetry. Departures $\xi > 1/2$ and $\xi < 1/2$ correspond to opposite directional biases. Under the adopted convention, these biases manifest as opposite radial deformations of $|\Gamma_g|$ at fixed distributed loss budget. When mapping to data (e.g., electrochemical polarization), we keep the global meaning of $\xi$ fixed and treat dataset-specific current/overpotential sign handling as a preprocessing map (Supplementary Methods S9) so that $\xi$ retains a consistent physical interpretation across cathodic and anodic systems.

**S0.3 Roadmap: main-text methods ↔ supplementary methods ↔ extended data**

**Table S0 | Navigation map for reproducibility.**

| Main-text methodological element | Where specified/proved | Extended Data reproduction |
|---|---|---|
| RLGC reduction; normalization; dispersion regimes | S1 (definitions, $\alpha L, \beta L, \Omega_c$) | Extended Data Fig. 1 |
| Boundary composition; Möbius map for $\Gamma_g$; disk reachability | S2 (terminal reflections, $\bar{\Gamma}_g$, unit-disk constraints) | Extended Data Fig. 2 (geometry inputs) |
| Intrinsic asymmetry $\xi$; phase snapshots; envelope bounds | S3 (deformation law, $E_{\min/\max}$, parameter dependence) | Extended Data Fig. 2 |
| Low-mass/high-speed expansions; truncation errors | S4 (expansions, error bounds, scaling) | Extended Data Fig. 3 |
| Rapidity formalism; Doppler factors; composition rules | S5 (definitions, identities, cascade composition) | Extended Data Fig. 4 |
| Unified feedback model; resonance archetypes; ring-down and $Q$ maps | S6 (definitions of $F, \kappa, \Theta$, ring-down asymptotics) | Extended Data Figs. 5-7, Fig. 9 |

| Four fundamental laws; proofs; limiting cases | S7 (formal statements and proofs) | Extended Data Fig. 8, Fig. 10 |
| --- | --- | --- |
| Optical CPA-EP reconstruction: fixed coherent setting versus SVD probe | S8 (Waveguide-invariant mapping revealing optical CPA-EP reconstruction) | Extended Data Figs. 11-13 |
| Waveguide-to-electrochemistry mapping; two-regime fit; optimization; lifetime inference | S9 (data map, objectives, fitting/diagnostics) | Extended Data Figs. 14-25 |

**Supplementary Methods S1 | Waveguide reduction, RLGC scaling, and dispersion regimes**

**S1.0 Overview**

This section derives the dimensionless RLGC electrical length $K(\Omega)$, which serves as the universal dispersion-control parameter for any finite, linear, time-invariant, single-mode, one-dimensional guided system that admits a telegrapher-type reduction. The key outcome is a closed-form mapping from the physical per-unit-length parameters $(R', L', G', C')$ and physical length $L$ to a dimensionless complex scalar $K(\Omega) = \gamma(\omega)L = R(\Omega) + i\Phi(\Omega)$, whose real part gives the cumulative attenuation $(\alpha L)$ and whose imaginary part gives the accumulated phase $(\beta L)$. The crossover frequency $\Omega_c = \sqrt{\delta_R \delta_G}$ emerges naturally from $K(\Omega)$ and separates dissipation-dominated from phase-dominated regimes, providing the classification used throughout the main text and Extended Data Fig. 1.

**S1.1 Telegrapher-type reduction and modal solutions**

We start from the standard telegrapher equations for a distributed, single-mode, axially uniform guide with per-unit-length series impedance $R' + i\omega L'$ and shunt admittance $G' + i\omega C'$. Let $v(x, t)$ and $i(x, t)$ denote the generalized voltage-like and current-like conjugate variables. In the time domain,

$$\frac{\partial v}{\partial x} = -R'i - L'\frac{\partial i}{\partial t}, \frac{\partial i}{\partial x} = -G'v - C'\frac{\partial v}{\partial t}. \qquad (S1-1)$$

Adopting the time-harmonic convention $e^{+i\omega t}$ and writing phasors $v(x,t) = \Re\{V(x)e^{i\omega t}\}$, $i(x,t) = \Re\{I(x)e^{i\omega t}\}$ gives

$$\frac{dV}{dx} = -(R' + i\omega L')I, \frac{dI}{dx} = -(G' + i\omega C')V. \qquad (S1-2)$$

Eliminating $I$ (or $V$) yields the Helmholtz-type second-order equations

$$\frac{d^2V}{dx^2} = \gamma^2(\omega)V, \frac{d^2I}{dx^2} = \gamma^2(\omega)I, \qquad (S1-3)$$

with the complex propagation constant

$$\gamma(\omega) = \sqrt{(R' + i\omega L')(G' + i\omega C')} = \alpha(\omega) + i\beta(\omega), \qquad (S1-4)$$

where $\alpha \geq 0$ is the attenuation constant and $\beta$ is the phase constant. The general solutions can be written in forward/backward form $V(x) = V^+e^{-\gamma x} + V^-e^{+\gamma x}$ and $I(x) = \frac{1}{Z_c}(V^+e^{-\gamma x} - V^-e^{+\gamma x})$, with the characteristic impedance

$$Z_c(\omega) = \sqrt{\frac{R' + i\omega L'}{G' + i\omega C'}}. \qquad (S1-5)$$

Terminal reflection coefficients $\Gamma_T(\omega)$ are defined with respect to $Z_c(\omega)$ (global convention in Supplementary Methods S0), so that the boundary conditions can be expressed entirely in terms of $\Gamma_S(\omega)$ and $\Gamma_L(\omega)$ and the propagation across the finite length $L$.

**S1.2 Reference scales and dimensionless loss budgets**

To expose a domain-independent, low-dimensional description, we introduce the reference frequency and impedance scales

$$\omega_c = \frac{1}{\sqrt{L'C'}}, Z_c = \sqrt{\frac{L'}{C'}}. \qquad (S1-6)$$

We then define the dimensionless frequency and the dimensionless loss budgets using the physical length $L$:

$$\Omega = \frac{\omega}{\omega_c}, \delta_R = \frac{R'L}{Z_c}, \delta_G = G'LZ_c. \qquad (S1-7)$$

Here $\delta_R$ measures the cumulative (length-integrated) series-like dissipation relative to $Z_c$, and $\delta_G$ measures the cumulative shunt-like leakage relative to $1/Z_c$. These two non-negative scalars are the minimal loss parameters required to classify the dispersion and feedback behaviour of a finite guide under the RLGC reduction. Two identities follow directly from Eq. (S1-6): $\omega_c L' = \sqrt{L'/C'} = Z_c$ and $\omega_c C' = 1/Z_c$. Substituting $\omega = \Omega \omega_c$ into Eq. (S1-4) and multiplying by $L$ therefore yields a dimensionless complex electrical length $K(\Omega)$.

**S1.3 Dimensionless electrical length and separation into attenuation and phase**

Define the (one-pass) complex electrical length

$$K(\Omega) = \gamma(\omega)L = \sqrt{(\delta_R + i\Omega)(\delta_G + i\Omega)} = R(\Omega) + i\Phi(\Omega), \qquad (S1-8)$$

where

$$R(\Omega) = \Re[K(\Omega)] = \alpha(\omega)L, \Phi(\Omega) = \Im[K(\Omega)] = \beta(\omega)L. \qquad (S1-9)$$

Thus, for a fixed physical length $L$, all frequency dependence of cumulative attenuation and phase accumulation is captured by the single dimensionless scalar $K(\Omega)$, and all platform dependence enters only through $(\delta_R, \delta_G)$ after normalization.

For explicit evaluation, it is useful to write $K^2 = X + iY$ with

$$X(\Omega) = \delta_R \delta_G - \Omega^2, Y(\Omega) = \Omega(\delta_R + \delta_G). \qquad (S1-10)$$

Choosing the principal branch consistent with $\alpha \geq 0$ and $\Phi \geq 0$ for $\Omega > 0$, we obtain

$$R(\Omega) = \sqrt{\frac{\sqrt{X^2 + Y^2} + X}{2}}, \Phi(\Omega) = \sqrt{\frac{\sqrt{X^2 + Y^2} - X}{2}}. \quad (S1-11)$$

Equation (S1-11) provides a closed-form, numerically stable route from $(\delta_R, \delta_G, \Omega)$ to $(\alpha L, \beta L)$ and underpins the dispersion plots in Extended Data Fig. 1.

**S1.4 Universal crossover and regime classification**

A central feature of Eq. (S1-8) is the existence of a universal crossover frequency

$$\Omega_c = \sqrt{\delta_R \delta_G}, \quad (S1-12)$$

at which $X(\Omega_c) = 0$ and therefore $R(\Omega_c) = \Phi(\Omega_c)$. Indeed, substituting $\Omega = \Omega_c$ into Eq. (S1-10) gives $K^2(\Omega_c) = i\,\Omega_c(\delta_R + \delta_G)$ and hence

$$K(\Omega_c) = \frac{1+i}{\sqrt{2}}\sqrt{\Omega_c(\delta_R + \delta_G)}, R(\Omega_c) = \Phi(\Omega_c) = \sqrt{\frac{\Omega_c(\delta_R + \delta_G)}{2}}. \quad (S1-13)$$

This crossover separates two limiting behaviours that are directly relevant for finite-length feedback: (i) a dissipation-dominated regime in which $R(\Omega)$ is large relative to $\Phi(\Omega)$, so that multiple echoes are strongly suppressed, and (ii) a phase-dominated regime in which $\Phi(\Omega)$ grows rapidly while $R(\Omega)$ varies weakly, enabling strong interference and narrow resonance features when boundaries provide sufficient reflectivity.

A convenient diagnostic is the ratio $\Phi/R$. When $\Phi/R \ll 1$, the response is predominantly controlled by exponential decay across the guide (weak recirculation), whereas $\Phi/R \gg 1$ indicates strong phase winding per unit attenuation (strong recirculation and high sensitivity to detuning). Near $\Omega_c$, where $\Phi/R \approx 1$, attenuation and phase compete on equal footing, and resonance/critical-coupling conditions become most sensitive to boundary composition.

**S1.5 Asymptotic regimes and physical interpretation**

The following expansions clarify the limiting behaviours and provide intuition for how $(\delta_R, \delta_G)$ shape $R(\Omega)$ and $\Phi(\Omega)$.

**Low-frequency regime, $\Omega \ll \min(\delta_R, \delta_G)$**

In this regime $(\delta_R + i\Omega)(\delta_G + i\Omega) = \delta_R\delta_G + i\Omega(\delta_R + \delta_G) + \mathcal{O}(\Omega^2)$, so

$$K(\Omega) = \sqrt{\delta_R\delta_G} + i\Omega \frac{\delta_R + \delta_G}{2\sqrt{\delta_R\delta_G}} + \mathcal{O}(\Omega^2), \quad (S1-14)$$

implying $R(\Omega) \approx \sqrt{\delta_R\delta_G}$ (nearly frequency-independent cumulative attenuation) and $\Phi(\Omega) \propto \Omega$ (weak phase accumulation).

**High-frequency regime, $\Omega \gg \max(\delta_R, \delta_G)$**

Writing $K(\Omega) = \sqrt{-\Omega^2 + i\Omega(\delta_R + \delta_G) + \delta_R\delta_G}$ and expanding in $1/\Omega$ yields

$$K(\Omega) = \frac{\delta_R + \delta_G}{2} + i\left[\Omega - \frac{4\delta_R\delta_G - (\delta_R + \delta_G)^2}{8\Omega}\right] + \mathcal{O}\left(\frac{1}{\Omega^2}\right). \quad (S1-15)$$

Thus, at high frequency the cumulative attenuation approaches a constant $R(\Omega) \to (\delta_R + \delta_G)/2$, while the phase grows approximately linearly, $\Phi(\Omega) \sim \Omega$, producing rapidly winding trajectories in any boundary-composed response.

**Series-only and shunt-only limits**

If $\delta_G = 0$ (series-only dissipation), then $K(\Omega) = \sqrt{(i\Omega)(\delta_R + i\Omega)} = \sqrt{-\Omega^2 + i\delta_R\Omega}$; the crossover $\Omega_c = \sqrt{\delta_R\delta_G}$ collapses to $0$, and attenuation/phase are comparable at sufficiently small $\Omega$ while $\Phi(\Omega)$ dominates at large $\Omega$. If $\delta_R = 0$ (shunt-only leakage), then $K(\Omega) = \sqrt{(i\Omega)(\delta_G + i\Omega)}$ with analogous behaviour under the exchange $\delta_R \leftrightarrow \delta_G$. These limits bound the general two-budget case and help interpret extreme parameter choices in platform-specific realizations.

**S1.6 Practical recipe: computing $K(\Omega)$, $(\alpha L, \beta L)$, and the crossover**

Given measured or effective $(R', L', G', C')$ and a chosen physical length $L$, the workflow is: (i) compute $\omega_0$ and $Z_0$ via Eq. (S1-6); (ii) form $(\delta_R, \delta_G)$ via Eq. (S1-7); (iii) evaluate $K(\Omega)$ from Eq. (S1-8) on the desired frequency grid; (iv) extract $\alpha L = R(\Omega)$ and $\beta L = \Phi(\Omega)$ via Eq. (S1-9) or Eq. (S1-11); (v) compute the crossover $\Omega_c$ via Eq. (S1-12), which provides a regime marker independent of boundary conditions. All subsequent finite-length resonance and decay diagnostics depend on propagation only through $K(\Omega)$ (and therefore through $(\delta_R, \delta_G)$ under this normalization), with boundaries entering separately through the terminal reflection coefficients (Supplementary Methods S2 and S6).

### S1.7 Parameter sets and frequency sweep for extended data fig. 1

Extended Data Fig. 1 is generated by plotting $R(\Omega) = \alpha L$ and $\Phi(\Omega) = \beta L$ from Eq. (S1-11) for representative loss budgets: (i) symmetric: $\delta_R = \delta_G = 10^{-2}$; (ii) asymmetric: $\delta_R = 10^{-2}$, $\delta_G = 5 \times 10^{-1}$; (iii) series-only: $\delta_R = 5 \times 10^{-1}$, $\delta_G = 0$. The crossover marker is $\Omega_c = \sqrt{\delta_R \delta_G}$ (Eq. S1-12) and the reference normalization point is $\Omega = 1$. A log-spaced sweep in $\Omega$ (spanning several decades) is used to reveal both low- and high-frequency asymptotics and the crossover neighbourhood.

### Supplementary Methods S2 | Boundary composition and the generalized reflection coefficient

### S2.0 Overview

This section defines the terminal reflection coefficients $(\Gamma_S, \Gamma_L)$ referenced to the characteristic impedance $Z_c(\omega)$, derives the boundary-composed generalized reflection coefficient $\bar{\Gamma}_g(\Omega)$ via a multiple-reflection composition with the complex electrical length $K(\Omega)$, and constructs the axial two-wave fields that lead to the exact invariant $\mathcal{U}^2 - \mathcal{S}^2 = |\Gamma_g(\zeta)|^2$. These results provide the mathematical backbone of the Cai–Smith disk representation and the frequency-sweep trajectories used throughout the main text and Extended Data.

**S2.1 Terminal reflection coefficients and boundary conditions**

At any terminal plane referenced to the characteristic impedance $Z_c(\omega)$, we define the (complex) reflection coefficient as the backward-to-forward wave ratio at that plane,

$$\Gamma(\omega) = \frac{V^-(\omega)}{V^+(\omega)}. \tag{S2-1}$$

For an impedance termination $Z_T(\omega)$ at that plane, the corresponding reflection coefficient is

$$\Gamma_T(\omega) = \frac{Z_T(\omega) - Z_c(\omega)}{Z_T(\omega) + Z_c(\omega)}, \tag{S2-2}$$

which reduces to the standard voltage-wave definition when $Z_c$ is real; in a lossy guide, Eq. (S2-2) is understood as the reflection coefficient of the reduced single-mode port referenced to $Z_c(\omega)$, so that passivity of the terminal implies $|\Gamma_T| \leq 1$ under the adopted normalization. We denote the source-side and load-side terminal reflection coefficients by $\Gamma_S(\omega)$ and $\Gamma_L(\omega)$, respectively. All boundary conditions used in the manuscript (fixed load, reactive load, source impedance, contact resistance, Thevenin/Norton equivalence, etc.) are reduced to $(\Gamma_S, \Gamma_L)$ at the two terminal planes.

**S2.2 Propagation across a finite length and the one-round-trip factor**

Let $K(\Omega) = \gamma(\omega)L$ be the dimensionless complex electrical length defined in Supplementary Methods S1, with $\gamma = \alpha + i\beta$ and $K = R + i\Phi$ ($R = \alpha L$, $\Phi = \beta L$). One-way propagation over length $L$ multiplies the wave amplitude by $e^{-\gamma L}$; equivalently $e^{-K} = e^{-R}e^{-i\Phi}$, where $R = \alpha L$ and $\Phi = \beta L$. The backward wave accumulates the same attenuation $e^{-R}$ over a path of length $L$, with the phase sign determined by the chosen spatial convention for the backward-travelling component. The net factor accumulated per full round trip between the terminals is therefore

$$E(\Omega) = e^{-2K(\Omega)} = e^{-2R(\Omega)}e^{-i2\Phi(\Omega)}. \quad (S2-3)$$

For passive propagation $R(\Omega) \geq 0$, hence $|E(\Omega)| = e^{-2R(\Omega)} \leq 1$, with strict inequality $|E| < 1$ whenever distributed dissipation is nonzero over the finite length.

**S2.3 Multiple-reflection sum and Möbius (Bilinear) composition**

Consider a forward wave incident on the load. The load reflects a backward component with factor $\Gamma_L$, which returns to the source with factor $e^{-K}$, reflects at the source with factor $\Gamma_S$, and returns to the load with another factor $e^{-K}$. The net multiplicative factor per full round trip between successive load reflections is $\Gamma_S\Gamma_L E$. Summing the geometric series for the backward/forward ratio at the chosen reference plane yields a bilinear (Möbius) combination of $(\Gamma_S, \Gamma_L, E)$. In the convention used throughout the manuscript, we define the boundary-composed generalized reflection coefficient as

$$\bar{\Gamma}_g(\Omega) = \frac{\Gamma_L(\Omega)E(\Omega) + \Gamma_S(\Omega)}{1 + \Gamma_S(\Omega)\Gamma_L(\Omega)E(\Omega)}. \quad (S2-4)$$

Equation (S2-4) is the key reduction: all boundary and propagation information enters the Cai–Smith state through $(\Gamma_S, \Gamma_L, K)$ only via the scalar $E = e^{-2K}$. The resonance-sensitive denominator

$$D(\Omega) = 1 + \Gamma_S(\Omega)\Gamma_L(\Omega)E(\Omega) \quad (S2-5)$$

controls the buildup of internal recirculation; $|D|$ small corresponds to strong feedback proximity and underpins the narrowband resonance/critical-coupling phenomenology developed in Results and Supplementary Methods S7.

**S2.4 ABCD/Scattering equivalence and derivation from transfer matrices**

The propagation segment of length $L$ possesses the standard transfer (ABCD)

matrix

$$\begin{pmatrix} v(0) \\ i(0) \end{pmatrix} = \begin{pmatrix} \cosh(\gamma L) & Z_c \sinh(\gamma L) \\ Z_c^{-1} \sinh(\gamma L) & \cosh(\gamma L) \end{pmatrix} \begin{pmatrix} v(L) \\ i(L) \end{pmatrix}, \qquad (S2-6)$$

where $(v, i)$ are the generalized voltage-/current-like phasors in the reduced model (Supplementary Methods S1). Imposing terminal relations at the boundaries (e.g., $v(L) = Z_L i(L)$ and $v(0) = Z_S i(0)$, or their Thevenin/Norton equivalents), converting these impedances to reflection coefficients via Eq. (S2-2), and eliminating the internal amplitudes yields Eq. (S2-4) after straightforward algebra. Thus $\bar{\Gamma}_g$ can be obtained either by a multiple-reflection sum (cavity viewpoint) or by two-port elimination (network viewpoint); both are exactly equivalent within the single-mode RLGC reduction.

**S2.5 Passivity and admissibility on the Cai–Smith disk**

Under passive propagation ($|E| \leq 1$) and passive terminals (contractive one-port reductions, $|\Gamma_S| \leq 1$ and $|\Gamma_L| \leq 1$ under the adopted normalization), the composed state $\bar{\Gamma}_g$ remains admissible. In particular, for strictly lossy propagation ($|E| < 1$) and passive terminals, the scalar Redheffer star-product underlying Eq. (S2-4) is contractive, implying

$$|\bar{\Gamma}_g(\Omega)| < 1 \text{ for } |E(\Omega)| < 1, \ |\Gamma_S(\Omega)| \leq 1, \ |\Gamma_L(\Omega)| \leq 1. \qquad (S2-7)$$

Equality $|\bar{\Gamma}_g| = 1$ can occur only in limiting marginal cases (lossless propagation $|E| = 1$ with fully reflecting terminals and phase-aligned feedback). This admissibility motivates the Cai–Smith unit-disk representation used in the main text: $\bar{\Gamma}_g$ provides a compact, bounded state coordinate for resonance, decay, and power-transfer optimization. Experimentally inferred values that lie outside the unit disk indicate active elements (gain), calibration mismatch, or breakdown of the passive single-mode reduction assumptions.

### S2.6 Axial state variable, two-wave fields, and construction of $(\mathcal{U}, \mathcal{S})$

We introduce the dimensionless axial coordinate $s = x/L \in [0,1]$ and the complex axial state

$$\zeta(s; \Omega) = s\, K(\Omega), \Re(\zeta) = sR(\Omega), \Im(\zeta) = s\Phi(\Omega). \qquad (S2-8)$$

To accommodate intrinsic asymmetry while preserving a two-wave superposition structure, we use the direction-weighted axial representation (with asymmetry coefficient $\xi \in [0,1]$ defined globally in Supplementary Methods S0 and developed in Supplementary Methods S3). For a given boundary-composed reflection $\bar{\Gamma}_g(\Omega)$ at the chosen reference plane, we write the normalized voltage-/current-like phasors as

$$v(\zeta) = \frac{1}{\sqrt{2}}(e^{-2\xi\zeta} + \bar{\Gamma}_g(\Omega)\, e^{2(1-\xi)\zeta}), i(\zeta) = \frac{1}{\sqrt{2}}(e^{-2\xi\zeta} - \bar{\Gamma}_g(\Omega)\, e^{2(1-\xi)\zeta}). \qquad (S2-9)$$

Equation (S2-9) is normalized so that in the absence of backward content ($\bar{\Gamma}_g = 0$) the forward contribution is unity at $s = 0$, while $\xi$ continuously reweights the forward/backward attenuation along $\Re(\zeta)$ without altering the phase coordinate $\Im(\zeta)$. We then define the dimensionless energy-like and power-flow-like state variables (used in the main text) by

$$\mathcal{U}(\zeta) = \frac{|v(\zeta)|^2 + |i(\zeta)|^2}{2}, \mathcal{S}(\zeta) = \Re\,[v(\zeta)i(\zeta)^*]. \qquad (S2-10)$$

Direct substitution of Eq. (S2-9) into Eq. (S2-10) cancels the interference cross-terms in $\mathcal{U}$ and isolates the dissipative dependence through $\Re(\zeta)$, giving

$$\mathcal{U}(\zeta) = \frac{1}{2}\left(e^{-4\xi\,\Re(\zeta)} + |\bar{\Gamma}_g(\Omega)|^2\, e^{4(1-\xi)\Re(\zeta)}\right),$$
$$\mathcal{S}(\zeta) = \frac{1}{2}(e^{-4\xi\,\Re(\zeta)} - |\bar{\Gamma}_g(\Omega)|^2\, e^{4(1-\xi)\,\Re(\zeta)}). \qquad (S2-11)$$

### S2.7 Exact invariant and identification of the generalized reflection "Mass"

Combining the two expressions in Eq. (S2-11) yields

$$\mathcal{U}^2(\zeta) - \mathcal{S}^2(\zeta) = |\bar{\Gamma}_g(\Omega)|^2 \exp[-4(2\xi - 1)\Re(\zeta)]. \qquad (S2-12)$$

This motivates the axial generalized reflection state variable

$$\Gamma_g(\zeta) = \bar{\Gamma}_g(\Omega) \exp\left[-2(2\xi - 1)\Re(\zeta)\right], \quad (S2-13)$$

so that Eq. (S2-12) takes the exact invariant form used in the main text,

$$\mathcal{U}^2(\zeta) - \mathcal{S}^2(\zeta) = |\Gamma_g(\zeta)|^2. \quad (S2-14)$$

For the symmetric case $\xi = 1/2$, $\Gamma_g(\zeta) = \bar{\Gamma}_g(\Omega)$ is axially constant and Eq. (S2-14) reduces to a spatially uniform mass shell. For $\xi \neq 1/2$, $\Gamma_g$ acquires a controlled radial drift along the dissipative coordinate $\Re(\zeta)$ while preserving the same invariant structure; the resulting envelope bounds, deformation interpretations, and end-to-end relations are detailed in Supplementary Methods S3.

**S2.8 Polar decomposition and Cai–Smith trajectories**

For each frequency point, we evaluate $\bar{\Gamma}_g(\Omega)$ from Eq. (S2-4) and define its polar decomposition

$$\bar{\Gamma}_g(\Omega) = \rho(\Omega)e^{i\varphi(\Omega)}, \rho(\Omega) = |\bar{\Gamma}_g(\Omega)|, \varphi(\Omega) = \arg \bar{\Gamma}_g(\Omega). \quad (S2-15)$$

Frequency sweeps $\Omega \mapsto \bar{\Gamma}_g(\Omega)$ generate trajectories on the Cai–Smith unit disk. In all trajectory-based diagnostics (resonance markers, phase detuning, group-delay proxies, and derivative-based quantities), $\varphi(\Omega)$ is unwrapped consistently on the same frequency grid to avoid spurious branch jumps; numerical details and approximation-controlled derivatives are given in Supplementary Methods S4.

**Supplementary Methods S3 | Intrinsic asymmetry $\xi$: deformation law and reachable envelopes**

**S3.0 Overview**

This section operationalizes the intrinsic asymmetry coefficient $\xi \in (0,1)$. Within the two-wave representation introduced in Supplementary Methods S2, $\xi$ re-weights the attenuation of forward and backward components along the dissipative coordinate, thereby deforming (i) the axial evolution of the generalized-reflection "mass" $|\Gamma_g(\zeta)|$ and (ii) the set of instantaneous field profiles reachable at fixed $(K, \bar{\Gamma}_g)$. We derive phase-independent upper/lower envelopes for the instantaneous voltage- and current-like fields, identify the corresponding "extreme" constructive/destructive snapshots, and show how $\xi$ produces a controlled radial drift in the Cai–Smith state while preserving the invariant structure $\mathcal{U}^2 - \mathcal{S}^2 = |\Gamma_g|^2$.

### S3.1 Direction-weighted two-wave form and the asymmetry-controlled attenuation coordinate

Let $s = x/L \in [0,1]$ be the normalized axial coordinate and $K(\Omega) = R(\Omega) + i\Phi(\Omega)$ the complex electrical length (Supplementary Methods S1). We use the axial complex coordinate $\zeta(s) = sK(\Omega)$ and define the dissipative coordinate

$$x(s;\Omega) = \Re[\zeta(s)] = s\,R(\Omega) = s\,\Re[K(\Omega)] \geq 0. \qquad (S3-1)$$

At fixed frequency (fixed $K$) and fixed boundary-composed reflection $\bar{\Gamma}_g(\Omega)$, we write the normalized two-wave phasors in a general complex-amplitude form

$$v(\zeta) = A(e^{-2\xi\zeta} + \rho\,e^{i\theta}\,e^{2(1-\xi)\zeta}),\, i(\zeta) = A(e^{-2\xi\zeta} - \rho\,e^{i\theta}\,e^{2(1-\xi)\zeta}), \qquad (S3-2)$$

where $A$ is a (real, non-negative) amplitude scale, $\rho = |\bar{\Gamma}_g(\Omega)| \in [0,1)$ is the boundary-composed reflection magnitude, and $\theta = \arg \bar{\Gamma}_g(\Omega)$ the boundary-composed reflection phase. Equation (S3-2) is equivalent to the normalized form in Eq. (S2-9) up to a constant scaling (for the S2 normalization, $A = 1/\sqrt{2}$). Crucially, $\xi$ enters only through the real attenuation weights $e^{-2\xi x}$ and $e^{2(1-\xi)x}$, while the oscillatory dependence comes from $\Im[\zeta] = s\Phi(\Omega)$ and the phase $\theta$.

### S3.2 Instantaneous profiles and phase snapshots

Define the real instantaneous fields under the global time-harmonic convention

$e^{+i\omega t}$ as

$$v(s,t) = \Re\{v(\zeta(s))\, e^{i\omega t}\}, i(s,t) = \Re\{i(\zeta(s))\, e^{i\omega t}\}. \qquad (S3-3)$$

At a fixed operating frequency, varying $t$ scans the relative phase between forward and backward contributions at each axial position. For visualization we present discrete phase snapshots $\vartheta \in [0, 2\pi]$ (e.g., $\vartheta \in \{0, \pi/2, \pi, 3\pi/2, 2\pi\}$), where $\vartheta$ is a global phase offset representing a fixed time $t$ modulo $2\pi/\omega$. Although the axial dependence involves $s\Phi(\Omega)$ and thus different $s$ experience different local interference phases at the same snapshot, the instantaneous amplitude at each $s$ is bounded by phase-independent extrema derived below.

**S3.3 Phase-independent reachable envelopes for $v$ and $i$**

At a fixed axial position, the complex magnitudes of $v(\zeta)$ and $i(\zeta)$ depend only on the relative phase between the two terms in Eq. (S3-2). Using $|e^{-2\xi\zeta}| = e^{-2\xi x}$ and $|e^{2(1-\xi)\zeta}| = e^{2(1-\xi)x}$, the triangle inequality yields sharp, phase-attainable bounds

$$|A||e^{-2\xi x} - \rho\, e^{2(1-\xi)x}| \le |v(\zeta)| \le |A|(e^{-2\xi x} + \rho\, e^{2(1-\xi)x}), \qquad (S3-4)$$

and identically for $|i(\zeta)|$ (since $i$ differs from $v$ only by a sign in the second term). We therefore define the phase-independent reachable envelopes

$$E_{\max}(s) = |A|(e^{-2\xi x(s)} + \rho\, e^{2(1-\xi)x(s)}), E_{\min}(s) = |A||e^{-2\xi x(s)} - \rho\, e^{2(1-\xi)x(s)}|. (S3-5)$$

At each axial position $s$, the instantaneous fields satisfy $v(s,t) \in [-E_{\max}(s), E_{\max}(s)]$ and $i(s,t) \in [-E_{\max}(s), E_{\max}(s)]$, while the smallest achievable absolute magnitude at that position is $E_{\min}(s)$. The upper envelope $E_{\max}$ is attained when the two contributions are locally in phase (constructive interference), and the lower envelope $E_{\min}$ when they are locally out of phase (destructive interference). These envelopes are independent of the snapshot set and thus reproducible without time stepping.

## S3.4 Axial Drift of the effective mass $|\Gamma_g(\zeta)|$ and its $\xi$-dependence

Within the intrinsic-asymmetry construction (Supplementary Methods S2), the invariant uses the axial generalized-reflection state

$$\Gamma_g(\zeta) = \bar{\Gamma}_g(\Omega)\exp[-2(2\xi-1)\,x(s)], \qquad (S3-6)$$

so the effective "mass" magnitude drifts radially as

$$|\Gamma_g(\zeta)| = \rho\exp[-2(2\xi-1)\,x(s)]. \qquad (S3-7)$$

Equation (S3-7) makes the sign meaning of $\xi$ explicit under the global convention: for $\xi = 1/2$, $|\Gamma_g|$ is axially constant; for $\xi > 1/2$ the exponent $-2(2\xi-1)x$ is negative and $|\Gamma_g|$ decreases monotonically with $s$ (radial contraction along propagation); for $\xi < 1/2$ the exponent is positive and $|\Gamma_g|$ increases with $s$ (radial expansion). This controlled radial drift is the geometric mechanism by which intrinsic asymmetry modifies reachable envelopes while preserving the invariant form.

At the two terminals $s = 0$ and $s = 1$,

$$|\Gamma_g(0)| = \rho,\ |\Gamma_g(1)| = \rho\exp[-2(2\xi-1)R(\Omega)], \qquad (S3-8)$$

so the end-to-end deformation factor depends solely on $(\xi, R)$.

## S3.5 Consequences for nodal structure and snapshot asymmetry

The destructive-interference envelope $E_{\min}(s)$ vanishes when the two magnitude contributions match, i.e.,

$$e^{-2\xi x} = \rho\,e^{2(1-\xi)x} \iff e^{-2x} = \rho \iff x_{\text{node}} = \frac{1}{2}\ln\left(\frac{1}{\rho}\right). \qquad (S3-9)$$

Notably, the node location in the dissipative coordinate $x$ is independent of $\xi$; intrinsic asymmetry does not shift the matching condition in $x$, but it does affect how rapidly $x(s) = sR$ is traversed with $s$ (via $R(\Omega)$) and alters the skewness of the two contributing magnitudes $e^{-2\xi x}$ and $\rho e^{2(1-\xi)x}$. A node within the physical interval exists when $0 < x_{\text{node}} < R(\Omega)$, equivalently $e^{-2R(\Omega)} < \rho < 1$. When $\rho$ is small or $R$ is large, equality may lie beyond the segment, and $E_{\min}(s)$ remains strictly positive across $s \in [0,1]$, indicating that complete local cancellation is unattainable under passive losses.

The constructive envelope $E_{\max}(s)$ reveals a different aspect: for $\xi < 1/2$ the factor $e^{2(1-\xi)x}$ grows rapidly with $x$, so the backward contribution can dominate near the far end even when $\rho < 1$; for $\xi > 1/2$ this growth is suppressed and the envelope becomes more strongly forward-dominated. These features explain the qualitative differences among rows in snapshot-and-envelope visualizations (e.g., $\xi = 0.30$ vs. $\xi = 0.80$) at identical $(\delta_R, \delta_G, \Omega, \Gamma_S, \Gamma_L)$.

**S3.6 Practical evaluation and envelope-plotting protocol**

Given $(\delta_R, \delta_G)$ and $\Omega$, compute $K(\Omega)$ using Supplementary Methods S1, then compute $x(s) = s\Re[K(\Omega)]$ on a uniform grid $s \in [0,1]$. For given terminal reflections $(\Gamma_S, \Gamma_L)$, compute the boundary-composed state $\bar{\Gamma}_g(\Omega)$ via Eq. (S2-4), set $\rho = |\bar{\Gamma}_g|$ and $\theta = \arg \bar{\Gamma}_g$, and choose $\xi$ to generate the two-wave phasors in Eq. (S3-2). Instantaneous snapshots $v(s,t)$ and $i(s,t)$ may be generated either by selecting times $t$ corresponding to global phase offsets $\vartheta$ (Eq. S3-3) or, equivalently, by applying a uniform phase shift to $A$ in Eq. (S3-2). The phase-independent envelopes are then computed directly from Eq. (S3-5) and plotted as $\pm E_{\max}(s)$ and $\pm E_{\min}(s)$.

**S3.7 Parameter set for extended data Fig. 2**

Extended Data Fig. 2 uses the snapshot set $\vartheta \in \{0, \pi/2, \pi, 3\pi/2, 2\pi\}$ and rows $\xi \in \{0.30, 0.50, 0.80\}$, with propagation budgets and boundaries chosen as $\delta_R = 10^{-1}$, $\delta_G = 3 \times 10^{-1}$, $\Omega = 10$, $\Gamma_L = 0.1 e^{i\pi/2}$, and $\Gamma_S = 0$. With these choices, $\bar{\Gamma}_g(\Omega)$ is computed from Eq. (S2-4), $x(s) = s\Re[K(\Omega)]$ from Eq. (S3-1), and the envelopes $E_{\max}, E_{\min}$ from Eq. (S3-5).

**Supplementary Methods S4 | Low-mass expansions and quantitative error bounds**

**S4.0 Overview**

This section quantifies the "low-mass/high-speed" regime of the waveguide mass-energy relation and provides practical, provably bounded approximations for the energy-like state $\mathcal{U}$ (and derived kinematic variables) when the effective mass $|\Gamma_g(\zeta)|$ is small compared to the power-flow magnitude $|\mathcal{S}|$. The results are used to (i) justify the high-speed approximation shown in the main text, (ii) construct fast, numerically stable evaluators for $(\mathcal{U}, \beta_w, \phi)$, and (iii) generate the error-scaling diagnostics in Extended Data Fig. 3.

**S4.1 Exact relations and the expansion parameter**

At any axial position $\zeta$ (and at any operating frequency), define the effective mass magnitude and the dimensionless speed proxy

$$\chi(\zeta) = \frac{|\mathcal{S}(\zeta)|}{|\Gamma_g(\zeta)|}. \tag{S4-1}$$

The exact mass-energy relation (Supplementary Methods S0 and S2) implies the closed-form identity

$$\mathcal{U}(\zeta) = \sqrt{\mathcal{S}^2(\zeta) + |\Gamma_g(\zeta)|^2} = |\Gamma_g(\zeta)|\sqrt{1+\chi^2(\zeta)} \ (\mathcal{U} \geq 0). \tag{S4-2}$$

Equivalently, introducing the small parameter $u = |\Gamma_g|^2/\mathcal{S}^2 = 1/\chi^2$ (defined when $\mathcal{S} \neq 0$), Eq. (S4-2) can be written as $\mathcal{U} = |\mathcal{S}|\sqrt{1+u}$. The low-mass/high-speed regime corresponds to $\chi \gg 1$ (or $u \ll 1$), i.e., the state lies near the light-cone boundary $|\beta_w| \to 1$ with $\beta_w = \mathcal{S}/\mathcal{U}$.

**S4.2 First- and second-order low-mass expansions for $\mathcal{U}$**

Using the Taylor expansion of $\sqrt{1+u}$ about $u=0$ and substituting $u=1/\chi^2$ yields the standard high-speed approximations

$$\mathcal{U}(\zeta) = |\mathcal{S}|(1+\frac{u}{2}-\frac{u^2}{8}+\mathcal{O}(u^3)) = |\mathcal{S}|+\frac{|\Gamma_g|^2}{2|\mathcal{S}|}-\frac{|\Gamma_g|^4}{8|\mathcal{S}|^3}+\mathcal{O}(\frac{|\Gamma_g|^6}{|\mathcal{S}|^5}). \quad (S4-3)$$

We denote the first-order (main-text) approximation and the second-order refinement by

$$\mathcal{U}_{(1)} = |\mathcal{S}|+\frac{|\Gamma_g|^2}{2|\mathcal{S}|}, \mathcal{U}_{(2)} = |\mathcal{S}|+\frac{|\Gamma_g|^2}{2|\mathcal{S}|}-\frac{|\Gamma_g|^4}{8|\mathcal{S}|^3}. \quad (S4-4)$$

These approximations are local in $\zeta$ and therefore apply equally to symmetric and intrinsically asymmetric guides: one simply evaluates $|\Gamma_g(\zeta)|$ (Supplementary Methods S2–S3) and $\mathcal{S}(\zeta)$ at the position of interest.

**S4.3 Guaranteed bracketing and absolute/relative error bounds**

Let $u = m^2/\mathcal{S}^2 \geq 0$. The function $f(u) = \sqrt{1+u}$ is concave for $u \geq 0$ (since $f''(u) < 0$), so it lies below its tangent at $u=0$, giving $\sqrt{1+u} \leq 1+u/2$. A second-order Taylor theorem with remainder yields

$$\sqrt{1+u} = 1+\frac{u}{2}-\frac{u^2}{8(1+\theta u)^{3/2}} \text{ for some } \theta \in (0,1). \quad (S4-5)$$

Combining these properties gives a practical and globally valid bracketing for $\mathcal{U}$:

$$\mathcal{U}_{(2)} \leq \mathcal{U} \leq \mathcal{U}_{(1)} \; (m \geq 0, \mathcal{U} \geq 0). \quad (S4-6)$$

From Eq. (S4-5), the first-order absolute error satisfies

$$0 \leq \mathcal{U}_{(1)}-\mathcal{U} = |\mathcal{S}|(1+\frac{u}{2}-\sqrt{1+u}) \leq |\mathcal{S}|\frac{u^2}{8} = \frac{|\Gamma_g|^4}{8|\mathcal{S}|^3}. \quad (S4-7)$$

Dividing by $\mathcal{U} = |\mathcal{S}|\sqrt{1+u} \geq |\mathcal{S}|$ yields the first-order relative-error bound

$$0 \leq \varepsilon_{(1)} = \frac{\mathcal{U}_{(1)}-\mathcal{U}}{\mathcal{U}} \leq \frac{u^2}{8} = \frac{1}{8\chi^4}. \quad (S4-8)$$

A third-order Taylor theorem applied to $f(u) = \sqrt{1+u}$ gives a non-negative remainder for truncation at second order because $f'''(u) > 0$. Concretely, there exists $\theta \in (0,1)$ such that

$$\sqrt{1+u} = 1 + \frac{u}{2} - \frac{u^2}{8} + \frac{u^3}{16(1+\theta u)^{5/2}}, \qquad (S4-9)$$

which implies

$$0 \le \mathcal{U} - \mathcal{U}_{(2)} = |\mathcal{S}|(\sqrt{1+u} - 1 - \frac{u}{2} + \frac{u^2}{8}) \le |\mathcal{S}|\frac{u^3}{16} = \frac{|\Gamma_g|^6}{16|\mathcal{S}|^5}, \qquad (S4-10)$$

and hence the second-order relative error satisfies

$$0 \le \varepsilon_{(2)} = \frac{\mathcal{U} - \mathcal{U}_{(2)}}{\mathcal{U}} \le \frac{u^3}{16} = \frac{1}{16\chi^6}. \qquad (S4-11)$$

These bounds provide a direct accuracy guarantee in terms of the single dimensionless ratio $\chi = |\mathcal{S}|/|\Gamma_g|$.

### S4.4 Rapidity, doppler factors, and high-speed asymptotics

Using the rapidity parameterization $\mathcal{U} = |\Gamma_g| \cosh \phi$ and $\mathcal{S} = |\Gamma_g| \sinh \phi$ (Supplementary Methods S5), the exact inversion is

$$\phi(\zeta) = \operatorname{asinh}\left(\frac{\mathcal{S}(\zeta)}{|\Gamma_g(\zeta)|}\right) = \operatorname{asinh}(\operatorname{sgn}(\mathcal{S})\chi). \qquad (S4-12)$$

For $\chi \gg 1$, the large-argument expansion $\operatorname{asinh}(\chi) = \ln(\chi + \sqrt{1+\chi^2})$ gives

$$\phi = \operatorname{sgn}(\mathcal{S})[\ln(2\chi) + \frac{1}{4\chi^2} - \frac{3}{32\chi^4} + \mathcal{O}(\chi^{-6})]. \qquad (S4-13)$$

The waveguide velocity proxy and Doppler factors follow as

$$\beta_w = \frac{\mathcal{S}}{\mathcal{U}} = \tanh \phi = \operatorname{sgn}(\mathcal{S})[1 - \frac{1}{2\chi^2} + \frac{3}{8\chi^4} + \mathcal{O}(\chi^{-6})], D_\pm = e^{\pm\phi}. \qquad (S4-14)$$

Equations (S4-13)–(S4-14) are used when a logarithmic or near-light-cone parameterization is numerically preferable (e.g., for regime separation and composition rules), while Eqs. (S4-4)–(S4-11) provide strict error control for $\mathcal{U}$ itself.

### S4.5 Practical accuracy criterion and the $\chi_{\text{th}}$-rule

For fast evaluation we use a threshold $\sigma_{th}$ on the ratio $\chi = |\mathcal{S}|/|\Gamma_g| : \chi \geq \chi_{th}$, use the low-mass approximation (first or second order); $\chi < \chi_{th}$, use the exact form $\mathcal{U} = \sqrt{\mathcal{S}^2 + |\Gamma_g|^2}$. With $\chi_{th} = 10$ (the value used in Extended Data Fig. 3), Eq. (S4-8) gives $\varepsilon_{(1)} \leq 1/(8\chi_{th}^4) = 1.25 \times 10^{-5}$, and Eq. (S4-11) gives $\varepsilon_{(2)} \leq 1/(16\chi_{th}^6) = 6.25 \times 10^{-8}$. Thus the first-order approximation already provides $\mathcal{O}(10^{-5})$ relative accuracy in the designated low-mass region, and the second-order form is typically indistinguishable from the exact relation at plotting resolution.

**S4.6 Numerical stability notes**

When $\chi \gg 1$, direct evaluation of $\sqrt{\mathcal{S}^2 + |\Gamma_g|^2}$ is stable in double precision for typical manuscript ranges; however, the expansion $\mathcal{U}_{(1)}$ avoids unnecessary square-root calls inside large parameter sweeps and makes the scaling with $x$ explicit. Conversely, near $\mathcal{S} = 0$ the ratio $u = |\Gamma_g|^2/\mathcal{S}^2$ is ill-conditioned and the expansion must not be used; the exact evaluation in Eq. (S4-2) is always employed there. In automated plotting we additionally mask any points that violate $|\beta_w| < 1$ due to numerical roundoff, where $\beta_w = \mathcal{S}/\mathcal{U}$ is computed from consistent $(\mathcal{S}, \mathcal{U})$.

**Supplementary Methods S5 | Waveguide relativity formalism: rapidity, doppler factors, and composition rules**

**S5.0 Overview**

This section develops the rapidity-based "waveguide relativity" parameterization used throughout the manuscript as a numerically stable and physically transparent coordinate system for the invariant state $(\mathcal{U}, \mathcal{S}, |\Gamma_g|)$. The central idea is to treat $(\mathcal{U}, \mathcal{S})$ as a Minkowski-like pair constrained by $\mathcal{U}^2 - \mathcal{S}^2 = |\Gamma_g|^2$, so the state can be reparameterized by a single signed rapidity $\phi$ and a bounded velocity proxy $\beta_w$.

Rapidity linearizes composition under cascaded segments (additive in $\phi$), while Doppler factors $D_\pm = e^{\pm\phi}$ provide multiplicative diagnostics that remain well-conditioned near the light-cone ($|\beta_w| \to 1$). These tools are used for regime separation, for interpreting feedback trajectories, and for the two-regime decomposition in electrochemical mapping.

**S5.1 Definitions: velocity proxy, lorentz factor, and rapidity**

At any axial position $\zeta$ (or for any reduced state extracted from data), define the effective-mass magnitude $m = |\Gamma_g| \geq 0$ and the signed waveguide-velocity proxy

$$\beta_w = \frac{\mathcal{S}}{\mathcal{U}}, |\beta_w| < 1 \text{ for } m > 0, \beta_w \to \pm 1 \text{ as } |\Gamma_g|/|\mathcal{S}| \to 0. \quad (S5-1)$$

Define the Lorentz-like factor $\gamma_w$ by

$$\gamma_w = \frac{\mathcal{U}}{|\Gamma_g|} = \frac{1}{\sqrt{1-\beta_w^2}}, \gamma_w \geq 1. \quad (S5-2)$$

The rapidity $\phi$ is defined as

$$\phi = \operatorname{atanh}(\beta_w) = \frac{1}{2}\ln\left(\frac{1+\beta_w}{1-\beta_w}\right) = \operatorname{asinh}\left(\frac{\mathcal{S}}{|\Gamma_g|}\right) = \operatorname{acosh}\left(\frac{\mathcal{U}}{|\Gamma_g|}\right), \quad (S5-3)$$

where the multiple equivalent forms follow from the invariant $\mathcal{U}^2 - \mathcal{S}^2 = |\Gamma_g|^2$ (Supplementary Methods S2). The sign convention is global: $\phi > 0$ corresponds to forward-dominated power flow ($\mathcal{S} > 0$), and $\phi < 0$ corresponds to backward-dominated power flow ($\mathcal{S} < 0$).

**S5.2 Hyperbolic parameterization and light-cone geometry**

Using $\gamma_w = \cosh \phi$ and $\gamma_w \beta_w = \sinh \phi$, the invariant state admits the exact parameterization

$$\mathcal{U} = |\Gamma_g|\cosh \phi, \mathcal{S} = |\Gamma_g|\sinh \phi, \mathcal{U}^2 - \mathcal{S}^2 = |\Gamma_g|^2. \quad (S5-4)$$

In the $(\mathcal{S}, \mathcal{U})$ plane, fixed $|\Gamma_g|$ corresponds to a hyperbola (a "mass shell"); the boundary $|\Gamma_g| \to 0$ collapses the shell to the light-cone lines $\mathcal{U} = \pm \mathcal{S}$ with $|\beta_w| \to 1$ and $|\phi| \to \infty$. This representation is exact within the single-mode reduction and does not require weak-loss assumptions; all platform dependence enters through how $|\Gamma_g|(\zeta) = |\Gamma_g(\zeta)|$ and $\mathcal{S}(\zeta)$ are generated from propagation and boundaries (Supplementary Methods S1–S3, S6).

### S5.3 Doppler factors and numerically stable invariants

Define the bidirectional Doppler factors

$$D_\pm = e^{\pm\phi}. \qquad (S5-5)$$

Using Eq. (S5-3), $D_\pm$ admit the equivalent forms

$$D_+ = \sqrt{\frac{1+\beta_w}{1-\beta_w}} = \frac{\mathcal{U}+\mathcal{S}}{|\Gamma_g|}, \quad D_- = \sqrt{\frac{1-\beta_w}{1+\beta_w}} = \frac{\mathcal{U}-\mathcal{S}}{|\Gamma_g|}, \quad D_+ D_- = 1. \qquad (S5-6)$$

Equation (S5-6) offers two practical advantages: (i) $(\mathcal{U} \pm \mathcal{S})$ are always non-negative for $\mathcal{U} \geq |\mathcal{S}|$, so $D_\pm$ avoid catastrophic cancellation that can occur when forming $\beta_w$ near $|\beta_w| \to 1$; (ii) the ratio

$$\frac{\mathcal{U}+\mathcal{S}}{\mathcal{U}-\mathcal{S}} = D_+^2 = e^{2\phi} \qquad (S5-7)$$

provides a high-dynamic-range diagnostic of forward/backward dominance that remains well-conditioned in the low-mass regime. Specifically, large positive $\phi$ corresponds to $D_+ \gg 1$ (strong forward dominance), while large negative $\phi$ corresponds to $D_- \gg 1$ (strong backward dominance).

### S5.4 Boost representation and composition under cascaded segments

A central advantage of rapidity is that it linearizes composition. Consider two consecutive reduced segments (or two effective transformations) that, at the level of the invariant state, act as hyperbolic boosts in the $(\mathcal{U}, \mathcal{S})$ plane. Define the boost matrix

$$\mathbf{B}(\phi) = \begin{pmatrix} \cosh\phi & \sinh\phi \\ \sinh\phi & \cosh\phi \end{pmatrix}, \qquad (S5-8)$$

which preserves the Minkowski metric $\mathcal{U}^2 - \mathcal{S}^2$ and maps a rest-like state $(m, 0)$ to $(m\cosh\phi, m\sinh\phi)$. For two boosts with rapidities $\phi_1$ and $\phi_2$, matrix multiplication yields

$$\mathbf{B}(\phi_1)\mathbf{B}(\phi_2) = \mathbf{B}(\phi_1 + \phi_2), \qquad (S5-9)$$

so the net rapidity is additive:

$$\phi_{12} = \phi_1 + \phi_2. \qquad (S5-10)$$

Equivalently, the velocity proxy composes via the Einstein addition law

$$\beta_w^{12} = \tanh(\phi_1 + \phi_2) = \frac{\beta_w^1 + \beta_w^2}{1 + \beta_w^1 \beta_w^2}, \qquad (S5-11)$$

and the Doppler factors multiply:

$$D_{\pm,12} = e^{\pm(\phi_1 + \phi_2)} = D_{\pm,1} D_{\pm,2}. \qquad (S5-12)$$

These identities are used in Extended Data Fig. 4 to visualize how cascaded effects accumulate linearly in $\phi$ (and multiplicatively in $D_\pm$) while respecting the light-cone constraint $|\beta_w| < 1$ for $|\Gamma_g| > 0$.

**S5.5 Relation to measurable quantities and practical evaluation**

In the reduced waveguide model, $(\mathcal{U}, \mathcal{S})$ are computed from normalized two-wave fields (Supplementary Methods S2). For any state triplet $(\mathcal{U}, \mathcal{S}, |\Gamma_g|)$, one may compute $\phi$ using any of the equivalent forms in Eq. (S5-3). In numerical pipelines we recommend the $D_\pm$ route: compute $(\mathcal{U} \pm \mathcal{S})$ and form $D_\pm = (\mathcal{U} \pm \mathcal{S})/|\Gamma_g|$, then obtain $\phi = \ln D_+$ (or $\phi = -\ln D_-$). This avoids subtractive cancellation when $\mathcal{U} \approx |\mathcal{S}|$ in the low-mass regime. When $\mathcal{S}$ is near zero, use $\phi = \mathrm{asinh}(\mathcal{S}/|\Gamma_g|)$ to avoid forming $\beta_w$ explicitly. In all cases, the identities $\cosh\phi = \mathcal{U}/|\Gamma_g|$ and $\sinh\phi = \mathcal{S}/|\Gamma_g|$ (Eq. S5-4) provide a direct consistency check.

### S5.6 Low-mass asymptotics and consistency with supplementary methods S4

Let $\chi = |\mathcal{S}|/|\Gamma_g|$ (Supplementary Methods S4). Then $\phi = \mathrm{asinh}(\mathcal{S}/|\Gamma_g|) = \mathrm{sgn}(\mathcal{S})\,\mathrm{asinh}(\chi)$ and, for $\chi \gg 1$, $\phi = \mathrm{sgn}(\mathcal{S})[\ln(2\chi) + \frac{1}{4\chi^2} - \frac{3}{32\chi^4} + \mathcal{O}(\chi^{-6})]$, $\beta_w = \tanh\phi = \mathrm{sgn}(\mathcal{S})[1 - \frac{1}{2\chi^2} + \frac{3}{8\chi^4} + \mathcal{O}(\chi^{-6})]$, consistent with the error-controlled low-mass expansions for $\mathcal{U}$ and derived quantities in Supplementary Methods S4. Thus, the rapidity formalism provides a smooth bridge between the exact invariant state and the high-speed approximation regime.

### Supplementary Methods S6 | Unified feedback model for resonance, absorption, useful power, and ring-down

### S6.0 Overview

This section provides a minimal feedback description for any finite, linear, single-mode, lossy waveguide with terminal reflection coefficients $\Gamma_S(\Omega)$ and $\Gamma_L(\Omega)$ defined with respect to the characteristic impedance. The model reduces resonance, absorptivity, objective optimization, and ring-down to two scalar feedback parameters: the round-trip magnitude $\kappa(\Omega)$ and feedback phase $\Theta(\Omega)$. These quantities generate (i) the build-up/proximity factor $\Lambda$, (ii) the boundary-composed Cai–Smith state $\bar{\Gamma}_g$, (iii) one-port absorptivity $\mathcal{A}$, (iv) useful load power $\mathcal{T}_L$, (v) absorbed-power landscapes $\mathcal{P}_{\mathrm{abs}}$ (including intrinsic asymmetry $\xi$), and (vi) ring-down and $Q$-factor proxies.

### S6.1 Round-trip feedback factor and two-parameter reduction

Let $K(\Omega) = R(\Omega) + i\Phi(\Omega)$ be the dimensionless complex electrical length (Supplementary Methods S1), and define the round-trip propagation factor $E(\Omega) = e^{-2K(\Omega)} = e^{-2R(\Omega)}e^{-i2\Phi(\Omega)}$. The unified feedback factor is

$$F(\Omega) = \Gamma_S(\Omega)\,\Gamma_L(\Omega)\,e^{-2K(\Omega)} = \Gamma_S \Gamma_L\, E. \tag{S6-1}$$

Writing $\quad F(\Omega) = \kappa(\Omega)e^{i\Theta(\Omega)} \quad$ gives

$$\kappa(\Omega) = |F(\Omega)| = |\Gamma_S(\Omega)\Gamma_L(\Omega)|\,e^{-2R(\Omega)},\; \Theta(\Omega) = \arg\left(\Gamma_S(\Omega)\Gamma_L(\Omega)\right) - 2\Phi(\Omega). \tag{S6-2}$$

Here $\kappa \in [0,1)$ is the round-trip amplitude survival factor (controlling lifetime), while $\Theta$ governs interference, selecting constructive (odd-$\pi$) versus destructive (even-$\pi$) recirculation.

### S6.2 Boundary-composed Cai–Smith state and resonance denominator

The boundary-composed generalized reflection coefficient at the driven port is

$$\bar{\Gamma}_g(\Omega) = \frac{\Gamma_L(\Omega)e^{-2K(\Omega)} + \Gamma_S(\Omega)}{1 + \Gamma_S(\Omega)\Gamma_L(\Omega)e^{-2K(\Omega)}} = \frac{\Gamma_L E + \Gamma_S}{D(\Omega)}, \tag{S6-3}$$

where the feedback denominator is

$$D(\Omega) = 1 + F(\Omega),\; |D(\Omega)|^2 = |1 + \kappa(\Omega)e^{i\Theta(\Omega)}|^2 = 1 + \kappa(\Omega)^2 + 2\kappa(\Omega)\cos\Theta(\Omega). \tag{S6-4}$$

We define the build-up/proximity factor

$$\Lambda(\Omega) = \frac{1}{|D(\Omega)|} = \frac{1}{|1 + F(\Omega)|}, \tag{S6-5}$$

and a detuning variable (relative to the odd-$\pi$ resonance condition)

$$\Delta\Theta(\Omega) = \Theta(\Omega) - \pi \;(\mathrm{mod}\; 2\pi). \tag{S6-6}$$

### S6.3 One-port absorptivity, useful-power partition, and $\eta_{\mathrm{use}}$

The one-port absorptivity is

$$\mathcal{A}(\Omega) = 1 - |\bar{\Gamma}_g(\Omega)|^2. \tag{S6-7}$$

Using Eq. (S6-3),

$$|\bar{\Gamma}_g(\Omega)|^2 = \frac{|\Gamma_S(\Omega) + \Gamma_L(\Omega)e^{-2K(\Omega)}|^2}{|1 + \Gamma_S(\Omega)\Gamma_L(\Omega)e^{-2K(\Omega)}|^2} = \frac{|\Gamma_S + \Gamma_L E|^2}{|D|^2}. \tag{S6-8}$$

We define the dimensionless useful load power fraction (power delivered into the load termination) as

$$\mathcal{T}_L(\Omega) = \frac{(1-|\Gamma_S(\Omega)|^2)(1-|\Gamma_L(\Omega)|^2)\,e^{-2R(\Omega)}}{|1+\Gamma_S(\Omega)\Gamma_L(\Omega)e^{-2K(\Omega)}|^2} = \frac{(1-|\Gamma_S|^2)(1-|\Gamma_L|^2)e^{-2R}}{|D|^2}. \quad (S6-9)$$

This motivates two partition diagnostics:

$$\eta_{\text{use}}(\Omega) = \frac{\mathcal{T}_L(\Omega)}{\mathcal{A}(\Omega)}, \mathcal{T}_{\text{int}}(\Omega) = \mathcal{A}(\Omega) - \mathcal{T}_L(\Omega), \quad (S6-10)$$

where $\eta_{\text{use}}$ is the fraction of absorbed power delivered to the load, while $\mathcal{T}_{\text{int}}$ is the complementary internal dissipation share (distributed loss, leakage, or other non-load absorption channels within the reduced model).

**S6.4 Resonance class, two optimal couplings, and their duality**

Critical coupling corresponds to perfect one-port absorption at the driven port, i.e., $\bar{\Gamma}_g(\Omega) = 0$, which is equivalent to

$$\Gamma_S(\Omega) + \Gamma_L(\Omega)e^{-2K(\Omega)} = 0. \quad (S6-11)$$

This splits into magnitude matching and phase tuning:

$$|\Gamma_S(\Omega)| = |\Gamma_L(\Omega)|e^{-2R(\Omega)}, \arg\Gamma_L(\Omega) - 2\Phi(\Omega) = \arg\Gamma_S(\Omega) + \pi \ (\text{mod } 2\pi). (S6-12)$$

Equation (S6-12) makes the feasibility constraint explicit: passivity requires $|\Gamma_L(\Omega)| \leq 1$, so exact critical coupling becomes impossible when the required $|\Gamma_L|$ exceeds unity.

To compare absorptivity and useful-power optimization on the same interference class, we restrict to the constructive odd-$\pi$ resonance direction $\Theta(\Omega) = \pi$, which can always be enforced by selecting the load phase $\arg\Gamma_L$ such that $\Theta(\Omega) = \arg(\Gamma_S\Gamma_L) - 2\Phi = \pi \ (\text{mod } 2\pi)$. On this resonance class, Eq. (S6-8) reduces to

$$|\bar{\Gamma}_g(\Omega)|^2 = \frac{(|\Gamma_S(\Omega)| - |\Gamma_L(\Omega)|e^{-2R(\Omega)})^2}{(1 - |\Gamma_S(\Omega)||\Gamma_L(\Omega)|e^{-2R(\Omega)})^2}, \quad (S6-13)$$

and Eqs. (S6-7) and (S6-9) simplify to

$$\mathcal{A}(\Omega) = \frac{(1 - |\Gamma_S(\Omega)|^2)(1 - |\Gamma_L(\Omega)|^2 e^{-4R(\Omega)})}{(1 - |\Gamma_S(\Omega)| |\Gamma_L(\Omega)| e^{-2R(\Omega)})^2},$$

$$\mathcal{T}_L(\Omega) = \frac{(1 - |\Gamma_S(\Omega)|^2)(1 - |\Gamma_L(\Omega)|^2) e^{-2R(\Omega)}}{(1 - |\Gamma_S(\Omega)| |\Gamma_L(\Omega)| e^{-2R(\Omega)})^2}. \quad (S6-14)$$

The two optimization problems then yield two distinct optimal couplings on the same resonance class:

$$|\Gamma_L(\Omega)|_{cc} = |\Gamma_S(\Omega)| \, e^{2R(\Omega)} \ (\bar{\Gamma}_g = 0, \text{ maximizes } \mathcal{A}),$$
$$|\Gamma_L(\Omega)|_{opt} = |\Gamma_S(\Omega)| \, e^{-2R(\Omega)} \ (\text{maximizes } \mathcal{T}_L \text{ over } |\Gamma_L| \leq 1). \quad (S6-15)$$

Their separation is controlled solely by attenuation,

$$|\Gamma_L|_{cc} \, |\Gamma_L|_{opt} = |\Gamma_S|^2, \frac{|\Gamma_L|_{cc}}{|\Gamma_L|_{opt}} = e^{4R}, \quad (S6-16)$$

which explains why the useful-power optimum typically lies in an undercoupled regime even when the absorptivity optimum is (or would be) critically coupled. Under $\Theta = \pi$, maximizing $\mathcal{T}_L$ in Eq. (S6-14) over $|\Gamma_L| \in [0,1]$ yields the analytic maximum useful load power

$$\mathcal{T}_{L,\max}(\Omega_{res}) = \frac{(1 - |\Gamma_S(\Omega_{res})|^2) e^{-2R(\Omega_{res})}}{1 - |\Gamma_S(\Omega_{res})|^2 e^{-4R(\Omega_{res})}}. \quad (S6-17)$$

By contrast, the critical-coupling requirement in Eq. (S6-15) becomes infeasible when $|\Gamma_L|_{cc} > 1$, i.e., for $R > R_{cc} = \frac{1}{2} \ln(1/|\Gamma_S|)$.

Finally, on the resonance class $\Theta = \pi$, Eq. (S6-10) admits the closed form

$$\eta_{use}(\Omega_{res}) = \frac{(1 - |\Gamma_L(\Omega_{res})|^2) e^{-2R(\Omega_{res})}}{1 - |\Gamma_L(\Omega_{res})|^2 e^{-4R(\Omega_{res})}}, \mathcal{T}_{int}(\Omega_{res}) = \mathcal{A}(\Omega_{res}) - \mathcal{T}_L(\Omega_{res}), (S6-18)$$

showing that, along the same constructive interference class, the division between load-delivered power and internally dissipated power is governed primarily by attenuation and the load reflectivity.

**S6.5 Intrinsic asymmetry $\xi$ and resonance-aware absorbed-power landscapes**

To encode intrinsic asymmetry, we use the resonance-aware absorbed-power objective (Supplementary Methods S2.6) written in terms of the boundary state $\bar{\Gamma}_g$:

$$\mathcal{P}_{abs}(\Omega;\xi) = e^{-4\xi R(\Omega)} - |\bar{\Gamma}_g(\Omega)|^2 \, e^{4(1-\xi)R(\Omega)}. \tag{S6-19}$$

Substituting Eq. (S6-13) into Eq. (S6-19) on the odd-$\pi$ resonance class ($\Theta = \pi$) yields the closed-form resonant landscape

$$\mathcal{P}_{abs}(\Omega_{res};\xi) = e^{-4\xi R(\Omega_{res})} - e^{4(1-\xi)R(\Omega_{res})} \frac{(|\Gamma_S(\Omega_{res})| - |\Gamma_L(\Omega_{res})|e^{-2R(\Omega_{res})})^2}{(1 - |\Gamma_S(\Omega_{res})| \, |\Gamma_L(\Omega_{res})|e^{-2R(\Omega_{res})})^2}. \tag{S6-20}$$

Its upper envelope is achieved at the critical-coupling match $|\Gamma_L|_{cc} = |\Gamma_S|e^{2R}$ when feasible, giving $\mathcal{P}_{abs,max}(\Omega_{res};\xi) = e^{-4\xi R(\Omega_{res})}$. When this requires $|\Gamma_L|_{cc} > 1$, the optimum becomes feasibility-limited and saturates on the passive boundary $|\Gamma_L| = 1$.

### S6.6 Coupling between $\mathcal{P}_{abs}$, $\mathcal{A}$, and $\mathcal{T}_L$ under asymmetry

Although $\mathcal{T}_L$ is defined independently of $\xi$, $\mathcal{P}_{abs}$ and $\mathcal{T}_L$ are coupled through the common state variable $|\bar{\Gamma}_g|^2$. Eliminating $|\bar{\Gamma}_g|^2$ between Eqs. (S6-7) and (S6-19) gives

$$\begin{aligned} |\bar{\Gamma}_g(\Omega)|^2 &= (e^{-4\xi R(\Omega)} - \mathcal{P}_{abs}(\Omega;\xi))e^{-4(1-\xi)R(\Omega)}, \\ \mathcal{A}(\Omega) &= 1 - e^{-4R(\Omega)} + \mathcal{P}_{abs}(\Omega;\xi)e^{-4(1-\xi)R(\Omega)}. \end{aligned} \tag{S6-21}$$

Using $\mathcal{T}_L = \eta_{use}\mathcal{A}$ (Eq. S6-10) yields the feedback-coupled mapping from the resonance-aware absorbed-power metric to useful delivered power:

$$\mathcal{T}_L(\Omega) = \eta_{use}(\Omega)[\,1 - e^{-4R(\Omega)} + \mathcal{P}_{abs}(\Omega;\xi)e^{-4(1-\xi)R(\Omega)}], \tag{S6-22}$$

which makes explicit how intrinsic asymmetry re-weights the conversion from absorbed-power landscapes to the absorptivity and useful-load-power diagnostics.

### S6.7 Ring-down time and quality-factor proxies from $\kappa$

Since $\kappa(\Omega)$ is the round-trip amplitude survival factor, it directly controls ring-down. Define the dimensionless amplitude and energy e-folding times (in units of round trips):

$$\tilde{\tau}_A(\Omega) = \frac{1}{-\ln \kappa(\Omega)}, \tilde{\tau}_E(\Omega) = \frac{1}{-\ln \kappa^2(\Omega)} = \frac{\tilde{\tau}_A(\Omega)}{2}. \qquad (S6-23)$$

Equivalently, the number of round trips required for the amplitude to decay to a ratio $A_Q \in (0,1)$ is

$$n_Q(\Omega) = \frac{\ln A_Q}{\ln \kappa(\Omega)}. \qquad (S6-24)$$

In the high-$Q$ limit $\kappa \to 1$, with $\delta = 1 - \kappa \ll 1$, the approximation $-\ln \kappa \simeq \delta$ gives

$$\tilde{\tau}_A(\Omega) \approx \frac{1}{1-\kappa(\Omega)}, \tilde{\tau}_E(\Omega) \approx \frac{1}{2[1-\kappa(\Omega)]} \quad (\kappa \to 1). \qquad (S6-25)$$

To connect to a conventional quality factor, introduce the physical round-trip time $t_{\rm rt}(\Omega)$ (e.g., $t_{\rm rt} = 2L/v_g$ in a single-mode guide). Then $\tau_E = t_{\rm rt}\tilde{\tau}_E$ and

$$Q(\Omega) = \frac{\omega \tau_E(\Omega)}{2} = \frac{\omega\, t_{\rm rt}(\Omega)}{4[-\ln \kappa(\Omega)]} \approx \frac{\omega\, t_{\rm rt}(\Omega)}{4[1-\kappa(\Omega)]} \quad (\kappa \to 1). \qquad (S6-26)$$

**S6.8 Phase detuning, interference classes, and resonance lines**

The feedback phase partitions universal interference classes: constructive build-up occurs near $\Theta \approx (2m+1)\pi$ (odd-$\pi$ resonance), while destructive suppression occurs near $\Theta \approx 2m\pi$. From Eq. (S6-5), loci of constant $\Lambda$ in the $(\kappa, \Theta)$ plane satisfy

$$|1 + \kappa e^{i\Theta}|^2 = \Lambda^{-2} \iff 1 + \kappa^2 + 2\kappa\cos\Theta = \Lambda^{-2}. \qquad (S6-27)$$

At fixed $\kappa$, the peak build-up is $\Lambda_{\max} = 1/|1-\kappa|$ at $\Theta = \pi$, and the minimum is $\Lambda_{\min} = 1/(1+\kappa)$ at $\Theta = 0$, explaining why similar lifetimes can correspond to strongly different resonance strengths depending on phase alignment.

**S6.9 Feedback-centered approximation used for compact diagnostics**

Near resonance ($\Theta \approx \pi$), using $|D(\Omega)|^2 = (1-\kappa(\Omega))^2 + \kappa(\Omega)(\Delta\Theta(\Omega))^2 + \mathcal{O}(\Delta\Theta^4)$ and expanding for small $\Delta\Theta$ gives a compact feedback-centered approximation useful for visualization:

$$\frac{1}{|D(\Omega)|^2} \approx \frac{1}{(1-\kappa(\Omega))^2 + \kappa(\Omega)(\Delta\Theta(\Omega))^2} \quad (\Delta\Theta \to 0). \qquad (S6-28)$$

**Supplementary Methods S7 | Proofs and extensions for the four fundamental Laws**

**S7.0 Overview**

Here we provide a compact operator-level derivation of the four fundamental laws for power absorption and emission in finite, linear, time-invariant, passive waveguide systems, and then connect the modal statements to the one-dimensional Cai–Smith geometry used in the main text. The derivation follows three steps: (i) formulate absorption and emission as quadratic forms of positive-semidefinite operators acting on input and output modal spaces; (ii) use fluctuation–dissipation to relate the emission operator to the absorption operator (unitary equivalence up to a scalar normalization), which fixes the nonzero eigenvalue spectra; and (iii) invoke reciprocity to establish mode pairing for directional equivalence.

**S7.1 Input–output scattering operator and the absorption operator**

Consider a passive system at fixed angular frequency $\omega$ connecting an input modal Hilbert space $\mathcal{H}_{\text{in}}$ to an output modal Hilbert space $\mathcal{H}_{\text{out}}$. Let $\mathbf{S}: \mathcal{H}_{\text{in}} \to \mathcal{H}_{\text{out}}$ be the (frequency-dependent) scattering operator mapping normalized incoming modal amplitudes to outgoing modal amplitudes. For a normalized input mode $|i\rangle \in \mathcal{H}_{\text{in}}$ with unit incident power, the absorbed power fraction (modal absorptivity) is

$$\alpha_i = 1 - \|\mathbf{S}|i\rangle\|^2 = \langle i|\mathbf{A_b}|i\rangle, \quad \mathbf{A_b} = \mathbf{I}_{\text{in}} - \mathbf{S}^\dagger \mathbf{S}. \qquad (S7-1)$$

Passivity implies $\mathbf{S}^\dagger \mathbf{S} \leq \mathbf{I}_{\text{in}}$ (in the operator sense), hence

$$0 \leq \mathbf{A_b} \leq \mathbf{I}_{\text{in}} \text{ (positive semidefinite and bounded).} \qquad (S7-2)$$

**S7.2 Emission operator and fluctuation–dissipation equivalence**

Define the modal emissivity into a normalized output mode $|o\rangle \in \mathcal{H}_{\text{out}}$ as

$$\epsilon_o = \langle o|\mathbf{E_m}|o\rangle, \quad (S7-3)$$

where $\mathbf{E_m}$ is the (frequency-resolved) emission operator acting on $\mathcal{H}_{\text{out}}$. For thermal (or more general fluctuational) emission in linear media, the fluctuation–dissipation theorem implies that emission is determined by the same dissipative degrees of freedom that set absorption. In the normalization adopted in Results (unit incident power for absorption versus unit emitted spectral density for emission), this reduces to the statement that $\mathbf{A_b}$ and $\mathbf{E_m}$ share the same nonzero eigenvalues; equivalently, there exists a partial isometry (unitary on the support) $\mathbf{U}: \text{supp}(\mathbf{A_b}) \to \text{supp}(\mathbf{E_m})$ such that $\mathbf{E_m} = \mathbf{UAU^\dagger}$ on the nonzero-eigenvalue subspace suggests

$$\text{spec}_+(\mathbf{E_m}) = \text{spec}_+(\mathbf{A_b}), \quad (S7-4)$$

where $\text{spec}_+$ denotes the multiset of nonzero eigenvalues. Laws 1–3 follow directly from Eq. (S7-4); Law 4 further uses reciprocity to identify the relevant mode pairing.

**S7.3 Singular-value decomposition and intrinsic (Eigenchannel) basis**

Let the singular-value decomposition of $\mathbf{S}$ be

$$\mathbf{S} = \sum_p \sigma_p |u_p\rangle\langle v_p|, \quad 0 \leq \sigma_p \leq 1, \quad (S7-5)$$

where $\{|v_p\rangle\}$ is an orthonormal set in $\mathcal{H}_{\text{in}}$ (right singular vectors) and $\{|u_p\rangle\}$ is an orthonormal set in $\mathcal{H}_{\text{out}}$ (left singular vectors). Then

$$\mathbf{S^\dagger S} = \sum_p \sigma_p^2 |v_p\rangle\langle v_p|, \quad \mathbf{A_b} = \mathbf{I_{in}} - \mathbf{S^\dagger S} = \sum_p (1-\sigma_p^2) |v_p\rangle\langle v_p| + \mathbf{A_{b_\perp}}, \quad (S7-6)$$

where $\mathbf{A_{b_\perp}}$ acts on the null space of $\mathbf{S}$ (perfect absorption channels if present). Thus the absorption eigenchannels are precisely $\{|v_p\rangle\}$ with eigenvalues

$$\mathbf{A_b}|v_p\rangle = \alpha_p|v_p\rangle, \quad \alpha_p = 1 - \sigma_p^2, \quad 0 \leq \alpha_p \leq 1. \quad (S7-7)$$

By Eq. (S7-4), the emission operator $E_m$ has the same nonzero eigenvalues $\epsilon_p = \alpha_p$ with corresponding emission eigenchannels $\{|u_p\rangle\}$ (up to unitary mixing inside degenerate subspaces).

**S7.4 Law 1: Modal power-absorption–emission equivalence (Eigenchannel equality)**

In the matched eigenchannel basis (denoted $M_p$ in Results), the channel-resolved absorptivity equals emissivity:

$$\alpha_{M_p} = \epsilon_{M_p} = \alpha_p = \epsilon_p = 1 - \sigma_p^2, p = 1,2,\dots \qquad (S7-8)$$

which is Law 1. In the Cai–Smith representation, each eigenchannel behaves as an effective one-dimensional guide with its own boundary-composed state $\Gamma_{g,p}$ on the disk; exchanging absorption with emission corresponds to reversing the net flux direction ($\mathcal{S} \to -\mathcal{S}$) at fixed $|\Gamma_{g,p}|$, leaving the invariant shell $\mathcal{U}^2 - \mathcal{S}^2 = |\Gamma_{g,p}|^2$ unchanged and therefore enforcing $\alpha_{M_p} = \epsilon_{M_p}$ channel by channel.

**S7.5 Law 2: Equal power-profile equivalence (mode mixtures with identical internal weights)**

Let $|i\rangle \in \mathcal{H}_{\text{in}}$ be any normalized input, expanded in the absorption eigenbasis: $|i\rangle = \sum_p c_p |v_p\rangle + |i_\perp\rangle$ with $\langle v_p | i_\perp \rangle = 0$. Using Eq. (S7-1) and Eq. (S7-7),

$$\alpha_i = \langle i|\mathbf{A_b}|i\rangle = \sum_p |c_p|^2 \alpha_p + \langle i_\perp|\mathbf{A_{b_\perp}}|i_\perp\rangle. \qquad (S7-9)$$

Now consider a normalized output mode $|o\rangle \in \mathcal{H}_{\text{out}}$ expanded in the corresponding emission eigenbasis $|o\rangle = \sum_p d_p |u_p\rangle + |o_\perp\rangle$. By the eigenvalue matching implied by Eq. (S7-4), we have

$$\epsilon_o = \langle o|\mathrm{E_m}|o\rangle = \sum_p |d_p|^2 \epsilon_p + \langle o_\perp|\mathbf{E_{m_\perp}}|o_\perp\rangle, \epsilon_p = \alpha_p. \qquad (S7-10)$$

If $|i\rangle$ and $|o\rangle$ induce identical internal power distributions in the system, then they excite the same eigenchannels with the same weights, i.e. $|d_p|^2 = |c_p|^2$ (and the same corresponding null-space weight when present). Substituting into Eqs. (S7-9)–(S7-10) yields

$$\alpha_i = \epsilon_o, \qquad (S7-11)$$

which is Law 2. In the Cai–Smith picture, this condition means that the two modes sample the same distribution of shell states (the same internal partition of $(\mathcal{U}, \mathcal{S}, |\Gamma_g|)$), hence must yield identical integrated absorption/emission outcomes even though they may correspond to different external decompositions.

**S7.6 Law 3: Integrated power-absorption–emission balance (trace invariance)**

Let $\{|i_n\rangle\}_{n=1}^{N}$ be any complete orthonormal basis of $\mathcal{H}_{\text{in}}$ and $\{|o_m\rangle\}_{m=1}^{M}$ any complete orthonormal basis of $\mathcal{H}_{\text{out}}$. Summing Eq. (S7-1) over the input basis gives

$$\sum_{n=1}^{N} \alpha_{i_n} = \sum_{n=1}^{N} \langle i_n | \mathbf{A_b} | i_n \rangle = \mathbf{Tr}(\mathbf{A_b}). \qquad (S7-12)$$

Similarly, summing Eq. (S7-3) over the output basis yields

$$\sum_{m=1}^{M} \epsilon_{o_m} = \sum_{m=1}^{M} \langle o_m | \mathbf{E_m} | o_m \rangle = \mathbf{Tr}(\mathbf{E_m}). \qquad (S7-13)$$

Because $\mathbf{A_b}$ and $\mathbf{E_m}$ share the same nonzero eigenvalues (Eq. S7-4) and are trace-equal under the matched normalization, $\mathbf{Tr}(\mathbf{A_b}) = \mathbf{Tr}(\mathbf{E_m})$, hence

$$\sum_{n} \alpha_{i_n} = \sum_{m} \epsilon_{o_m}, \qquad (S7-14)$$

which is Law 3. This statement is basis-independent: changing modal decompositions redistributes weight across eigenchannels (and across Cai–Smith disk states) but preserves the total absorption/emission budget fixed by passivity.

**S7.7 Law 4: Reciprocal waveguide directionality (phase-conjugate pairing)**

Assume the system is reciprocal (time-reversal symmetric in the absence of bias). In an appropriate modal normalization, reciprocity implies a symmetry constraint on the scattering operator (matrix symmetry in a reciprocal basis), which enforces a pairing between absorption and emission channels under phase conjugation. Operationally, this implies that the emission into the phase-conjugated (backward) version of an input mode equals the absorption of the original input mode:

$$\alpha_i = \epsilon_{i^*}. \qquad (S7-15)$$

One way to see this is through the SVD in Eq. (S7-5): reciprocity allows the left and right singular vectors to be chosen as phase-conjugate pairs (up to phase factors) within each singular subspace, so that the matched emission channel for an input excitation corresponds to its phase-conjugated direction-reversal partner. In the Cai–Smith representation, phase conjugation maps $\Gamma_g \mapsto \Gamma_g^*$ (mirror about the real axis) and reverses the net flux direction ($\mathcal{S} \to -\mathcal{S}$) while preserving $|\Gamma_g|$, hence preserving the shell and enforcing Eq. (S7-15).

**S7.8 Connection to the one-dimensional waveguide reduction**

In the single-mode waveguide reduction used in the main text, each intrinsic eigenchannel $p$ can be represented as an effective one-dimensional guide characterized by a boundary-composed generalized reflection state $\Gamma_{g,p}(\Omega)$ and the associated invariant shell $\mathcal{U}^2 - \mathcal{S}^2 = |\Gamma_{g,p}|^2$. Laws 1–4 therefore act as channel-resolved constraints on how any macroscopic absorption/emission/transfer optimum (set by boundary control and the feedback proximity encoded by $F(\Omega)$) must be partitioned across independent channels: (i) equality per eigenchannel, (ii) equality for any two modes with identical internal power profiles, (iii) a basis-independent sum rule, and (iv) directional pairing under reciprocity. Extended Data Fig. 8 provides numerical verification and visualization of each law in the modal operator language.

# Supplementary Section 8 | Waveguide-invariant mapping reveals optical CPA-EP reconstruction

## 8.0 Overview

This section describes a fully reproducible workflow to reconstruct coherent perfect absorption at exceptional points (CPA-EPs) from a two-resonator temporal coupled-mode theory (TCMT) scattering model, and to map the two-port scattering response onto an equivalent single-loop waveguide feedback representation. The workflow consists of three steps: (i) compute the TCMT two-port scattering matrix $\mathbf{S}(\omega)$ and identify CPA/CPA-EP conditions via singular-value decomposition (SVD) eigenchannels; (ii) evaluate two experimentally relevant probing protocols, i.e., fixed coherent input versus SVD-probe (eigenchannel tracking), to distinguish coherent-mismatch leakage from intrinsic absorbing-channel collapse; and (iii) reconstruct effective waveguide parameters $\{z(\omega), K_{\text{eff}}(\omega), \Gamma_{S,L}^{\text{eff}}(\omega)\}$ that reproduce $\mathbf{S}(\omega)$ with numerical precision, thereby interpreting the CPA-EP as a single-loop feedback geometry. This section reproduces Extended Data Figs. 11-13.

## 8.1 Two-resonator TCMT scattering model and CPA-EP constraints

We consider a reciprocal two-port, two-resonator TCMT model with resonance frequencies $\omega_{1,2}$, intrinsic loss rates $\gamma_{1,2}$, external coupling rates $\gamma_{c1,c2}$ to the ports, and inter-resonator coupling $\kappa_{\text{TCMT}}$. Under the time-harmonic convention $e^{+i\omega t}$, the $2 \times 2$ scattering matrix $\mathbf{S}_{\text{TCMT}}(\omega)$ is evaluated on a real-frequency grid $\omega = \omega_0 + \Delta\omega$. Parameter sets are chosen to realize either nongeneric (symmetric) or generic (asymmetric) CPA-EPs, while satisfying real-axis CPA feasibility and absorbing-EP thresholds. For equal resonance frequencies $\omega_1 = \omega_2 = \omega_0$, the CPA-EP conditions are: (i) real-zero balance $\gamma_{c1} + \gamma_{c2} = \gamma_1 + \gamma_2$ and (ii) absorbing-EP coupling threshold $\kappa_{\text{TCMT}} = |\gamma_1 - \gamma_{c1}|/2$. These yield coalesced scattering zeros on the real axis and a defective absorbing channel. At $\Delta\omega = 0$, the TCMT CPA eigenvector satisfies $|\mathbf{a_1}/\mathbf{a_2}| = \sqrt{\gamma_{c1}/\gamma_{c2}}$ and $\arg(\mathbf{a_1}) - \arg(\mathbf{a_2}) = \pm\pi/2$.

### 8.2 SVD eigenchannels and operational CPA criterion

For any two-port scatterer with scattering matrix $\mathbf{S}(\omega)$, we compute the SVD $\mathbf{S}(\omega) = \mathbf{U}(\omega)\mathbf{\Sigma}(\omega)\mathbf{V}^\dagger(\omega)$ with singular values $\sigma_{\max}(\omega) \geq \sigma_{\min}(\omega) \geq 0$ and right singular vectors $\mathbf{v_{\max}}(\omega), \mathbf{v_{\min}}(\omega)$. For a unit-norm coherent input $a$ (incident power $P_{\text{in}} = \|\mathbf{a}\|^2 = 1$), the output is $\mathbf{b} = \mathbf{S}(\omega)\mathbf{a}$, total output power $P_{\text{out}}(\omega) = \|\mathbf{b}\|^2$, and port-resolved outputs $P_j(\omega) = |\mathbf{b_j}|^2$ with $P_{\text{tot}} = \sum_j P_j$. Maximum absorption at frequency $\omega$ is achieved by minimizing output over all unit-norm inputs:

$$P^*_{\text{out}}(\omega) = \min_{\|a\|=1} \|\mathbf{S}(\omega)\mathbf{a}\|^2 = \sigma^2_{\min}(\omega), \mathbf{a}^*(\omega) = \mathbf{v_{\min}}(\omega). \qquad (S8-1)$$

CPA at real frequency $\omega_0$ thus occurs if and only if $\sigma_{\min}(\omega_0) = 0$, requiring coherent excitation $\mathbf{a_{CPA}} = \mathbf{v_{\min}}(\omega_0)$. We define the channel "mass" as output norm per unit incident power,

$$\rho(\omega;\mathbf{a}) \equiv \sqrt{\frac{P_{\text{out}}(\omega)}{P_{\text{in}}}} = \|\mathbf{S}(\omega)\mathbf{a}\|, \rho_{\text{svd}}(\omega) \equiv \min_{\|a\|=1} \rho(\omega;\mathbf{a}) = \sigma_{\min}(\omega), \quad (S8-2)$$

so CPA corresponds to $\rho_{\text{svd}}(\omega_0) = 0$.

### 8.3 Fixed coherent setting versus SVD-probe protocol

To isolate protocol dependence near CPA-EPs, we implement two complementary probing schemes: (i) Fixed coherent setting. We compute the optimal absorbing input at $\omega_0$ as $\mathbf{a_0} = \mathbf{v_{\min}}(\omega_0)$ and hold it fixed while scanning detuning:

$$P^{\text{fixed}}_{\text{out}}(\Delta\omega) = \|\mathbf{S}(\omega_0 + \Delta\omega)\mathbf{a_0}\|^2, \rho_{\text{fixed}}(\Delta\omega) = \|\mathbf{S}(\omega_0 + \Delta\omega)\mathbf{a_0}\|. \qquad (S8-3)$$

(ii) SVD probe (eigenchannel tracking). At each detuning we re-optimize the input as $\mathbf{a}^*(\omega) = \mathbf{v_{\min}}(\omega)$, yielding the eigenchannel bound:

$$P_{\text{out}}^{\text{svd}}(\Delta\omega) = \min_{\|a\|=1} \|\mathbf{S}(\omega_0 + \Delta\omega)\mathbf{a}\|^2 = \sigma_{\min}^2(\omega_0 + \Delta\omega), \rho_{\text{svd}}(\Delta\omega) = \sigma_{\min}(\omega_0 + \Delta\omega). \quad (S8-4)$$

Extended Data Fig. 11a-f curves are computed from these definitions, including port-resolved outputs $P_j(\Delta\omega) = |[\mathbf{S}(\omega_0 + \Delta\omega)\mathbf{a}(\Delta\omega)]_j|^2$ with $\mathbf{a}(\Delta\omega) = \mathbf{a_0}$ for fixed setting and $\mathbf{a}(\Delta\omega) = \mathbf{a}^*(\omega_0 + \Delta\omega)$ for SVD-probe.

### 8.4 Near-zero scaling extraction and universality classes

We quantify absorbing-channel collapse by fitting small-detuning scaling of $P_{\text{out}}(\Delta\omega)$ under each protocol. Within a prescribed fitting window $0 < |\Delta\omega| \leq \Delta\omega_{\text{fit}}$, we perform linear regression of $\log P_{\text{out}}$ versus $\log |\Delta\omega|$ to extract the local exponent $p$ in $P_{\text{out}} \propto |\Delta\omega|^p$ (Extended Data Fig. 11c). The fixed setting probes coherent-mismatch leakage from detuning-induced drift of the optimal singular vector, while SVD-probe isolates intrinsic collapse of the minimum-output eigenchannel. In nongeneric (symmetric) CPA-EP configurations, fixed-setting spectra exhibit an apparent quadratic floor $p \simeq 2$ from first-order coherent mismatch, whereas SVD-probe recovers higher-order suppression approaching quartic scaling. In generic (asymmetric) CPA-EP configurations, the defective absorbing channel enforces quartic universality under both protocols (Extended Data Fig. 11a-c).

### 8.5 Cai–Smith disk trajectories for driven CPA-EP channels

To represent the driven coherent channel within the complex unit-disk geometry used throughout our waveguide-invariant formulation, we compute a scalar complex proxy:

$$\Gamma(\Delta\omega) = \mathbf{a}^\dagger(\Delta\omega)\, \mathbf{S}(\omega_0 + \Delta\omega)\, \mathbf{a}(\Delta\omega), \quad (S8-5)$$

with $\mathbf{a}(\Delta\omega) = \mathbf{a_0}$ for fixed setting and $\mathbf{a}(\Delta\omega) = \mathbf{a}^*(\omega_0 + \Delta\omega)$ for SVD-probe. For passive two-port scattering with $\|\mathbf{S}(\omega)\|_2 = \sigma_{\max}(\omega) \leq 1$, we have $|\Gamma(\Delta\omega)| \leq 1$ for unit-norm inputs. Plotting $\Gamma(\Delta\omega)$ in the complex plane with the unit-disk boundary yields Cai–Smith trajectories that characterize approach to the disk center and protocol-dependent detuning drift (Extended Data Fig. 11g, h).

**8.6 Frequency-domain inversion to waveguide feedback form**

To express CPA-EP optics as single-loop feedback, we convert $\mathbf{S}(\omega)$ to an equivalent waveguide-form factorization at each frequency (Extended Data Fig. 12). With $\mathbf{r}_1(\omega) = \mathbf{S}_{11}(\omega)$, $\mathbf{r}_2(\omega) = \mathbf{S}_{22}(\omega)$, and $\mathbf{t}(\omega) = \mathbf{S}_{12}(\omega) = \mathbf{S}_{21}(\omega)$, we seek frequency-dependent effective boundaries $\Gamma_S^{\text{eff}}(\omega)$, $\Gamma_L^{\text{eff}}(\omega)$ and propagation constant $K_{\text{eff}}(\omega)$ satisfying

$$\mathbf{r}_1(\omega) = \frac{\Gamma_S^{\text{eff}}(\omega) + \Gamma_L^{\text{eff}}(\omega)z(\omega)}{D(\omega)}, \mathbf{r}_2(\omega) = \frac{\Gamma_L^{\text{eff}}(\omega) + \Gamma_S^{\text{eff}}(\omega)z(\omega)}{D(\omega)}, \mathbf{t}(\omega) = \frac{\sqrt{z(\omega)}\,\tau_S^{\text{eff}}\tau_L^{\text{eff}}}{D(\omega)}, (S8-6)$$

where $z(\omega) = e^{-2K_{\text{eff}}(\omega)}$ and $D(\omega) = 1 + \Gamma_S^{\text{eff}}(\omega)\Gamma_L^{\text{eff}}(\omega)z(\omega)$. We set $\tau_S^{\text{eff}} = \sqrt{\gamma_{c1}}$, $\tau_L^{\text{eff}} = \sqrt{\gamma_{c2}}$, and adopt $\sqrt{z} = e^{-K_{\text{eff}}}$. For trial $z(\omega)$, we compute

$$D(\omega) = \frac{\sqrt{z(\omega)}\,\tau_S^{\text{eff}}\tau_L^{\text{eff}}}{t(\omega)}, \qquad (S8-7)$$

and obtain boundaries via back-substitution:

$$\Gamma_S^{\text{eff}}(\omega) = \frac{D(\omega)(\mathbf{r}_1(\omega) - z(\omega)\mathbf{r}_2(\omega))}{1 - z^2(\omega)}, \Gamma_L^{\text{eff}}(\omega) = \frac{D(\omega)(\mathbf{r}_2(\omega) - z(\omega)\mathbf{r}_1(\omega))}{1 - z^2(\omega)}. (S8-8)$$

The loop-closure condition $D(\omega) = 1 + \Gamma_S^{\text{eff}}(\omega)\Gamma_L^{\text{eff}}(\omega)z(\omega)$ provides a complex scalar equation for $z(\omega)$, solved frequency-by-frequency. We parameterize $z(\omega) = e^{-u(\omega)+iv(\omega)}$ with $u(\omega) \geq 0$, optimize $(u,v)$ at each $\omega$ using warm-start initialization, and compute $K_{\text{eff}}(\omega) = -\frac{1}{2}\log z(\omega)$ with phase unwrapping of $v(\omega)$. The reconstructed waveguide-form matrix $\hat{\mathbf{S}}_{\mathbf{WG}}(\omega)$ agrees elementwise with $\mathbf{S}(\omega)$, and their singular values coincide across the band (Extended Data Fig. 12a). We report $\log_{10}|z(\omega)|$, $\arg z(\omega)$, $\text{Re}(K_{\text{eff}})$, $\text{Im}(K_{\text{eff}})$, $|\Gamma_{S,L}^{\text{eff}}|$, and $\arg(\Gamma_{S,L}^{\text{eff}})$ (Extended Data Fig. 12b-e). For feedback diagnostics we compute

$$\kappa_{\text{eff}}(\omega) = |\Gamma_S^{\text{eff}}(\omega)\Gamma_L^{\text{eff}}(\omega)|\, e^{-2\text{Re}[K_{\text{eff}}(\omega)]}, \Delta\Theta(\omega) = \arg\left(\Gamma_S^{\text{eff}}\Gamma_L^{\text{eff}}\right) - 2\,\text{Im}[K_{\text{eff}}], (S8-9)$$

representing effective loop-factor magnitude and phase-closure (Extended Data Fig. 12f).

**8.7 Local coherent-input strategies around CPA frequency**

To connect protocol dependence to experimental input control, we compare (i) fixed-at-$\Omega_0$ and (ii) per-frequency optimal excitations near CPA (Extended Data Fig. 13). With normalized frequency $\Omega = \omega/\omega_{\text{ref}}$ and CPA point $\Omega_0$, the fixed strategy uses $\mathbf{a}_{\text{fixed}} = \mathbf{v}_{\text{min}}(\Omega_0)$ for all $\Omega$; the per-frequency strategy uses $\mathbf{a}^*(\Omega) = \mathbf{v}_{\text{min}}(\Omega)$. We present: (a) singular-value ratio $\sigma_{\text{min}}/\sigma_{\text{max}}$ and effective output norm $\|\mathbf{S}(\Omega)\mathbf{a}_{\text{fixed}}\|$; (b) detuning-dependent $|a_1/a_2|$ and $\Delta\psi = \arg(a_1) - \arg(a_2)$ for both strategies; (c) port-resolved $P_1(\Omega)$, $P_2(\Omega)$ and $P_{\text{tot}}(\Omega)$; and (d) near-zero scaling of $P_{\text{tot}}$ versus $\Delta\Omega = \Omega - \Omega_0$ from log-log regression. These diagnostics show how a frozen coherent state reduces an intrinsically higher-order CPA-EP collapse to an apparent lower-order floor via coherent mismatch, while per-frequency optimization recovers the eigenchannel-defined universality class.

**Supplementary Methods S9 | Electrochemical polarization mapping and parameter-estimation pipeline**

**S9.0 Overview**

This section specifies the reproducible pipeline used to map steady-state electrochemical polarization data to the waveguide-invariant state space and to estimate intrinsic-asymmetry parameters from experiments. The workflow consists of: (i) dataset-specific sign/axis unification into a common driving coordinate $\eta_{\text{eff}}$ and response $j$; (ii) evaluation of an invariant-compatible two-channel kinetic form under Ohmic drop through an implicit solver; (iii) parameter estimation via staged, bounded nonlinear least-squares with model selection; and (iv) post-fit construction of polarization-derived state diagnostics, including a complex generalized reflection factor $\Gamma_g(\eta_{\text{eff}})$, storage/transfer proxies, and density-optimal operating points.

**S9.1 Data preprocessing and unified sign convention**

Polarization measurements provide pairs $\{(\eta_m, j_m)\}_{m=1}^{M}$ (overpotential and current density, with sign depending on cathodic/anodic convention). We include an effective series Ohmic drop $R_\Omega \geq 0$ and define the effective interfacial overpotential as

$$\eta_{\text{eff}} = \eta - R_\Omega j, \qquad (S9-1)$$

where $R_\Omega$ is area-normalized in units consistent with $j$. We model the total current as a sum of two orthogonal channels $p \in \{A, B\}$,

$$j(\eta) = \sum_{p \in \{A,B\}} j_p(\eta_{\text{eff}}), \qquad (S9-2)$$

where each channel follows an intrinsically asymmetric Butler-Volmer kinetics with cathodic/anodic branch exponents $(\alpha_p, \beta_p)$,

$$x_p(\eta_{\text{eff}}) = j_p^*[\exp(\alpha_p \eta_{\text{eff}}) - \exp(\beta_p \eta_{\text{eff}})], \qquad (S9-3)$$

together with a transport-limited saturation applied to the net channel current:

$$j_p(\eta_{\text{eff}}) = \frac{x_p(\eta_{\text{eff}})}{\sqrt{1 + \left(\frac{x_p(\eta_{\text{eff}})}{j_{\text{lim},p}}\right)^2}}. \qquad (S9-4)$$

We parameterize $(a_p, b_p)$ using the intrinsic-asymmetry variables $(\xi_p, \lambda_p)$ as $a_p = \xi_p \lambda_p f$, $b_p = -(1-\xi_p)\lambda_p f$, with $F = \mathcal{F}/(R_g T)$ (where $\mathcal{F}$ is the Faraday constant (96485 C/mol), $R_g$ is the universal gas constant (8.314 J/mol·K), and $T$ is the temperature), enforcing physical admissibility $j_p^* > 0$, $\xi_p \in (0,1)$, $\lambda_p \geq 0$, $j_{\text{lim},p} >$

0, and $R_\Omega \geq 0$.

## S9.2 Implicit invariant-mapped kinetics with Ohmic drop and transport saturation

In dimensionless form, defining $\tilde{j}_p = j_p/j_{\text{lim},p}$ and $\tilde{x}_p = x_p/j_{\text{lim},p}$, the expression (S9-4) simplifies to:

$$\tilde{j}_p = \frac{\tilde{x}_p}{\sqrt{1+\tilde{x}_p^2}}, \qquad (S9-5)$$

where

$$\tilde{x}_p = \frac{j_p^*}{j_{\text{lim},p}}\left[\exp(\xi_p\tilde{\eta}) - \exp(-(1-\xi_p)\tilde{\eta})\right], \tilde{\eta} = \lambda_p f \eta_{\text{eff}}. \qquad (S9-6)$$

This hyperbolic mapping ensures passivity ($|\Gamma_g| \leq 1$) and reveals distinct regimes as a function of $|\tilde{\eta}|$ and $\xi_p$. Because $\tilde{\eta}$ depends on $\tilde{J}$ through Eq. (S9-5), the model evaluation at each $\tilde{\eta}$ requires solving the scalar implicit equation

$$\tilde{J} = \sum_{p\in\{A,B\}} \tilde{J}_p(\tilde{\eta} - R_\Omega\tilde{J}). \qquad (S9-7)$$

## S9.3 Implicit model evaluation via safeguarded one-dimensional root finding

For each $\tilde{\eta}$ and parameter set $\Theta$, we solve Eq. (S9-6) by a robust 1D method with warm starts along a monotonic sweep in $\tilde{\eta}$: the initial guess for $\tilde{\eta}_m$ is $\tilde{J}_{\text{model}}(\tilde{\eta}_{m-1};\Theta)$; the first point is initialized using the explicit $R_\Omega = 0$ approximation. Each step uses safeguarded Newton updates with bracketing and backtracking; if the Newton step becomes ill-conditioned or exits the bracket, we fall back to bisection until convergence. Solver tolerances and iteration caps are fixed across all conditions before fitting.

## S9.4 Parameter-estimation objective and staged fitting

Parameters are estimated by minimizing an asinh-transformed least-squares objective to balance low- and high-current regimes:

$$r_m(\Xi) = \operatorname{asinh}\left(\frac{\tilde{J}_{\text{model}}(\tilde{\eta}_m; \Xi)}{J_{\text{ref}}}\right) - \operatorname{asinh}\left(\frac{\tilde{J}_m}{J_{\text{ref}}}\right), \quad \min_{\Xi} \sum_{m=1}^{M} w_m \, r_m(\Xi)^2, \quad (S9-8)$$

with fixed $J_{\text{ref}}$ and within-dataset weights $\{w_m\}$. Fitting proceeds in two stages: Stage 1 fits the explicit $R_\Omega = 0$ form on a reduced grid to obtain stable initial values and a residual-informed two-channel split; Stage 2 refines all free parameters using the full implicit solver (S9.3) on the full grid under bounded constraints. Weak regularization terms may be included to prevent single-channel collapse and to discourage unrealistically large $R_\Omega$, without imposing hard caps.

**S9.5 Model selection (single-mode vs two-mode)**

For each condition we fit three candidate models: single-mode A, single-mode B, and the two-mode orthogonal sum in Eq. (S9-3). The primary selection criterion is the Akaike information criterion (AIC) computed from the Stage 2 residual sum of squares and the number of free parameters; the minimum-AIC model is reported and used for subsequent diagnostics.

**S9.6 Parameter bounds and reporting**

All optimizations enforce physical admissibility: $j_p^* > 0$, $\xi_p \in (0,1)$, $\lambda_p \geq 0$, $j_{\text{lim},p} > 0$, and $R_\Omega \geq 0$. Reported parameter sets include best-fit values for each condition, the selected model identity (A, B, or A+B), and the corresponding $(j_{\text{lim},A}, j_{\text{lim},B})$.

**S9.7 From fitted kinetics to invariant diagnostics and objective markers**

Given fitted parameters, we evaluate $\tilde{J}_{\text{tot}}(\eta_{\text{eff}})$ along the experimental grid and compute polarization-derived directional fluxes using the unsaturated exponential branches:

$$J_{\text{ox},p} = j_p^* \exp(a_p \eta_{\text{eff}}), J_{\text{red},p} = j_p^* \exp(b_p \eta_{\text{eff}}), \quad (S9-9)$$

and totals $J_{\text{ox}}^{\text{tot}} = \sum_p J_{\text{ox},p}$, $J_{\text{red}}^{\text{tot}} = \sum_p J_{\text{red},p}$.

We then form exchange and transfer proxies

$$\mathcal{U} = J_{\text{ox}}^{\text{tot}} + J_{\text{red}}^{\text{tot}}, \mathcal{S} = J_{\text{ox}}^{\text{tot}} - J_{\text{red}}^{\text{tot}}, \beta_w = \frac{\mathcal{S}}{\mathcal{U}}, \quad (S9-10)$$

together with the branch-balance (coherence) metric

$$\beta_w = \frac{J_{\text{ox}}^{\text{tot}} - J_{\text{red}}^{\text{tot}}}{J_{\text{red}}^{\text{tot}} + J_{\text{ox}}^{\text{tot}}}, \varphi = \arccos(\beta_w) \in [0, \pi], \quad (S9-11)$$

and an amplitude ratio on the square-root scale,

$$\rho = \sqrt{\frac{\min(J_{\text{red}}^{\text{tot}}, J_{\text{ox}}^{\text{tot}})}{\max(J_{\text{red}}^{\text{tot}}, J_{\text{ox}}^{\text{tot}})}} = \sqrt{\frac{1 - |\beta_w^2|}{1 + |\beta_w^2|}} \in [0,1]. \quad (S9-12)$$

The polarization-derived generalized complex reflection factor is finally

$$\Gamma_g(\eta_{\text{eff}}) = \rho(\eta_{\text{eff}}) e^{i\varphi(\eta_{\text{eff}})}, |\Gamma_g| \leq 1 \text{ by construction.} \quad (S9-13)$$

We also compute a retention-like proxy

$$\tilde{\tau}_{\text{ec}}(\eta_{\text{eff}}) = \frac{1}{-\ln \rho(\eta_{\text{eff}})}, \quad (S9-14)$$

and a standing-wave/storage intensity

$$I_{\text{store}}(\eta_{\text{eff}}) = \frac{2\sqrt{J_{\text{ox}}^{\text{tot}} J_{\text{red}}^{\text{tot}}}}{J_{\text{ox}}^{\text{tot}} + J_{\text{red}}^{\text{tot}}} \in [0,1]. \quad (S9-15)$$

**S9.8 Useful-output density and density-optimal operating point (HER)**

For cathodic reduction (HER), we define a useful output proxy from the fitted total current and the inferred feedback magnitude:

$$\Pi_{\text{use}}(\eta_{\text{eff}}) = \max(0, -\tilde{J}_{\text{tot}}(\eta_{\text{eff}}))[1 - \rho(\eta_{\text{eff}})^2], \quad (S9-16)$$

and the (regularized) useful-output density

$$\Pi_{\text{dens}}(\eta_{\text{eff}}) = \frac{\Pi_{\text{use}}(\eta_{\text{eff}})}{|\eta_{\text{eff}}| + \eta_0}, \eta_0 > 0, \quad (S9-17)$$

which avoids the formal divergence at $\eta_{\text{eff}} \to 0$ and is equivalent to imposing a small numerical cutoff in practice. The density-optimal operating point is

$$\eta_{\text{eff}}^* = \arg \max_{\eta_{\text{eff}} < 0} \Pi_{\text{dens}}(\eta_{\text{eff}}), \quad (S9-18)$$

together with $\Pi_{\text{use}}^* = \Pi_{\text{use}}(\eta_{\text{eff}}^*)$ and $\Pi_{\text{dens}}^* = \Pi_{\text{dens}}(\eta_{\text{eff}}^*)$.